\documentstyle[12pt,epsf]{article}
\newcommand{\diag}{\mbox{diag}}
\newcommand{\tr}{\mbox{Tr}}
\newcommand{\re}{\mbox{Re}}
\newcommand{\beq}{\begin{eqnarray}}    
\newcommand{\eeq}{\end{eqnarray}}    

\newcommand{\np}{Nucl.\ Phys.\ } 
\newcommand{\npp}{Nucl.\ Phys.\ [Proc.\ Suppl.]}   
\newcommand{\pl}{Phys.\ Lett.\ }    
\newcommand{\pr}{Phys.\ Rev.\ }
\newcommand{\prl}{Phys.\ Rev.\ Lett.\ }
\begin{document}    

\begin{titlepage}
\begin{flushright}
CERN-TH.7413/94\\
WUB 94-24~~~~~~~~~~\\
August 1994~~~~~~~~~\\
\end{flushright}
\begin{center}
{\bf
\Large
Observing  Long  Colour Flux Tubes \\
in SU(2) Lattice Gauge Theory\footnote{Work
supported in part by DFG grant Schi 257/3-2 and EC 
contract CHRX-CT92-00551.} \\
}
\end{center}
\normalsize
\medskip

\begin{center}
G.S.\ Bali$^{b}$, K. Schilling$^{a,b}$, Ch.\ Schlichter$^{b}$\footnote{
Email: bali/schillin/schlicht@wpts0.physik.uni-wuppertal.de}\\
\end{center}
\vskip .5cm

\noindent{}$^a${\it CERN, Theory Division, CH-1211 Geneva-23, Suisse}\\
$^b${\it Physics Department, University of Wuppertal,
D-42097 Wuppertal, Germany\\}
\vskip 2.5 cm

\centerline{{\bf Abstract}}
\noindent 
We present results of a high statistics study of the 
chromo field
distribution between static quarks in $SU(2)$
gauge theory on lattices of volumes $16^4$, $32^4$, and $48^3\times 64$,
with physical extent ranging from $1.3$ fm up to $2.7$ fm
at $\beta=2.5$, $\beta=2.635$, and $\beta=2.74$. We 
establish string formation over physical distances 
as large as $2$ fm. The results are tested against Michael's sum
rules. A detailed investigation of the
transverse action and energy flux tube profiles is provided.
As a by-product, we obtain the static lattice potential in
unpreceded accuracy.

\vspace{1.2cm}
\centerline{{\em (extended version)}}
\centerline{September 7, 1994}
\end{titlepage}

\section{Introduction} 
\label{sec:intro}     
The issue of verifying the confinement mechanism in Quantum
Chromodynamics has been a great challenge ever since the phenomenon of
string formation between a static $Q\bar{Q}$ pair has been conceived
by t'Hooft and Mandelstam~\cite{thooft}
to be a dual Meissner effect in the scenario of a Type II
superconducting vacuum.  Early lattice gauge
theory attempts to compute the colour field distribution, without
recourse to modelling, were necessarily limited by the available
compute power and lattice methods of the period.
They rendered qualitative rather than
quantitative results, with lattice resolutions, $a$, and 
quark-antiquark separations, $r$, restrained to $a > .15$ fm and 
$r = aR < 1$ fm, respectively~\cite{sommer1,sommer,haymaker3,haymaker5}.

In recent high precision studies of
$SU(2)$ and $SU(3)$ gauge theo\-ries~\cite{ukqcdpot,wuppot} 
the static quark-antiquark
potential has been found to be consistent with a linearly rising part
and a subleading $-\pi/(12R)$ correction as predicted by the bosonic string
picture~\cite{luscher1,luscher2} for separations above $r_t \approx .5$ fm.
Moreover, there are numerical indications for hybrid potentials, with gluonic
excitations separated by energy gaps $n\pi /R$~\cite{michperan,phdbali}
as expected from effective string theories~\cite{luscher1}.

A compelling evidence about the nature of the confining string from
lattice gauge theory (LGT) is still lacking.
It would require measurements of
field distributions at quark separations well beyond $r_t$.  To study
the geometry of the colour flux tube between $Q$ and $\bar{Q}$
sources, one needs an increase both in {\it resolution} of the
underlying lattice and in the {\it linear extent}, $r$, of the
string.

These requirements are not so easily met, since (a) the energy density
carries  dimension $a^{-4}$ and therefore imposes a lower limit onto the
lattice spacing due to statistical noise and (b)
one is forced to work with very large lattices to attain
large quark-antiquark separations $r = Ra$. On top of this,
one is of course faced with the
ubiquitous problem of filtering ground state signals out of an
excited state background.

Thus, in order to really determine the structure of strings in the
heavy quark-antiquark interaction, one cannot avoid a
systematic high precision study, ensuring (a) good scaling behaviour
as well as (b) sufficient control on finite size effects (FSE) and
(c) reliable signals for the ground state.

The superconducting picture for QCD has been modelled
in terms of a dual effective Langrangian
some time ago by Baker and collaborators
and worked out subsequently~\cite{ball}. Lattice
gauge theory in principle 
offers the laboratory to test such confinement models,
as it  allows for {\it ab initio} studies from the QCD Lagrangian.
Within the lattice community,
there has recently been revived interest to study the r\^ole  of
monopole condensation in the confinement mechanism, by recourse to the
maximal Abelian gauge projection, in $SU(2)$ gauge
theory~\cite{suzukirev,suzuki2,stack1}. 
Some first encouraging
evidence for the dual Meissner effect has also been reported~\cite{haymaker1}.
Nevertheless, all this pioneering lattice work on the confinement
mechanism has been carried out either at rather smallish quark-antiquark
separations, where the flux tube is not yet really developed, or at
rather large lattice spacing.

In fact, there definitely remains a gap: so far flux tubes of
sufficient  physical lengths have never been observed on the lattice!
In this paper, we intend to bridge this gap: exploiting state-of-the-art
lattice
techniques --- for noise reduction and ground state enhancement --- as
well as the compute power (and memory!) of ``small'' Connection Machines
CM-2 and CM-5, we will be able to demonstrate unambiguously that
quenched $SU(2)$ gauge theory does imply flux tube formation over distances
well above the $\pi$ Compton wavelength. 

Reliable lattice calculations can only be based on trustworthy  error
estimates. For this reason, we will expose
the underlying  lattice techniques in quite some detail (Section 2).
There is a shorter Section~3, augmented by three appendices,
on (a) weak coupling and (b) string model
issues that are helpful to appreciate certain qualitative features
of the field distributions and (c) sum rules for energy and action densities
that provide an important cross check of the lattice results.
The numerical results are presented in Section~4
which includes very precise potential data,
determination of the Symanzik $\beta$ function and many
pictures of the flux tubes\footnote{We regret
being  unable to expose the colour flux tube in colour in
this medium! A data base of colour images
is under construction. It can be accessed via anonymous ftp from
wpts0.physik.uni-wuppertal.de. The (compressed)
.rgb and .ps files are deposited in
the directory pub/colorflux.}.
Detailed checks on finite size effects, discretization effects and
ground state dominance are provided, substantiating the
interpretation of lattice correlators in terms of continuum fields.
Section 5 contains a discussion of the shape of the flux tube
and the status of Michael's sum rules.

\section{Lattice techniques}
The numerical calculations are performed on lattices with hypercubic
geometry and periodic boundary conditions in all four directions with
volumes $L_S^3\times L_T$ ranging from $16^4$ up to $48^3\times 64$.
Throughout the simulation the standard Wilson action
\begin{equation}
S_W=-\beta\sum_{n,\mu>\nu}U_{\mu\nu}(n)
\end{equation}
with
\begin{equation}
\label{Udef}
U_{\mu\nu}(n)=\frac{1}{2}\tr\left(U_{\mu}(n)U_{\nu}(n+\hat{\mu})
U_{\mu}^{\dagger}(n+\hat{\nu})U_{\nu}^{\dagger}(n)\right)
\end{equation}
and $\beta=4/g^2$ has been used.

For the updating of the gauge fields a hybrid of heatbath and
overrelaxation algorithms has been implemented~\cite{hor}. The 
Fabricius-Haan heatbath sweeps~\cite{fh}
have been randomly mixed with the overrelaxation step with probability
ranging from
$1/8$ at $\beta=2.5$ up to $1/14$ at $\beta=2.74$.
The links have been visited in lexicographical ordering within $2^4$
hypercubes, i.e. within each such hypercube, first all links pointing
into direction $\hat{1}$ are visited site by
site, then all links in direction $\hat{2}$ etc..

\subsection{Prerequisites}
In order to substantiate 
continuum results from lattice calculations
it is of utmost importance to investigate the impact of the finite
lattice volume as well as the scaling behaviour with the lattice
spacing $a$. This requires simulations both at (a) fixed lattice coupling
$\beta$ (i.e.\ spacing $a$) with a varying number of lattice sites
and (b) (approximately) fixed physical volume but
different lattice resolutions.

Of course one wants to work ---  within the computational means ---
as close as possible to the continuum
limit, i.e.\ at as large $\beta$-values
as possible. The bottleneck  is set by the 
memory requirements 
due to the increase of the number of lattice 
sites (needed to compensate for a smaller $a$)
as well as by the computer time required to suppress  the
statistical noise. Since the operators under investigation scale
with the fourth power of the lattice resolution
(up to $\ln a$ terms from anomalous dimensions, see below), the latter
limitation is the more serious one,
restricting all preceding lattice studies to $\beta\leq 2.5$.

Though simulations at small $\beta$ values
allow for rather large physical volumes,
lattice artefacts are expected to spoil results at physically
interesting scales.  Moreover, we should mention that our smearing
procedure provides inferior ground state overlaps at large lattice
spacings. There is more reason to stay away from too coarse
lattices: one needs a sufficiently large $T$-range for verification
of $T$-plateaus in the {\it bona fide} physical quantities. In
addition, at
smaller values of $\beta$, the lower limit $T\geq 3$,
implied by the minimal temporal extent of the
$\langle\Box\rangle_{\cal W}$ operator, amounts to overly large
physical separations and leads to small signals of the Wilson loops.

\begin{table}[htb]
\centerline{
\begin{tabular}{|c|c|c|c|c|}\hline
&$\beta = 2.50$ & $\beta = 2.50$ & $\beta = 2.635$ & $\beta = 2.74$\\\hline
$L_S^3\times L_T$& $16^4$ &$32^4$&$48^3\times 64$&  $32^4$ \\
$K$&.0350(4)&.0350(12)&.01458(8)&.00830(6)\\
$a/$fm                   &.0826(5)& .0826(14) &.0541(2)&.0408(2)\\
$a^{-1}/$GeV               &2.35 (1)& 2.35(4)&3.64(1)  &4.83(2)\\
$aL_S$/fm&1.32(1)  & 2.64(4)&2.60(1) &1.31(1)\\
MC sweeps& 868000  & 255600 & 53800&152000  \\
thermalization sweeps& 2000  & 51000& 4200&4000  \\
{\small meas., sweeps between meas.}&&&&\\
Wilson loops                & 8680, 100& 2046, 100 & 248, 200&1480, 100 \\
colour flux distribution    & 8680, 100& 2046, 100 & 248, 200&670, 100\\
plaquette                   &86800, 10 & 20460, 10 & 992,  50 &14800,
10  \\\hline
\end{tabular}
}
\caption{\em Simulation parameters. The physical scales have been
computed from the value $\protect\sqrt{K}a=440$~MeV.}
\label{Tab1}
\end{table}

In short, one has to find a way between Scylla and
Charybdis: i.e. to compromise between the shortcomings  of both small
and large lattice spacings $a$. We have chosen to simulate at $\beta=2.5,
2.635$, and 2.74 at various lattice volumes, ranging up to the unpreceded
volume  $48^3\times 64$. Our simulation parameters
are summarized in
Tab.~\ref{Tab1}.

As a by-product we compute the static potential and
obtain the most precise set of $SU(2)$ string
tension values that has ever been computed on the lattice.
Details of the fitting procedure are
explained in Section~\ref{PF}. The (lattice) string tension $K$
relates the lattice spacing $a$ to a physical scale: $\kappa=Ka^2$.
We ascribe the ``canonical'' value $\sqrt{\kappa}=440$~MeV
to the square root of the string
tension. Needless to say that this scale, taken from real world QCD
Regge trajectories, only serves as an orientation for its poor man's
quenched two colour version: $SU(2)$ gauge theory. Nonetheless, the
quantitative agreement between the $SU(2)$ and $SU(3)$ potentials is
remarkable. We also point out that the effective string model with
which we are going to compare our results does not depend on the
underlying gauge group.

{}From the string tension measurement we find the following
(approximate) relation between the present lattice spacings:
$a^{-1}_{2.5}:a^{-1}_{2.635}:a^{-1}_{2.74}\approx 1:1.5:2$.
Thus, the $16^4$ lattice at $\beta=2.5$ has approximately the same
physical volume as the $32^4$ lattice at $\beta=2.74$. The same holds
true for the $32^4$ lattice at $\beta=2.5$ and the $48^3\times 64$
lattice at $\beta=2.635$. These pairs of lattices can be used to
investigate the $a$ (in)dependence of the results.
In order to reveal possible volume effects, the
$16^4$ and $32^4$ lattice outcomes at $\beta=2.5$ will
be compared with each other.

As a prerequisite to the present investigation, let us consider the static
$Q\bar{Q}$ potential which can be computed from Wilson loops, $W(R,T)$.
A Wilson loop, i.e.\ an ordered product of
link variables along a closed rectangular path with spatial separation $R$
and temporal extent $T$, can be interpreted as a world sheet of a
$Q\bar{Q}$ pair: at Euclidean time $\tau=0$ a creation operator
\begin{equation}
\label{create}
\Gamma^{\dagger}_R=Q(0)U(0\rightarrow R)Q^{\dagger}(R)
\end{equation}
with a gauge
covariant transporter $U(0\rightarrow R)$ is applied to the vacuum
state $|0\rangle$. The $Q\bar{Q}$ pair is then propagated to $\tau=T$ by static
Wilson lines in presence of the gauge field background, and finally
annihilated by application of $\Gamma_R$. A spectral decomposition of the
Wilson loop exhibits the following behaviour ($\cal T=e^{-aH}$ denotes the
transfer matrix, ${\cal T}|n\rangle=e^{-E_n}|n\rangle$):
\begin{eqnarray}
\langle W(R,T)\rangle&=&
\frac{\tr\left(\Gamma_R{\cal T}^T\Gamma_R^{\dagger}{\cal T}^{L_T-T}\right)}
     {\tr\left({\cal T}^{L_T}\right)}\nonumber\\\label{expa1}
&=&\frac{1}{\sum_m e^{-E_mL_T}}\sum_{m,n}\left|\langle
m|\Gamma_R|n,R\rangle\right|^2 e^{-V_n(R)T}e^{-E_m(L_T-T)}\\\nonumber
&=&
\sum_n |d_n(R)|^2e^{-V_n(R)T}\times\left(1+{\cal O}
\left(e^{-E_1(L_T-T)}\right)\right)\quad,
\end{eqnarray}
where $d_n(R)=\langle 0|\Gamma_R|n,R\rangle$. $|n,R\rangle$ is the
$n$th eigenstate in the charged sector of the Hilbert space with
non-vanishing overlap to the creation operator $\Gamma_R^{\dagger}$, while
$|n\rangle$ is the $n$th eigenstate of the zero charge sector.
$V_n(R)$ denotes the $n$th excitation of the $Q\bar{Q}$ potential and
the vacuum energy $E_0$ has been set to {\em zero}. $E_1$ is the mass
gap, i.e.\ the mass of the $A_1^+$ glueball.

Actually, we are not restricted  to on-axis $Q\bar{Q}$ separations,
$\mathbf R =(R,0,0)$. Planar Wilson loops can be easily generalized to
off-axis separations by connecting sources that do not share a common
lattice axis. In the present investigation, the following 
off-axis directions have been realized:
\begin{eqnarray}
{\mathbf d}_1&=&(1,0,0)\quad,
\quad {\mathbf d}_2=(1,1,0)\quad,\quad {\mathbf d}_3=(2,1,0)\quad,
\nonumber\\\label{aa4}
{\mathbf d}_4&=&(1,1,1)\quad,
\quad {\mathbf d}_5=(2,1,1)\quad,\quad {\mathbf d}_6=(2,2,1)\quad,
\end{eqnarray}
with separations $m_i{\mathbf d}_i$ up to
$m_1, m_2, m_4\leq L_S/2$ and $m_3,m_5,m_6\leq L_S/4$.
For the largest lattice, $L_S=48$, this amounts to a measurement over
a set of 108 different
separations. All paths have been chosen as close to the
shortest linear
connection between the sources as the lattice permitted.

\subsection{Noise reduction}
In this section we will shortly discuss the implications
of noise reduction that we achieved by integrating out the temporal links
in the Wilson loops analytically~\cite{linkint}.

The  link integration amounts to 
the following substitution:
\begin{equation}
U_4(n)\longrightarrow V_4(n)=\frac{\int_{SU(2)}\!dU\,Ue^{\beta
S_{n,4}(U)}}
{\int_{SU(2)}\!dU\,e^{\beta S_{n,4}(U)}}
\end{equation}
with
\begin{equation}
S_{n,\mu}(U)=\frac{1}{2}\tr\left(UF_{\mu}^{\dagger}(n)\right)\quad,
\quad F_{\mu}(n)=\sum_{\nu\neq\mu}U_{\nu}(n)U_{\mu}(n+\hat{\nu})
U_{\nu}^{\dagger}(n+\hat{\mu})\quad.
\end{equation}
$V_4(n)$ is in general not an $SU(2)$ element anymore.

In this way, time-like links are replaced
by the mean field value they take in the neighbourhood of
(stroboscopically frozen) links that interact through
the staples $F_{\mu}(n)$. 
Only those links that do not share a common plaquette, can be
integrated independently.
This holds  in   particular  for  all temporal
links within our  spatially smeared Wilson loops, iff $R > 1$.

In case of $SU(2)$ gauge theory, $V_4(n)$ can be calculated analytically:
\begin{equation}
V_4(n)=\frac{I_2(\beta f_{\mu}(n))}{f_{\mu}(n)I_1(\beta
f_{\mu}(n))}F_{\mu}(n)\quad,
\quad f_{\mu}(n)=\sqrt{\det(F_{\mu}(n))}\quad.
\end{equation}
$I_n$ denote the modified Bessel functions.

The statistical error, $\Delta O$, of an observable,
$\langle O\rangle$, calculated
{\it without} link integration, is related by a constant $s<1$
to the corresponding error {\it with} link integration, $\Delta
O_{li}=s\Delta O$.  In order to discuss the impact  of link
integration on noise reduction, we start from the na\"{\i}ve
expectation that each integrated link contributes equally to $s$,
i.e.\ we assume $s=x^{2T}$ with $2T$ being the number of integrated
links used within the construction of $\langle O\rangle$.

In order to estimate the value of $s$, 
let us consider on-axis Wilson loops with integrated temporal links.
On symmetric lattices ($L_T=L_S$) we expect from
the relation $\langle W(R,T)\rangle=\langle W(T,R)\rangle$:
\begin{equation} \frac{\Delta W(R,T)}{\Delta W(T,R)}=x^{2(T-R)}\quad.
\end{equation} 

This leads to the estimate for $x^2$:
\begin{equation}
x^2=\exp\left(\frac{1}{T-R} \log\left(\frac{\Delta W(R,T)}{\Delta
W(T,R)}\right)\right)\quad.  \end{equation} 
Our data (with
bootstrapped errors of the errors) is consistent with a
factor\footnote{For small $R$ and $T$ where the statistical errors
have reached the same order of magnitude as the numerical machine
accuracy, we performed additional computations of $\Delta O$ and
$\Delta O_{li}$ on smaller subsamples to ensure that the errors still
follow the statistical $1/{\sqrt{N_{meas}}}$ expectation.}
$x=.889\pm.001$ ($x=.890\pm .001$) for $\beta=2.5$ ($\beta=2.635$). 
Thus, application of 
link integration at time $T=6$
(the largest temporal extent, used in the computation of the colour field
operators, see below) amounts to a reduction of computer time by a
factor $x^{-24}=16.8\pm 0.4$.

The improvement achieved by link integration tends to be smaller at
smaller lattice spacings. This is due to the fact 
that the physical extent of the neighbourhood to be integrated out
becomes smaller. On the other hand, the error of
a non-link integrated operator, measured on lattices with constant
physical volumes but different couplings, also decreases with the
lattice spacing (temperature $\beta^{-1}$).
At the bottom line, the two effects almost cancel each other
and the relative errors of
link integrated Wilson loops
appear to remain rather independent of the lattice resolution,
provided that the
physical lattice volumes and the number
of measurements are kept constant.

\subsection{Ground state enhancement}
\label{GSE}

The physically interesting ground state potential, $V(R)=V_0(R)$,
can be retrieved  in
the limit of large $T$:
\begin{equation}
\label{Wasym}
\langle W(R,T)\rangle
=\sum_nC_n(R)e^{-V_n(R)T}\longrightarrow C_0(R)e^{-V(R)T}\quad
(T\rightarrow\infty)\quad.
\end{equation}
The overlaps $C_n(R)=|d_n(R)|^2\geq 0$
obey the following normalization condition:
\begin{equation}
\sum_nC_n(R)=1\quad.
\end{equation}
The path of the transporter $U(0\rightarrow R)$ used for the
construction of the $Q\bar{Q}$ creation operator (Eq.~\ref{create})
does not affect the
eigenvalues of the transfer matrix and is by no means unique. One can
exploit this freedom 
 to maximize the ground state overlap by
a suitable  superposition  of such paths, aiming at
$C_0(R)\approx 1$. At any given value of $R$, the final deviation
of $C_0(R)$ from {\em one} can serve
 as a monitor for the suppression of excited
state contributions actually achieved in this way.

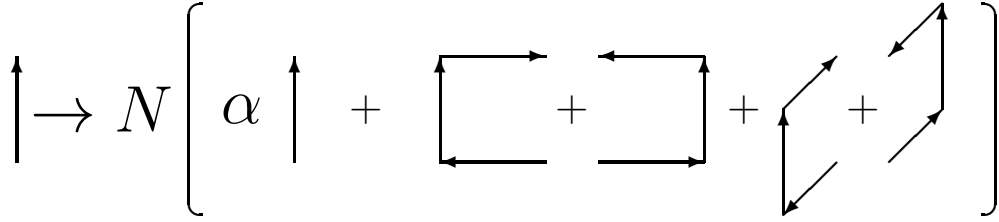
\begin{figure}
\begin{center}\begin{picture}(380,80)
\thicklines
\put(00,20){\vector(0,1){40}}
\put(32,40){\makebox(0,0){{\huge $\rightarrow N$}}}
\put(70,40){\oval(10,80)[l]}
\put(105,20){\vector(0,1){40}}
\put(85,40){\makebox(0,0){{\huge $\alpha$}}}
\put(132,40){\makebox(0,0){{\Large $+$}}}
\put(200,20){\vector(-1,0){40}}
\put(160,20){\vector(0,1){40}}
\put(160,60){\vector(1,0){40}}
\put(210,40){\makebox(0,0){{\Large $+$}}}
\put(220,20){\vector(1,0){40}}
\put(260,20){\vector(0,1){40}}
\put(260,60){\vector(-1,0){40}}
\put(275,40){\makebox(0,0){{\Large $+$}}}
\put(310,20){\vector(-1,-1){20}}
\put(290,0){\vector(0,1){40}}
\put(290,40){\vector(1,1){20}}
\put(320,40){\makebox(0,0){{\Large $+$}}}
\put(330,20){\vector(1,1){20}}
\put(350,40){\vector(0,1){40}}
\put(350,80){\vector(-1,-1){20}}
\put(365,40){\oval(10,80)[r]}
\end{picture}\end{center}
\caption{\em Visualization of a smearing iteration.}
\label{Fig0} 
\end{figure}

\begin{figure}[htb]
\begin{center}
\leavevmode
\epsfxsize=12cm\epsfbox{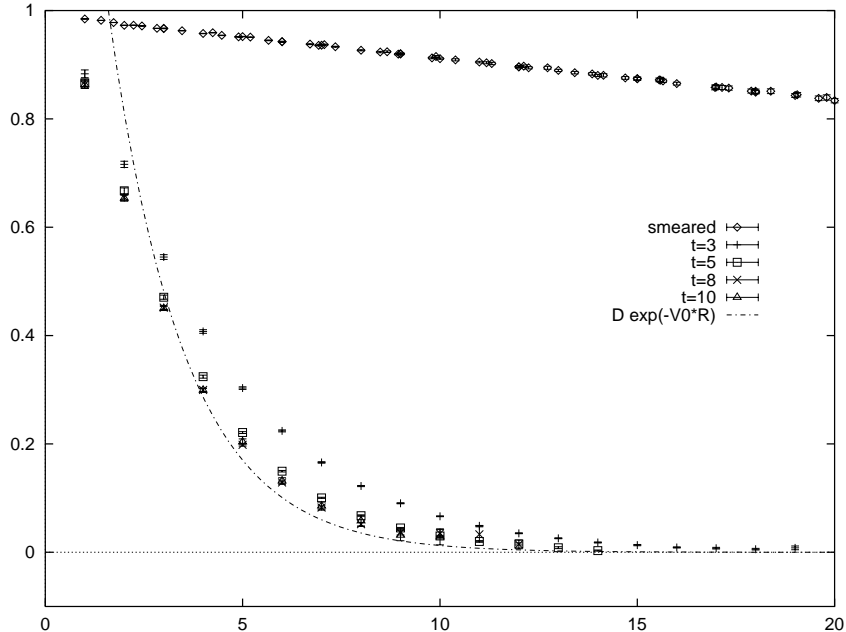}
\end{center}
\caption
{\em The ground state overlaps $C_0(R)$ versus $R$ at $\beta=2.635$.
In addition to the overlaps of smeared operators (diamonds), finite $T$
approximants, $C_0(R,T)$ for the overlaps from unsmeared 
(on-axis) operators are
plotted. The dashed line denotes the extrapolated large $T$ limit for
the unsmeared overlaps and should be a valid approximation to the
large $R$ behaviour.}
\label{Fig1}
\end{figure}

In the present simulation an iterative procedure (with $n_{iter}$
iteration steps) has been
applied~\cite{baliref,APE}:
each {\it spatial} link
\(U_i(n)\), occurring in the transporter, is substituted by a ``fat'' link,
\begin{equation}
\label{sme}
U_i(n)\longrightarrow N\left(\alpha U_i(n)+\sum_{j\neq
i}U_j(n)U_i(n+\hat{j})U_j^{\dagger}(n+\hat{i})\right)
\end{equation}
with the appropriate normalization constant $N$
and free parameter $\alpha$. One such
iteration step is visualized in Fig.~\ref{Fig0}.
For this smearing, the links are
visited in the  lexicographical ordering of the  updating sweep.

We find satisfactory ground state enhancement  with the parameter choice
$n_{iter}=150$ and $\alpha=2$.

One can define approximants to the asymptotic
potential values and overlaps,
$V(R,T)\rightarrow V(R)$ and
$C_0(R,T)\rightarrow C_0(R)\,\, (T\rightarrow\infty)$.
Due to the
positivity of the transfer matrix ${\cal T}$, these quantities decrease 
monotonically (in $T$) to their asymptotic limits:
\begin{eqnarray}
\label{hh1}
V(R,T)
&=&\log\left(\frac{\langle W(R,T)\rangle}{\langle
W(R,T+1)\rangle}\right)=V(R)+\frac{C_1(R)}{C_0(R)}h(R,T)+\cdots\\\label{hh7}
C_0(R,T)&=&\frac{\langle W(R,T)\rangle^{T+1}}{\langle W(R,T+1)\rangle^T}
=C_0(R)+C_1(R)h(R,T)+\cdots
\end{eqnarray}
with
\begin{equation}
\label{Vgap}
h(R,T)=e^{-\Delta V(R)T}\left(1-e^{-\Delta V(R)}\right)\quad,\quad
\Delta V(R)=V_1(R)-V(R)\quad.
\end{equation}
In our analysis, we follow
these approximants
until they reach a plateau.

As we wish to maximize $C_0(R)$, we would like to acquire a
qualitative understanding of the underlying physics. For this purpose,
we consider
unsmeared on-axis Wilson loops. 
Combining Eq.~\ref{Wasym} with the
$R$-$T$ symmetry\footnote{This symmetry is only exact on lattices with
$L_S=L_T$. However, within statistical accuracy it also holds true on
the $48^3\times 64$ lattice.}
$\langle W(R,T)\rangle=\langle W(T,R)\rangle$ and the
parametrization 
of the potential
$V(R)=V_0-e/R+KR$ we obtain
\begin{eqnarray}
\label{abf}
\ln\langle W(R,T)\rangle&\approx&
\ln C_0(R,T)-V_0T+eT/R-KRT\\\nonumber
&=&\ln D-V_0(R+T)+e(R/T+T/R)-KRT
\end{eqnarray}
for large $R$ and $T$ and arrive at the estimate (with constants $V_0$
and $e$ obtained from the potential analysis below),
\begin{equation}
\label{abg}
C_0(R,T)= De^{-(V_0-e/T)R}\quad\mbox{or}\quad C_0(R)=De^{-V_0R}\quad.
\end{equation}
This parametrization of the unsmeared overlaps
turns out to describe the off-axis data too
if we allow for a smaller (direction dependent) constant $V_0$.

The self-energy term $V_0/a$
diverges in the continuum limit, $a\rightarrow 0$. Thus, the
overlaps at fixed physical separation\footnote{At fixed (lattice) $R$, a
(slight) increase is observed and expected.},
$r=Ra$, decrease with increasing
$\beta$. This feature is in accord with the following
consideration:
in the scaling region the transverse size of the $Q\bar{Q}$ wave
function is expected to remain constant in physical units while the
transverse extent of the string like
creation operator remains on the scale of the
lattice resolution. Thus, the ground state overlap of this operator
decreases with increasing correlation length.

The ground state overlaps of smeared Wilson loops
at $\beta=2.635$ are shown  in Fig.~\ref{Fig1},
together with on-axis approximants to the
unsmeared overlaps and the asymptotic (large $R$) estimate 
of Eq.~\ref{abg} (dashed curve), where
the coefficient $D\approx 2.3$ was obtained by fitting the large $T$ data
to Eq.~\ref{abg}. There is
obviously a dramatic improvement from the use of an
extended creation operator.
In case of the smeared links, the 
(unsmeared) self-energy contribution $V_0$ appears
to be reduced to a number $f\ll V_0$
that is sufficiently small to allow for an expansion of the
exponential factor $C_0(R)=D\exp(-fR)\approx D(1-fR)$:
the ground state
overlaps of smeared Wilson loops exhibit a {\it linear} $R$ dependence
throughout the observed $R$ region.  Moreover,
rotational invariance in terms of the overlaps
is restored for all on- and off-axis separations.

For $\beta=2.5$ the overlaps vary between $C_0(\sqrt{2}a)=.95$ and
$C_0(r_m)=.73$ on the $32^4$ lattice and from
$C_0(a)=.98$ to $C_0(r_m)=.81$ on the $16^4$ lattice.
Within the same physical region the
$\beta=2.635$ overlaps range from  $C_0(\sqrt{2}a)=.98$
to $C_0(r_m)=.81$.
At $\beta=2.74$ we have used an inferior set of smearing parameters
($n_{iter}=40$ and $\alpha=0$); yet we achieve overlaps of
$C_0(a)=.96$ and $C(r_m)=.84$. We have chosen $r_m\approx 1.2$ fm for
the comparison. This scale corresponds to
$r_ma^{-1}=8\sqrt{3}, 12\sqrt{3}, 16\sqrt{3}$ for the three
$\beta$-values, respectively.
Even at fixed physical $r$ the overlaps tend to
increase with $\beta$, unlike in the situation with unsmeared
operators: the wave function becomes smoother at increased correlation
length and can be better modelled by the iterative smearing procedure.
For the largest
distance realized ($2.25$~fm at $\beta=2.635$)
we still achieve the value $C_0(24\sqrt{3}a)=.72$.

The success of smearing is twofold: (a) for rather small values of
$T$, extraction of the ground state potential becomes possible and (b)
the signal-to-noise ratio
is greatly improved as
$C_0(R)$ (and the signal) increases,
especially for large values of $R$.

\subsection{Lattice determination of colour fields}
\label{CF}
The central  observables in our present investigation are
the action and energy densities
in presence of two static quark sources (with separation $R$)
in the ground state of
the binding problem:
\begin{eqnarray}
\epsilon_R({\mathbf n})&=&\frac{1}{2}\left({\cal E}_R({\mathbf n})
+{\cal B}_R({\mathbf n})\right)\quad,\\\label{act1}
\sigma_R({\mathbf n})&=&\frac{1}{2}\left({\cal E}_R({\mathbf n})
-{\cal B}_R({\mathbf n})\right)\quad,
\end{eqnarray}
with
\begin{equation}
\label{act2}
{\cal E}_R({\mathbf n})=
\langle {\mathbf E}^2({\mathbf n})\rangle_{|0,R\rangle-|0\rangle}
\quad,\quad
{\cal B}_R({\mathbf n})=
\langle {\mathbf B}^2({\mathbf n})\rangle_{|0,R\rangle-|0\rangle}
\end{equation}
and
\begin{equation}
\langle O\rangle_{|0,R\rangle-|0\rangle}=
\langle 0,R|O|0,R\rangle-
\langle 0|O|0\rangle\quad.
\end{equation}
The sign convention corresponds to the Minkowski notation
with metric\\ $\eta=\diag(1,-1,-1,-1)$, in which
${\cal B}_R({\mathbf n})\leq 0$, ${\cal E}_R({\mathbf n})\geq 0$.
We point out that the 
Minkowski action density carries a different sign relative to the
(negative) Euclidean action, i.e.\ $S_W=-\sum_{\mathbf n}
\frac{1}{2}\langle {\mathbf E}^2-{\mathbf B}^2\rangle_{|0\rangle}$.

We shall extract these observables from the correlations between smeared
Wilson loops, ${\cal W}=W(R,T)$, and (unsmeared) plaquettes 
$\Box(\tau )=U_{\mu\nu}({\mathbf
n},\tau )$ (Eq.~\ref{Udef})\footnote{We do not follow the authors of
Ref.~\cite{haymaker3} who, in order to reduce statistical fluctuations,
advocate to subtract $\langle {\cal W}\Box({\mathbf
n})\rangle/\langle{\cal W}\rangle$ with the reference point, ${\mathbf
n}$, taken far away from the sources rather than the vacuum plaquette
expectation $\langle \Box\rangle$. In this way, we avoid possible
shifts of the normalization relative to the vacuum energy and action
densities. We would like to point out
that we found no reduction in statistical
errors for smeared Wilson loop operators by using the above
suggestion. However, we have been able to confirm this observation of
Ref.~\cite{haymaker3} for unsmeared Wilson loops.},
\begin{equation}
\label{wdef}
\langle\Box(S)\rangle_{\cal W}=
\frac{1}{2}\frac{\langle {\cal W}(\Box(T/2+S)+\Box(T/2-S))\rangle}{\langle
{\cal W}\rangle} -\langle\Box\rangle\quad.
\end{equation}
$S$ denotes the distance of the plaquette from the central time slice of
the Wilson loop and takes the values $S=0,1,\ldots$ ($S=1/2,3/2,\ldots$)
for even (odd) $T$. 

The plaquette insertion acts
as the chromo dynamical analogue of a Hall detector in 
electrodynamics\footnote{
We  note in passing, that the authors of Ref.~\cite{pisa} have
chosen to connect the plaquette to the Wilson loop via
two Wilson lines and take one overall trace instead of
two separate ones,
as this leads to an improved signal to noise ratio.
However, a prove that this observable indeed can be interpreted
as a colour field density in presence of a static $Q\bar{Q}$ pair
is missing. Moreover, the constraint through sum rules is lost.}.
For $0\leq S< T/2$, $\langle\Box(S)\rangle_{\cal W}$
can be decomposed into mass eigenstates as
follows
\begin{eqnarray}
\langle\Box(S)\rangle_{\cal W}&=&\frac{\tr\left(
\Gamma\left({\cal T}^{T/2+S}\Box{\cal T}^{T/2-S}+
{\cal T}^{T/2-S}\Box{\cal T}^{T/2+S}\right)\Gamma^{\dagger}{\cal
T}^{L_T-T}\right)}{2\tr\left(\Gamma{\cal T}^T\Gamma^{\dagger}{\cal
T}^{L_T-T}\right)}-\langle\Box\rangle\nonumber\\
&=&
\langle 0,R|\Box|0,R\rangle-\langle 0|\Box|0\rangle\nonumber\\
&+&2\re\left(\frac{d_1}{d_0}\langle 1,R|\Box|0,R\rangle\right)
e^{-\Delta VT/2}\cosh(\Delta VS)\nonumber\\\label{decom}
&+&\frac{|d_1|^2}{|d_0|^2}\left(\langle 1,R|\Box|1,R\rangle-
\langle 0,R|\Box|0,R\rangle\right)e^{-\Delta VT}\\
&+&2\re\left(\frac{d_2}{d_0}\langle 2,R|\Box|0,R\rangle\right)
e^{-(V_2-V)T/2}\cosh((V_2-V)S)\nonumber\\\nonumber
&+&{\cal O}(e^{-\Delta
V(\frac{3}{2}T-S)})\quad.
\end{eqnarray}
$\Delta V$ denotes the gap between the ground state and the first
excitation (Eq.~\ref{Vgap}).
In principle, $|d_n|^2=C_n(R)$ and $V_n(R)$ can
be determined from smeared Wilson loops. The non-diagonal
${\cal O}(e^{-(V_n-V)(T/2-S)})$ coefficients
can only be obtained from a fit to
the time dependence of the above operator. 
As we shall see in the next paragraph, $(V_2-V)/2\approx\Delta V$.
Thus, a measurement of the excited state colour field distribution
$\langle\Box\rangle_{|1,R\rangle-|0\rangle}$ appears to be unfeasible with the
present method.

In string model calculations the separations between ground and
excited state potentials without gluonic angular momentum, i.e.\
within the $A_{1g}$ representation of the cubic symmetry group, are
found to be multiples of $2\pi/R$~\cite{luscher1,michperan}. This
feature is in accord with numerical simulations of $SU(2)$ and $SU(3)$
gauge theories~\cite{michperan,phdbali}. Therefore, as a net result, we
expect the following asymptotic behaviour: \begin{eqnarray}
\label{largeT} \langle\Box(S)\rangle_{\cal
W}&=&\langle\Box\rangle_{|0,R\rangle-|0\rangle}\nonumber\\
&+&c_1e^{-\pi T/R}\cosh(2\pi S/R)\\\nonumber &+&c_2e^{-2\pi
T/R}\left(1+c_3\cosh(4\pi S/R)\right)+\cdots \end{eqnarray} with $c_i$
being free parameters. They are, contrary to the coefficients of the
spectral decomposition of the Wilson loop (Eq.~\ref{Wasym}), not
necessarily positive! Be aware, that $c_i$ vary with the $Q\bar{Q}$
separation as well as with the spatial position ${\mathbf n}$ of the
plaquette insertion.

The deviations from the asymptotic values are governed by ${\cal
O}(e^{-\Delta V(T/2-S)})$ terms, compared to
order $e^{-\Delta VT}$ corrections in case of the potential
(Eqs.~\ref{hh1}, \ref{Vgap}). So,
the issue of optimization for  ground state dominance 
is certainly more critical for field measurements.
While the reduction of systematic errors  would ask for large
$T$-values, the suppression of statistical uncertainties 
would lead one to the  contrary. Obviously, the
reasonable strategy is to ensure
that systematic and statistical errors are kept in balance.

A weak coupling expansion of the
plaquette yields the square of the Maxwell field strength tensor
${\cal F}_{\mu\nu}=F_{\mu\nu}^c{}\sigma_c/2$:
\begin{equation}
U_{\mu\nu}=1-\frac{a^4}{2\beta}
F_{\mu\nu}^cF_{\mu\nu}^c+{\cal O}(a^6)\quad.
\end{equation}
Thus, by an appropriate choice of $\Box=U_{i4}$ ($\Box=U_{jk}$)
expectation values of squared chromo electric (magnetic) field components
can be obtained\footnote{Remember, that we do not obtain any information
on the components 
of ${\mathbf E}$ and ${\mathbf B}$ themselves since,
in general, $\langle O^2\rangle\neq\langle O\rangle^2$.}, in the
limit of large $T$:
\begin{eqnarray}
\label{act3}
\frac{2\beta}{a^4}\langle U_{i4}(S)\rangle_{\cal W}&
\stackrel{T\rightarrow\infty}{\longrightarrow}&
\langle E_i^2({\mathbf n})\rangle_{|0,R\rangle-|0\rangle}
\quad,\\\label{act7}
-\frac{2\beta}{a^4}\langle U_{jk}(S)\rangle_{\cal W}
&\stackrel{T\rightarrow\infty}{\longrightarrow}&
\langle B_i^2({\mathbf n})\rangle_{|0,R\rangle-|0\rangle}
\quad,\quad|\epsilon_{ijk}|=1\quad.
\end{eqnarray}
The finite $T$ corrections to these relations have been elaborated in
Eq.~\ref{largeT}. Note, that $E_i^2=E_i^cE_i^c=2\tr E_i^2$.

\subsection{Implementation of colour field operators}
For measurement of the colour field distributions, the appropriate  plaquette
operators are suitably averaged in order
to obtain chromo magnetic or electric
insertions in symmetric position to a given lattice site, ${\mathbf n}$.
For the electric insertions two plaquettes are averaged:
\begin{equation}
U_{i4}({\mathbf n})\rightarrow
\frac{1}{2}\left(U_{i4}({\mathbf n}-{\mathbf e}_i)
+U_{i4}({\mathbf n})\right)\quad.
\end{equation}
For the magnetic fields four adjacent plaquettes are combined:
\begin{equation}
\label{act4}
U_{jk}({\mathbf n})\rightarrow
\frac{1}{4}\left(
U_{jk}({\mathbf n}-{\mathbf e}_j-{\mathbf e}_k)
+U_{jk}({\mathbf n}-{\mathbf e}_j)+U_{jk}({\mathbf n}-{\mathbf e}_k)+
U_{jk}({\mathbf n})\right)\quad.
\end{equation}

Notice that while $B_i^2$ is measured at integer values of $\tau$,
$E_i^2$ is measured between two time slices. To minimize contaminations
{}from excited states (Eq.~\ref{largeT}), $\tau$ is chosen as close as possible
to $T/2$. For even temporal extent of the Wilson loop, this
means $S=0$ for the magnetic field operator and $S=1/2$ for the
electric field insertion, while for odd $T$, $S=0(1/2)$ for electric
(magnetic) fields.

For measurement of the colour field distributions we have restricted
ourselves to on-axis separations of the two sources. All even distances
$R=2,4,\ldots,R_{\max}$ with $R_{\max} = 8,24,36$ for $L_S=16,32,48$,
respectively, have been realized. In order to identify the
asymptotic plateau, $T$ was varied up to\footnote{For historical
reasons, on our small lattice volumes ($16^4$ at $\beta=2.5$ and $32^4$
at $\beta=2.74$) only the odd values $T=1,3,5,7$ have been realized.}
$T=6$. The colour field distributions
have been measured up to a transverse distance $n_{\perp}=2$ along the
entire
$Q\bar{Q}$ axis. In between the two sources and up to 2 lattice
spacings outside the sources, the transverse distance was increased to
$n_{\perp,\max}=6,10,15$ for the three lattice extents $L_S=16,32,48$,
respectively.
In addition to ``on-axis'' positions, ${\mathbf
n}=(n_1,n_2,0)$, we chose plane-diagonal points ${\mathbf
n}=(n_1,n_2,n_2)$ with $n_2<n_{\perp,\max}/\sqrt{2}$.
We averaged over various coordinates ${\mathbf n}$, exploiting the
cylindrical and reflection symmetry of the problem.
All this  amounts to 9576 (26244) different combinations of $R$, $T$, and
${\mathbf n}$ on the $32^4$ ($48^3\times 64$) lattices, on which both 
$\langle U_{jk}\rangle_{\cal W}$ and $\langle U_{i4}\rangle_{\cal W}$
have been computed.

The temporal parts of the Wilson loops, appearing in the
colour field correlator, have been link integrated. 
Therefore, the electric
components 
have only been determined at distances larger than one lattice
spacing away from the sources. For the case of the $32^4$
lattice at $\beta=2.5$ we have substituted the missing values by the
corresponding entries, computed on a $16^4$ lattice without link
integration. Distances so close to the sources are not
relevant to continuum physics anyway, due to
contamination from the heavy quark
self-energy and lattice artefacts.

\section{Theoretical expectations}
\label{TE}
\subsection{Perturbative scenario}
\label{PS}
A perturbative order $g^4$ computation of the lattice potential can be
found in Ref.~\cite{karschheller}. Here, we recall the one gluon
exchange result only:
\begin{equation}
\label{a15}
V({\mathbf R})=-C_Fg^2\left(G_L({\mathbf R})-G_L({\mathbf 0})\right)
\stackrel{a\rightarrow
0}{\longrightarrow} -C_Fg^2\frac{1}{4\pi R}
\end{equation}
where we have dropped the (divergent) self energy in the continuum
expression.
The lattice gluon propagator $G_L({\mathbf R})$ 
(Eqs.~\ref{a10}, \ref{a11}) can be computed
numerically on finite lattices.
For $SU(2)$ one has  $C_F=(N^2-1)/(2N)=3/4$. For completeness, this
expression is derived in Appendix~\ref{ap1}.
A renormalization of the bare lattice coupling $g^2=2N/\beta$ 
turns out to be the main effect of the loop
diagrams that occur in the next order.

In order to investigate the nature of lattice artefacts, we have
performed a weak coupling expansion of the action and
energy densities. The lowest order term is a two gluon exchange. To
this order the action and energy densities turn out to identical,
both receiving just contributions from electric plaquettes.
Details of the calculation are contained in Appendix~\ref{ap1}.
The lattice integrals of the result (Eqs.~\ref{a8}, \ref{a9})
have been computed numerically.

\begin{figure}[htb]
\begin{center}
\leavevmode
\epsfxsize=12cm\epsfbox{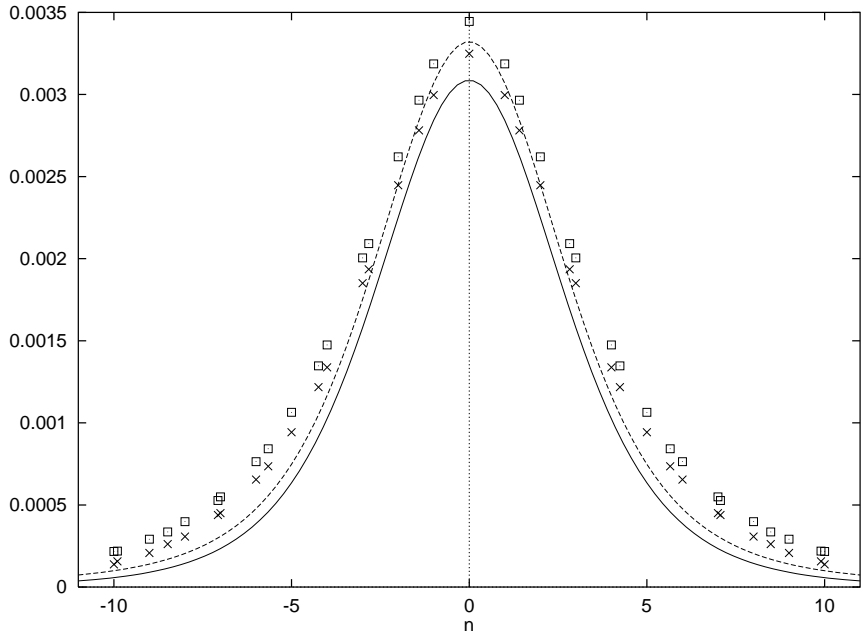}
\end{center}
\caption
{\em Comparison between continuum and lattice dipole fields
in the center plane between two sources, separated by $R=12$. The
ordinate, $n$, is the distance from the $Q\bar{Q}$ axis. The
values have been multiplied by a factor $(4\pi)^2/(C_Fg^2)$.
Crosses and the solid line correspond to the infinite
volume results. Squares and the dashed line are obtained
at finite volume, $R/L_S=12/32$.}
\label{Fig10}
\end{figure}

\begin{figure}[htb]
\begin{center}
\leavevmode
\epsfxsize=12cm\epsfbox{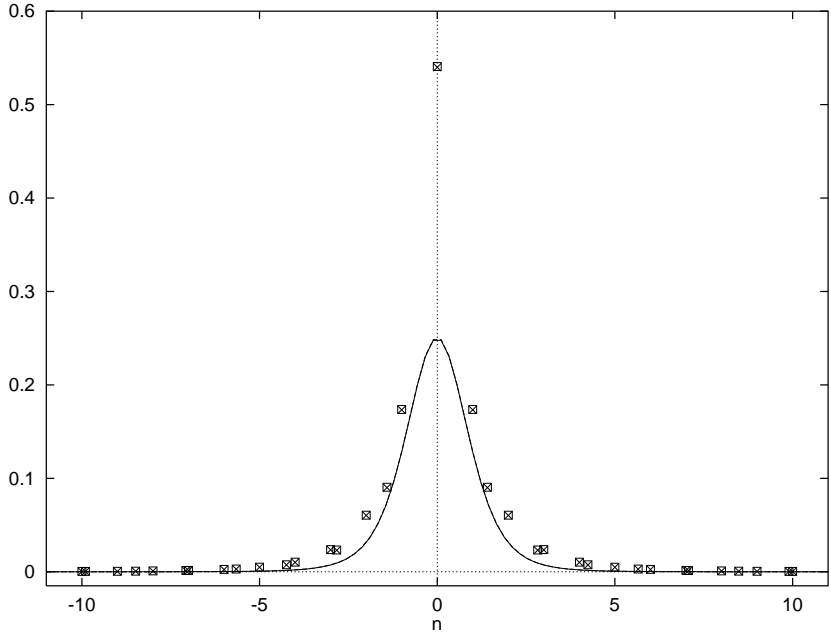}
\end{center}
\caption
{\em Same as Fig.~\protect\ref{Fig10} for $R=4$.}
\label{Fig11}
\end{figure}

In the continuum limit one finds the expression (from Eq.~\ref{a13}),
\begin{equation}
\label{a17}
\epsilon_R^{(c)}(0,n_{\perp})=
g^2C_F\frac{1}{(4\pi)^2}
\frac{R^2}{\left(R^2/4+n_{\perp}^2\right)^3}\quad,
\end{equation}
for the energy density distribution in the central transverse plane.
In Figs.~\ref{Fig10} and \ref{Fig11} we present 
a comparison of the dipole fields on finite ($32^3$) and infinite
lattices with their continuum forms\footnote{The continuum formula on
a finite lattice has been elaborated in Appendix~\ref{ap4}.}, for
separations $R=12$ and $R=4$,
respectively. The field positions are chosen
both along a transverse lattice axis and a plane-diagonal (multiples
of $\sqrt{2}$).

Up to order $g^4$ corrections,
perturbative lattice and continuum calculations
equally lead to\footnote{To obtain the continuum expression,
$\sum_{\mathbf n}\!a^3$ simply has to be replaced by
$\int\!d^3\!x$.}
\begin{equation}
\label{a16}
\sum_{\mathbf n}a^3\sigma_R({\mathbf n})
\approx \sum_{\mathbf n}a^3\epsilon_R({\mathbf n})
\approx \frac{V(R)}{a}\quad.
\end{equation}
As argued in the appendix, perturbation theory is expected to
describe the energy density better than the action density.

\subsection{Nonperturbative expectation}
In the limit of large $Q\bar{Q}$ separations, i.e.\ if the width of
the flux tube becomes small relative  to its length, an effective
relativistic string model is expected to describe the infra red
aspects of the interaction. Classical solutions of such string
Lagrangians predict, in agreement with the strong coupling 
expectation of pure
gauge theory, an area law of Wilson loops and, thus, a linearly rising
long range contribution to the potential. However, in reality, a quantum
mechanical string will fluctuate. An ultra violet cut-off
has to be imposed on the wave
length of such fluctuations, beyond which longitudinal degrees of
freedom become important and the (non renormalizable) string theory
looses its applicability. For a huge class of
string models the string fluctuations lead to a universal subleading
Coulomb type contribution~\cite{luscher1}, $-(d-2)\pi/(24R)$,
to the potential in the Gaussian
approximation ($d$ denotes the number of space-time dimensions).
For large $R$, excitations are expected to be separated from
the ground state by multiples of $\pi/R$~\cite{luscher1}.

The leading order expectation of string models for correlators of
smooth, large Wilson loops, ${\cal W}$ and $\Box$ with boundaries
$\partial{\cal W}$ and $\partial\Box$ is
\begin{equation}
\langle{\cal W}\Box\rangle-\langle{\cal W}\rangle\langle\Box\rangle
\propto
\exp\left(-KA(\partial{\cal W},\partial\Box)\right)\quad ,
\end{equation}
where $A(\partial{\cal W},\partial\Box)$ is the minimal area of a
surface with boundary $\partial{\cal W}\cup\partial\Box$ and $K$ is
the string tension.
Approximating the Wilson loop and the plaquette by circles,
the authors of
Ref.~\cite{luescher} obtained in the limit $n_{\perp}\ll R$
\begin{equation}
\langle E_1^2\rangle\propto\exp\left(-\frac{K\pi}{\ln{R}}
n_{\perp}^2\right)\quad.
\end{equation}
Thus, the central width of the fluctuating string
\begin{equation}
\label{width}
\delta_{\epsilon_R}^2=\langle n_{\perp}^2\rangle_{\epsilon_R}
=\frac{\int\!dn_{\perp}\,n_{\perp}^3\epsilon_R(0,n_{\perp})}
{\int\!dn_{\perp}\,n_{\perp}\epsilon_R(0,n_{\perp})}
\end{equation}
is expected to diverge logarithmically with the quark
separation:
\begin{equation}
\label{a18}
\delta_{\epsilon_R}^2=\delta_0^2\ln\frac{R}{R_0}\quad,
\end{equation}
where $R_0$ is an ultra violet
cut-off parameter. In a quantum mechanical
calculation, this relation has been
confirmed in the Gaussian approximation
for the probability of the fluctuating string, crossing the
central transverse plane $n_1=0$ at the position
$n_{\perp}$~\cite{luescher}.

For small distances, the perturbative result of Eq.~\ref{a17} suggests a
linear divergence of the width:
\begin{equation}
\label{a19}
\delta_{\epsilon_R}^2=\frac{R^2}{4}\quad.
\end{equation}

For $n_{\perp}\ll R$ (and large $r$) where the string picture 
is applicable a Gaussian transverse
profile of the flux tube is expected.
At large $n_{\perp}$, however, correlators of
(unsmeared) Wilson loops with plaquettes can be viewed as glueball
correlation functions in rotated space-time. Thus, for $n_{\perp}\gg
R, T$ an exponential form, governed by the
mass gap, might be expected\footnote{
However, the wave function of
the $Q\bar{Q}$ pair, created at $\tau=0$ has to be decayed into its
ground state, before the colour fields are measured at $\tau=T/2$.
Due to the structure of the action,
a Hamiltonian evolution in the strong coupling limit only
allows  hopping between neighbouring sites.
Thus, in this limit, communication occurs only 
between sites within the light
cone $n_{\perp}<T/2$ and the limit
$n_{\perp}\gg T$ is not justified.
As illustrated by the above example, the exponential
decay prediction for large $n_{\perp}$ has to be
taken with care.}.

\subsection{Sum rules}
Some important consistency conditions, relating the local chromo field
operators to the global $Q\bar{Q}$ potential have been derived by
Michael~\cite{sumrules}. In the following we will shortly recall these
sum rules.
More details and comments
related to contaminations from excited states can be found in
Appendices~\ref{ap2} and \ref{ap3}.

The action sum rule relates the action to the derivative of the
potential with respect to the inverse coupling
\begin{eqnarray}
\label{actsum3}
\sum_{\mathbf n}a^3\sigma_R({\mathbf n})&=&
-\frac{1}{a}\frac{\partial V(R)}{\partial\ln\beta}\\
&=&
-\frac{\partial\ln a}{\partial\ln\beta}v(R)-\frac{1}{a}\frac{\partial
V_0}{\partial\ln\beta}\quad ,
\end{eqnarray}
where we have decomposed the potential $V(R)=av(R)+V_0$ into a 
physical part, $v(R)$, that remains constant as $a\rightarrow 0$,
and a diverging self energy contribution, $V_0$.
The action sum rule is an exact relation.
It is in accord with the perturbative
expectation Eq.~\ref{a16}, which follows
by inserting the leading order expression for $V(R)$ 
(Eq.~\ref{a15}) into Eq.~\ref{actsum3}.

The energy sum rule involves derivatives with respect to
anisotropic spatial and temporal couplings.
After relating the latter to the isotropic lattice
coupling, $\beta$, via a weak coupling series,
one ends up with an approximate sum rule.
Thus, unlike the action sum rule,
the energy sum rule is not exact on the lattice. Here, we
just state the leading
order expectation
\begin{equation}
\label{ensum3}
\sum_{\mathbf n}a^3\epsilon_R({\mathbf
n})=\left(v(R)+\frac{V_0}{a}\right)\left(1+{\cal O}(\beta^{-1})\right)\quad .
\end{equation}
The correction to this energy conserving rule
reflects the fact that the local plaquette operator
undergoes a renormalization. However, mean field arguments
(Appendix~\ref{ap1}) suggest only small corrections.
The energy sum rule is also in accord with
Eq.~\ref{a16}.

The leading order contribution
to the self energy $V_0$, $C_FG_L({\mathbf 0})g^2$,
merely changes sign when differentiated with
respect to $\ln\beta$. As a consequence, this contribution to both
the action and the energy sums diverges like $1/(a\ln a)$.  Due to the
localization of the self energy to the vicinity of the two
sources, the peaks of the distributions diverge like $1/(a^4\ln a)$ in
physical units (or like $1/\ln a$ when measured in lattice units).
The physical part $v(R)$ on the r.h.s.\ of the action sum rule
is accompanied by an anomalous dimension
$\partial\ln a/\partial\ln\beta\propto\ln a$.
For this reason, the
measured lattice action density $\sigma a^4$ is expected to scale like
$a^4\ln a$ outside the peaks while the energy density vanishes like
$a^4$ (in lattice units). ${\cal E}$ and ${\cal B}$ mix
under renormalization group transformations since the sum of both
densities carries dimension $a^4$ while its difference is accompanied
by $a^4\ln a$. Thus, only the energy density and the combination
$(\partial\ln\beta/\partial\ln a)\sigma$
are relevant to the continuum limit.

\section{Results}

\subsection{Static potential}

\begin{figure}[htb]
\begin{center}
\mbox{\epsfxsize=12cm\epsfbox{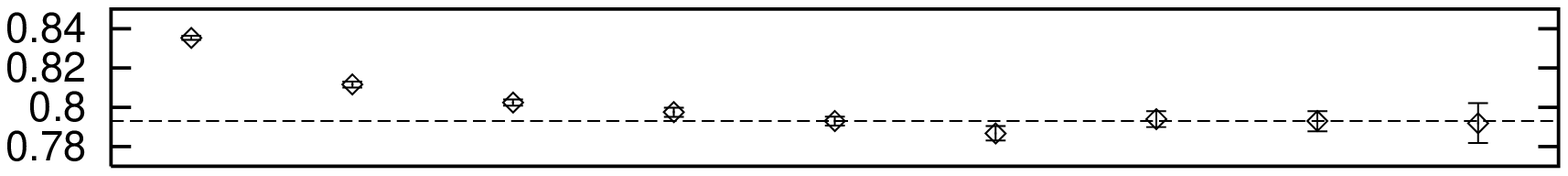}}\vspace{-.2cm}
\mbox{\epsfxsize=12cm\epsfbox{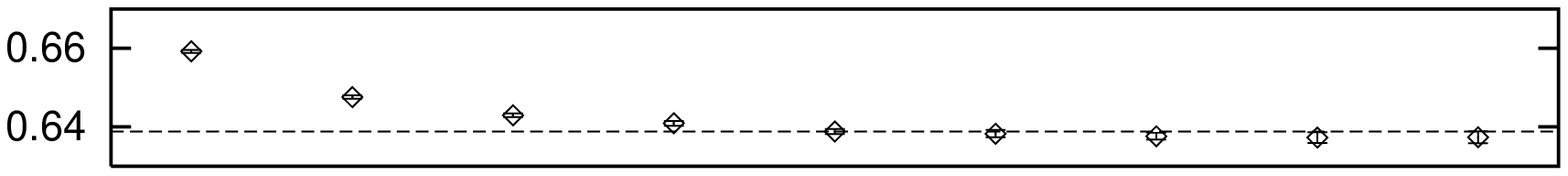}}\vspace{-.2cm}
\mbox{\epsfxsize=12cm\epsfbox{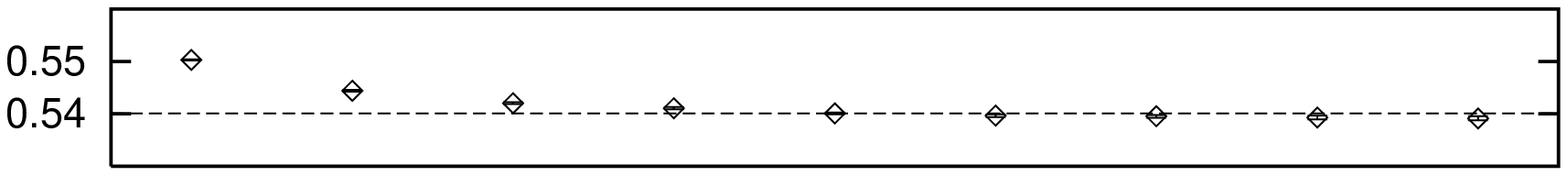}}\vspace{-.2cm}
\mbox{\epsfxsize=12cm\epsfbox{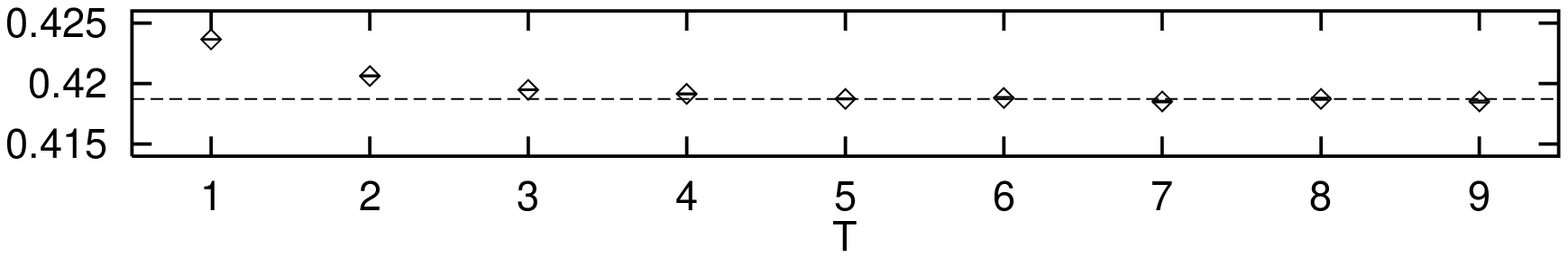}}\vspace{-.2cm}
\end{center}
\caption
{\em $V(R,T)$ as a function of $T$ for $R=20$, $R=10$, $R=5$ and
$R=2$, respectively, at $\beta=2.635$. The plateau values obtained at
$T=5$ are indicated by the horizontal dashed lines.}
\label{FigT}
\end{figure}

\begin{figure}[phtb]
\begin{center}
\leavevmode
\epsfxsize=15cm\epsfbox{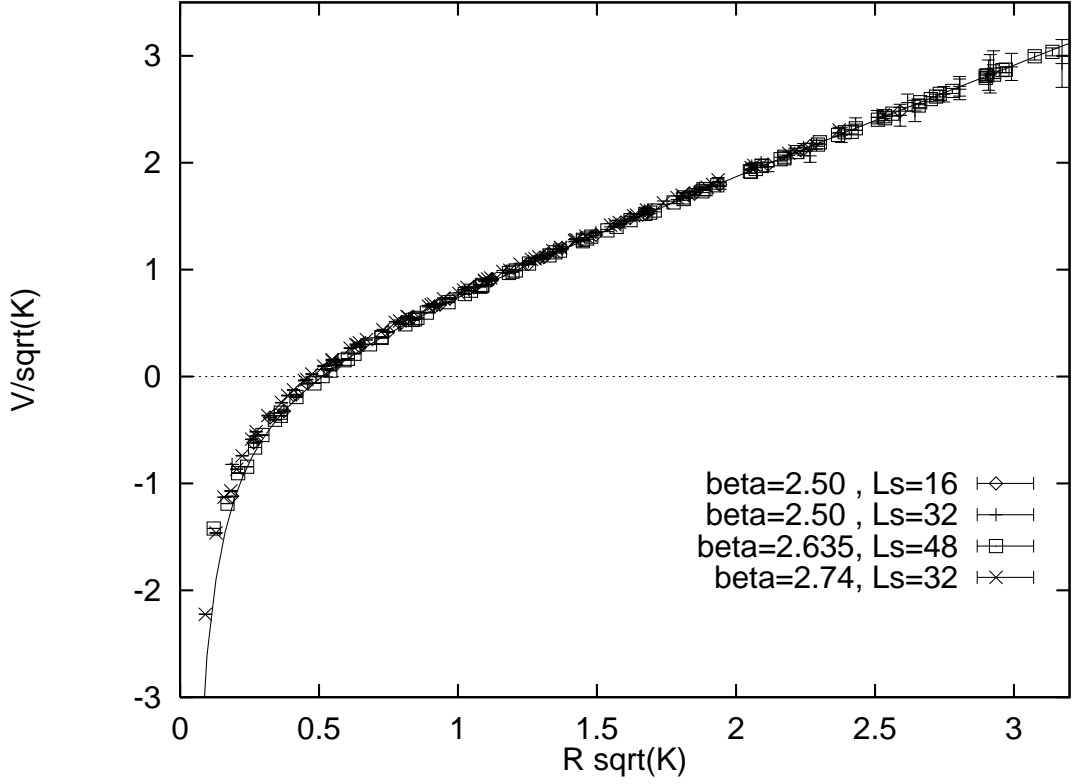}
\end{center}
\caption
{\em The potential, measured on the four lattices, scaled in units of
the string tension. The solid line refers to the string picture
expectation $V(R)=KR-\pi/(12R)$.}
\label{FIGP}
\end{figure}

To prepare the stage for the colour flux investigations,
we had to calculate the static
$Q\bar{Q}$ potential. It will render the  string tension,
which serves to fix the physical scale 
and relates the results, obtained at different $\beta$ values
to each other.
\subsubsection{Data}

The potential data has  been obtained by the method described in
Section~\ref{GSE}. Finite $T$ approximants $V(R,T)$ and
$C_0(R,T)$ to the ground state potential values and overlaps are
computed according to Eqs.~\ref{hh1} and \ref{hh7}. These are traced
until  a plateau (in $T$) is reached. 
The numerical situation  is illustrated in Fig.~\ref{FigT}
for a few typical quark
separations at $\beta=2.635$. For the $16^4$ lattice at
$\beta=2.5$ the $T_{\min}=3$
approximant has been found to agree with the plateau values while
for the $32^4$ lattices at $\beta=2.5$ and $\beta=2.74$,
$T_{\min}=4$ had to be taken
and at $\beta=2.635$ we went as far as $T_{\min}=5$. To exclude any
remaining systematic bias on the fitted parameters, all fits have been
performed for $T=T_{\min}$, and $T=T_{\min}+1$. Within statistical errors and
fixed $R$ range, the fit parameters remained stable. 
For larger $\beta$ values, the
reduced  physical $t=Ta$ separations
appear to be  partly compensated by better ground state overlaps.

We note, that the actual value of
$T_{\min}$ is not only affected by the ground state overlaps
but also influenced by
statistical errors that depend on
the particular number of measurements, physical volume and method
(link integration).

\begin{table}[p]
\begin{center}
\small
\begin{tabular}{|c|c|c|c||c|c|c|c|}\hline
$R$&Path&$V(R)$&$C_0(R)$&$R$&Path&$V(R)$&$C_0(R)$\\\hline
 1.00&1&.3356( 2)&.979(1)& 4.90&5&.6671(10)&.927(3)\\
 1.41&2&.4292( 4)&.975(1)& 5.00&1&.6706( 9)&.930(3)\\
 1.73&4&.4735( 5)&.970(2)& 5.20&4&.6817(15)&.928(5)\\
 2.00&1&.4844( 4)&.963(1)& 5.66&2&.7012(14)&.921(4)\\
 2.24&3&.5137( 4)&.963(2)& 6.00&1&.7137(11)&.915(3)\\
 2.45&5&.5330( 5)&.960(2)& 6.00&6&.7157(13)&.916(4)\\
 2.83&2&.5585( 7)&.954(2)& 6.71&3&.7418(13)&.902(4)\\
 3.00&1&.5660( 6)&.956(2)& 6.93&4&.7532(23)&.904(6)\\
 3.00&6&.5698( 5)&.952(2)& 7.00&1&.7537(12)&.904(4)\\
 3.46&4&.5969(10)&.945(3)& 7.07&2&.7586(17)&.904(5)\\
 4.00&1&.6229( 6)&.939(2)& 7.35&5&.7701(14)&.901(4)\\
 4.24&2&.6379(10)&.942(3)& 8.00&1&.7931(16)&.890(5)\\\cline{5-8}
 4.47&3&.6479( 8)&.934(3)\\\cline{1-4}
\end{tabular}
\end{center}
\caption{\em Potential and overlap values at $\beta=2.5$.}
\label{Tab2a}
\end{table}

\begin{table}[t]
\begin{center}
\small
\begin{tabular}{|c|c|c|c||c|c|c|c|}\hline
$R$&Path&$V(R)$&$C_0(R)$&$R$&Path&$V(R)$&$C_0(R)$\\\hline
 1.41&2& .37719( 4) & .9799( 1)& 6.93&4& .5826( 5) & .921(2)\\
 1.73&4& .41154( 7) & .9754( 3)& 7.00&1& .5839( 4) & .923(2)\\ 
 2.00&1& .41873( 6) & .9694( 2)& 7.07&2& .5853( 5) & .922(2)\\ 
 2.24&3& .44040( 8) & .9694( 3)& 7.35&5& .5909( 4) & .919(2)\\ 
 2.45&5& .45445( 9) & .9674( 4)& 8.00&1& .6033( 4) & .913(2)\\ 
 2.83&2& .47163(12) & .9627( 5)& 8.49&2& .6123( 7) & .908(3)\\
 3.00&1& .47604(11) & .9620( 4)& 8.66&4& .6154( 7) & .907(3)\\
 3.00&6& .47937(15) & .9610( 6)& 8.94&3& .6202( 7) & .902(3)\\
 3.46&4& .49685(16) & .9563( 7)& 9.00&1& .6211( 5) & .904(2)\\
 4.00&1& .51239(14) & .9508( 6)& 9.00&6& .6215( 7) & .902(3)\\
 4.24&2& .52051(24) & .9512(10)& 9.80&5& .6352( 7) & .894(3)\\
 4.47&3& .52678(18) & .9465( 7)& 9.90&2& .6367( 8) & .895(4)\\
 4.90&5& .53802(24) & .9421(10)&10.00&1& .6388( 7) & .895(3)\\
 5.00&1& .54008(20) & .9435( 8)&10.39&4& .6453(10) & .888(5)\\
 5.20&4& .54525(33) & .9403(14)&11.00&1& .6556( 6) & .887(3)\\
 5.66&2& .55598(29) & .9342(12)&11.18&3& .6580(10) & .881(4)\\
 6.00&1& .56349(26) & .9326(13)&11.31&2& .6608(11) & .881(5)\\
 6.00&6& .56369(34) & .9303(14)&12.00&1& .6726(10) & .880(4)\\
 6.71&3& .57818(36) & .9248(16)&12.00&6& .6717(12) & .872(5)\\\hline 
\end{tabular}
\end{center}
\caption{\em Potential and overlap values at $\beta=2.635$.}
\label{Tab2}
\end{table}

\begin{table}[t]
\begin{center}
\small
\begin{tabular}{|c|c|c|c||c|c|c|c|}\hline
$R$&Path&$V(R)$&$C_0(R)$&$R$&Path&$V(R)$&$C_0(R)$\\\hline
 1.00&1&.2786(1)&.959(4)& 8.94&3&.5325( 5)&.913(4)\\
 1.41&2&.3480(2)&.963(4)& 9.00&1&.5327( 5)&.921(4)\\
 1.73&4&.3780(2)&.953(4)& 9.00&6&.5329( 5)&.917(4)\\
 2.00&1&.3837(2)&.941(4)& 9.80&5&.5413( 5)&.909(4)\\
 2.24&3&.4021(2)&.945(4)& 9.90&2&.5430( 6)&.926(4)\\
 2.45&5&.4137(2)&.943(4)&10.00&1&.5434( 6)&.910(4)\\
 2.83&2&.4273(3)&.933(4)&10.39&4&.5474( 7)&.907(4)\\
 3.00&1&.4308(2)&.941(4)&11.00&1&.5533( 6)&.913(4)\\
 3.00&6&.4340(3)&.937(4)&11.18&3&.5550( 6)&.912(4)\\
 3.46&4&.4483(4)&.930(4)&11.31&2&.5571( 6)&.906(4)\\
 4.00&1&.4588(3)&.927(4)&12.00&1&.5635( 6)&.904(4)\\
 4.24&2&.4654(3)&.943(4)&12.00&6&.5633( 6)&.901(4)\\
 4.47&3&.4698(3)&.926(4)&12.12&4&.5653( 9)&.915(5)\\
 4.90&5&.4780(3)&.924(4)&12.25&5&.5663( 8)&.911(4)\\
 5.00&1&.4788(3)&.932(4)&12.73&2&.5711( 8)&.917(4)\\
 5.20&4&.4831(4)&.935(4)&13.00&1&.5730( 7)&.908(4)\\
 5.66&2&.4904(4)&.922(4)&13.42&3&.5771( 7)&.898(4)\\
 6.00&1&.4950(3)&.921(4)&13.86&4&.5811( 9)&.895(4)\\
 6.00&6&.4955(4)&.919(4)&14.00&1&.5825( 9)&.898(4)\\
 6.71&3&.5055(4)&.927(4)&14.14&2&.5842(10)&.897(5)\\
 6.93&4&.5089(5)&.919(4)&14.70&5&.5888( 9)&.892(5)\\
 7.00&1&.5085(3)&.925(4)&15.00&1&.5920( 9)&.903(5)\\
 7.07&2&.5104(4)&.933(4)&15.00&6&.5916( 9)&.896(4)\\
 7.35&5&.5135(5)&.926(5)&15.56&2&.5980(10)&.908(5)\\
 8.00&1&.5214(5)&.916(4)&15.59&4&.5973(10)&.902(5)\\
 8.49&2&.5273(5)&.915(4)&15.65&3&.5987( 9)&.900(5)\\
 8.66&4&.5284(7)&.923(4)&16.00&1&.6014( 8)&.891(4)\\\hline
\end{tabular}
\end{center}
\caption{\em Potential and overlap values at $\beta=2.74$.}
\label{Tab3}
\end{table}

In Tabs.~\ref{Tab2a}--\ref{Tab3} we have collected results
on the potential values, $V(R)$, and overlaps, $C_0(R)$,
up to a physical distance
of about $.7$~fm. This scale has been obtained from the relation 
$\kappa=\sqrt{K}a=440$~MeV.
For larger separations we refer to the
parametrizations presented below since no systematic
deviations are observed from the interpolating curve
(that is dominated by the
linear part of the potential).
Remember, that all estimates
for the potential and overlaps
constitute strict upper limits to their asymptotic
($T\rightarrow\infty$) values.
The paths, displayed in the second column, are numbered according to
Eq.~\ref{aa4}.

In Fig.~\ref{FIGP} we
show the familiar scaling plot for the potential in form
of the combination $(V(R\sqrt{K})-V_0)/\sqrt{K}$
with $V_0$ and $K$ as  obtained from the four-
parameter fits, described below.
Notice, that we can trace the potential up to the impressively 
large separation of 2.3~fm! The curve represents
the string picture prediction $R-\frac{\pi}{12R}$.
The nice scaling between the potentials illustrates the
restoration of continuum rotational invariance at remarkably small
lattice separations.

\subsubsection{Potential fits}
\label{PF}
Our potential values have been fitted to the parametrizations
\begin{equation}
\label{3par}
V(R)=V_0+KR-\frac{e}{R}
\end{equation}
and
\begin{equation}
\label{4par}
V({\mathbf
R})=V_0+KR-\frac{e}{R}+f\left(\frac{1}{R}-\left[\frac{1}{\mathbf
R}\right]\right)
\end{equation}
with
\begin{equation}
\left[\frac{1}{\mathbf R}\right]=4\pi G_L({\mathbf R})\quad,
\end{equation}
where $G_L({\mathbf R})$ is the lattice gluon propagator
(Eqs.~\ref{a10}, \ref{a11}), computed on an infinite\footnote{The
lattice sums have been computed numerically on $1024^3$, $2048^3$, and
$4096^3$ lattices and extrapolated in $1/L_S$ to their infinite volume
limits.} lattice. $V_0$, $K$, $e$, and $f$ are the fit parameters.

In the analysis we followed the fitting procedure, described in
Ref.~\cite{balifijue}. Four different fit algorithms have been applied
to the data:
\begin{itemize}
\item{uncorrelated fits with the errors of potential values obtained
on the original sample (UN),}
\item{uncorrelated fits with errors calculated for each bootstrap
separately with a subbootstrap (UB),}
\item{correlated fits with the covariance matrix computed on the
original sample (CN),}
\item{correlated fits with covariance matrices calculated on each
bootstrap separately (CB).}
\end{itemize}
The fit range has been adapted automatically. For each range, a
quality parameter,
\begin{equation}
Q=\alpha\frac{N_{DF}}{N_{DF,\max}}\frac{K}{\Delta K}\quad,
\end{equation}
of the fit has been computed from the confidence level, $\alpha$.
The largest quality corresponds to the ``best'' fit range. As a
systematic error we have taken the scatter between the fit parameters
{}from fits with $Q\geq\frac{3}{4}Q_{\max}$. 

\begin{table}[t]
\begin{center}
\footnotesize
\begin{tabular}{|c|c|c|c|c|c|}
\multicolumn{6}{c}{$\beta=2.5\quad,\quad L_S\times L_T =16^4$}\\\hline
&$V_0$   & $K$       & $e$    &
${\scriptstyle R_{min},R_{max}}$&$\chi^2/N_{DF}$\\\hline
UN&.545(3)( 4) &.0348(3)(4)&.240(9)(32)&3.00,13.86&$17/26$\\
UB&.547(3)( 8) &.0346(3)(7)&.244(8)(21)&3.00,13.86&$19/26$\\
CN&.547(3)( 8) &.0346(3)(7)&.244(8)(20)&3.00,13.86&$20/26$\\ 
CB&.546(4)(14)&.0348(4)(9)&.241(9)(44) &3.00,13.86&$19/26$\\\hline
\multicolumn{6}{c}{$\beta=2.5\quad,\quad L_S\times L_T =32^4$}\\\hline
&$V_0$   & $K$       & $e$    &
${\scriptstyle R_{min},R_{max}}$&$\chi^2/N_{DF}$\\\hline
UN&.542(3)(12)&.0354(3)(12)&.24(1)(3)&3.00,24.25&$50/59$\\
UB&.553(4)(20)&.0345(3)(14)&.27(1)(5)&4.24,24.25&$34/56$\\
CN&.553(4)(30)&.0345(4)(14)&.27(1)(7)&4.24,24.25&$48/56$\\
CB&.560(7)(20)&.0340(5)(18)&.29(2)(8)&4.00,17.32&$35/44$\\\hline
\multicolumn{6}{c}{$\beta=2.635\quad,\quad L_S\times L_T =48^3\times 64$}\\\hline
&$V_0$   & $K$       & $e$    &
${\scriptstyle R_{min},R_{max}}$&$\chi^2/N_{DF}$\\\hline
UN&.523(1)(6) &.01451( 4)(14)&.269( 2)(15)&4.00,41.57&$70/95$\\
UB&.519(1)(3) &.01466( 5)(15)&.250(10)(15)&4.00,41.57&$91/95$\\
CN&.519(1)(2) &.01466( 5)(14)&.251(12)(16)&4.00,41.57&$92/95$\\
CB&.518(7)(8) &.01467(25)(31)&.242(48)(62)&4.00,41.57&$95/95$\\\hline
\multicolumn{6}{c}{$\beta=2.74\quad,\quad L_S\times L_T =32^4$}\\\hline
&$V_0$   & $K$       & $e$    &
${\scriptstyle R_{min},R_{max}}$&$\chi^2/N_{DF}$\\\hline
UN&.4816(12)(10)&.00834(6)( 5)&.217(3)( 5)&6.00,27.71&$36/51$\\
UB&.4816( 9)(27)&.00834(4)(12)&.217(4)(14)&6.00,27.71&$47/51$\\
CN&.4816( 9)(31)&.00834(4)(13)&.217(4)(17)&6.00,27.71&$47/51$\\
CB&.4817(14)(55)&.00834(6)(37)&.218(6)(89)&6.00,27.71&$48/51$\\\hline
\end{tabular}
\end{center}
\caption{\em Three-parameter fits according to Eq.~\protect\ref{3par}.
The first column labels the fit algorithm.
In the last two columns, the ``best'' fit range and corresponding
$\chi^2/N_{DF}$ 
values are stated. The first errors are statistical only, the second
errors include systematic uncertainties.}\label{fitresults1}
\end{table}

\begin{table}[t]
\begin{center}
\footnotesize
\begin{tabular}{|c|c|c|c|c|c|c|}
\multicolumn{7}{c}{$\beta=2.5\quad,\quad L_S\times L_T =16^4$}\\\hline
&$V_0$   & $K$       & $e$    &
$f$&${\scriptstyle R_{min},R_{max}}$&$\chi^2/N_{DF}$\\\hline
UN&.545(1)(3)&.0349(2)(4)&.242(2)(5)&.244(7)(14)&1.73,13.86&$19/30$\\
UB&.545(1)(3)&.0349(2)(3)&.242(2)(5)&.248(7)(26)&1.73,13.86&$19/30$\\
CN&.545(1)(3)&.0350(2)(4)&.242(2)(5)&.248(7)(28)&1.73,13.86&$20/30$\\
CB&.545(1)(3)&.0350(3)(4)&.242(2)(5)&.245(8)(27)&1.73,13.86&$20/30$\\
\hline
\multicolumn{7}{c}{$\beta=2.5\quad,\quad L_S\times L_T =32^4$}\\\hline
&$V_0$   & $K$       & $e$    &
$f$&${\scriptstyle R_{min},R_{max}}$&$\chi^2/N_{DF}$\\\hline
UN&.550(2)( 6)&.0347(2)( 5)&.256(5)(14)&.30(2)( 9)&2.83,24.25&$45/60$\\
UB&.547(2)(10)&.0350(2)( 9)&.245(4)(23)&.34(2)(14)&3.00,17.32&$37/46$\\
CN&.547(2)( 9)&.0350(2)(12)&.245(5)(24)&.34(2)(15)&3.00,17.32&$37/46$\\
CB&.549(3)(14)&.0348(3)(12)&.251(9)(36)&.33(4)(14)&3.00,17.15&$37/45$\\
\hline
\multicolumn{7}{c}{$\beta=2.635\quad,\quad L_S\times L_T =48^3\times 64$}\\\hline
&$V_0$   & $K$       & $e$    &
$f$&${\scriptstyle R_{min},R_{max}}$&$\chi^2/N_{DF}$\\\hline
UN&.519(1)(2)&.01467(3)(14)&.254(2)(9)&.29(2)(11)&3.00,39.84&$67/95$\\
UB&.521(1)(2)&.01459(2)( 8)&.263(2)(6)&.23(3)( 9)&4.24,41.57&$74/93$\\
CN&.521(1)(2)&.01458(2)( 8)&.263(1)(7)&.15(2)( 8)&4.00,36.00&$71/90$\\
CB&.523(1)(3)&.01451(7)(13)&.267(4)(9)&.16(7)(15)&4.00,27.00&$56/76$\\\hline
\multicolumn{7}{c}{$\beta=2.74\quad,\quad L_S\times L_T =32^4$}\\\hline
&$V_0$   & $K$       & $e$    &
$f$&${\scriptstyle R_{min},R_{max}}$&$\chi^2/N_{DF}$\\\hline
UN&.4823(4)( 7)&.00832(3)(5)&.2220( 9)(21)&.21(1)(5)&2.45,26.00&$38/62$\\
UB&.4828(3)( 8)&.00830(2)(4)&.2235( 9)(26)&.19(1)(3)&2.83,27.71&$51/62$\\
CN&.4828(3)( 5)&.00830(2)(3)&.2235( 9)(15)&.19(1)(2)&2.83,27.71&$51/62$\\
CB&.4827(5)(10)&.00830(3)(6)&.2234(14)(34)&.20(2)(5)&2.83,26.00&$51/61$\\\hline
\end{tabular}
\end{center}
\caption{\em Four-parameter fits according to Eq.~\protect\ref{4par}.
The first column labels the fit algorithm.
In the last two columns, the ``best'' fit range and corresponding
$\chi^2/N_{DF}$ values
are stated. The first errors are statistical only, the second
errors include systematic uncertainties.}\label{fitresults2}
\end{table}

In Tabs.~\ref{fitresults1} and
\ref{fitresults2} the results are listed for three- and four-
parameter fits, respectively, together with their statistical and
systematic uncertainties. In addition, the ``best'' fit ranges and
corresponding $\chi^2$ values are included. 
Correlated and uncorrelated fits show
little difference in $\chi^2$ values, which gives evidence that
correlations between potential data at different $R$-values
are small. In the
further analysis we use the results of the CN four-
parameter fits for the string
tension $K$ and self energy $V_0$.

As expected from perturbation theory the self energy, $V_0$,
decreases slightly with $\beta$ from $V_0=.547$ at $\beta=2.5$ down to
$V_0=.483$ at $\beta=2.74$. On the two large lattice volumes, the
Coulomb coefficients, $e$,
are in agreement with the value $\pi/12\approx .262$,
as expected in the string picture. On the smaller lattice volumes
they come out somewhat smaller. This might be a result of the different fit
ranges: on the large lattices the fit result is dominated by the
potential values from separations where the string picture is expected
to be applicable. The parameter $f$, used to account for the lattice
symmetry turns out to be of approximately the same size as $e$ in all
cases, indicating that violations of rotational symmetry can be
understood in terms of the lattice one gluon exchange, contrary to
$SU(3)$ where values $f\approx .6e$ have been found~\cite{wuppot}. 

\subsubsection{Determination of the $\beta$-function}
{}From Eq.~\ref{actsum} it is evident that the action density diverges like
\begin{equation}
\tilde{B}^{-1}(\beta)=
\frac{\partial \ln a}{\partial\ln\beta}=
-\frac{\partial g}{\partial\ln\beta}B^{-1}(g)
=\frac{g}{2}B^{-1}(g)
\end{equation}
as a function of $a$ (see Eq.~\ref{actsum3}),
where
\begin{equation}
\label{bdef}
B(g)=-\frac{\partial g}{\partial\ln a}
\end{equation}
denotes the Callan-Symanzik $\beta$-function.
For this reason, we
shall set out in this section to
determine the $\beta$-function
within the $g^2=2N/\beta$ region covered by our simulations.

The $\beta$-function can be expanded in terms of the coupling:
\begin{equation}
\label{bdef2}
B(g)=-b_0g^3-b_1g^5-\cdots
\end{equation}
with $b_0=11N/(48\pi^2)$ and $b_1=34N^2/(3(16\pi^2)^2)$. From
Eqs.~\ref{bdef} and \ref{bdef2} one finds the familiar formula
\begin{equation}
\label{bdef3}
a=\Lambda_L^{-1}f(\beta)\left(1+{\cal O}(\beta^{-1})\right)
\quad\mbox{with}\quad
f(\beta)=e^{-\beta/4Nb_0}(\beta/2Nb_0)^{b_1/2b_0^2}\quad.
\end{equation}

\begin{figure}[h]
\begin{center}
\leavevmode
\epsfxsize=12cm\epsfbox{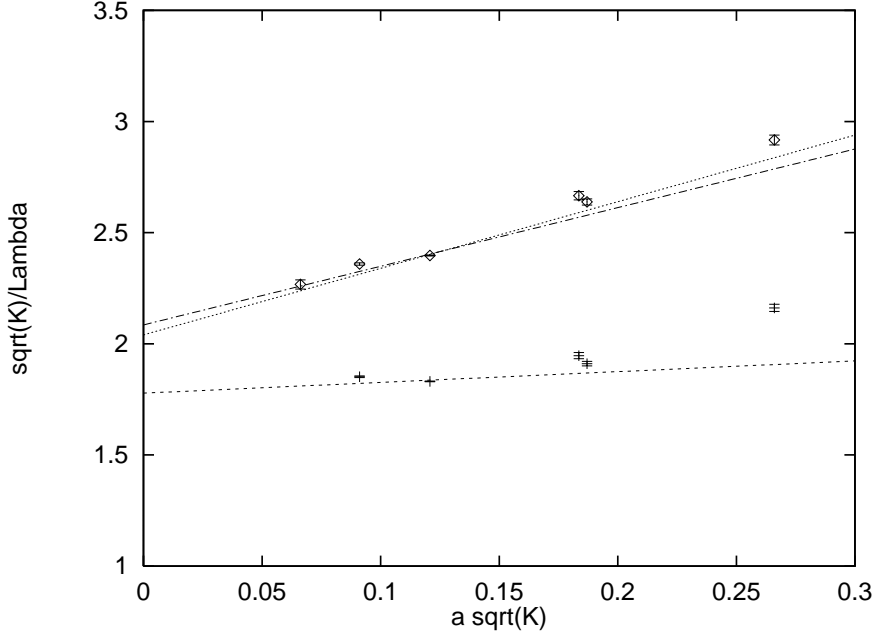}
\end{center}
\caption[ ]{\em 
$\Lambda_{\overline{MS}}^{-1}(a)\sqrt{\kappa}$ versus the lattice spacing
$a\sqrt{\kappa}$, both measured in units of the string tension.
The estimates for the $\Lambda$ values have been obtained from
the perturbative two loop formula by use of the bare coupling (upper
values) and the $\beta_E$ scheme (lower values). Linear
fits to the data are indicated.}
\label{Figscale}
\end{figure}

\begin{table}[t]
\begin{center}
\begin{tabular}{|c|c|c|c|c|}\hline
$\beta$&$\sqrt{K}$&$\langle\Box\rangle$&$-\tilde{B}^{-1}_{2loop}(\beta)$
&$-\tilde{B}^{-1}(\beta)$\\\hline
2.3  &  .3690(30) & .39746 (1)&5.77&7.20(13)\\
2.4  &  .2660(20) & .36352 (1)&6.04&7.28( 9)\\
2.5  &  .1870(10) & .34802 (1)&6.31&7.36( 8)\\
2.5115& .1836(13) & .34564 (1)&6.34&7.38( 8)\\
2.635&  .1208( 1) & .324308(2)&6.67&7.51( 6)\\
2.74 &  .0911( 2) & .308721(2)&6.95&7.63( 5)\\
2.85 &  .0662( 6) &    ---    &7.25&7.81( 4)\\\hline
\end{tabular}
\end{center}
\caption{\em The ``$\beta$-function'' $\tilde{B}^{-1}(\beta)={\partial(\ln
a)}/{\partial(\ln\beta)}$, obtained by use of the perturbative two loop
approximation, and by the interpolation procedure, described in the text.
The $\beta=2.3$ and $\beta=2.4$ values are taken from
Ref.~\protect\cite{MT,PHC}, the $\beta=2.5115$ value is from
Ref.~\protect\cite{balifi}, the $\beta=2.85$ value from
Ref.~\protect\cite{ukqcdpot}.}
\label{Tabscale}
\end{table}

In Tab.~\ref{Tabscale}, results for the square root of the string
tension, $\sqrt{K}$, and the plaquette expectation value are collected
for various $\beta$.
The results are taken from the present
simulation and Refs.~\cite{MT,PHC,balifi,ukqcdpot}.
{}From Eq.~\ref{bdef3} we obtain
$f(\beta)\sqrt{K}=\sqrt{\kappa}\Lambda^{-1}_L(a)$
by using
$a^2=K/\kappa$. 
In Fig.~\ref{Figscale}
the $a$ dependence of $\Lambda_{\overline{MS}}(a)=
19.82\Lambda_L(a)$ is shown. As can be seen from the non-vanishing
slope, within the $\beta$ region accessible by present computers,
higher order contributions to the asymptotic formula
Eq.~\ref{bdef3} are important.
A cut-off parameter
$\Lambda_E(a)$ from the effective
coupling $\beta_E=\frac{3}{4}(1-\langle\Box\rangle)^{-1}$, introduced
in Ref.~\cite{karschpet}, has also been computed.
This cut-off parameter is
translated into the $\overline{MS}$ scheme by the relation
$\Lambda_{\overline{MS}}(a)=11.51\Lambda_E(a)$.
As can be seen, the slope is substantially reduced
but asymptotic scaling remains violated.

 The difference between the $\Lambda^{-1}(a)$ sets
indicates the size of
higher order (perturbative and non-perturbative)
contributions to the $\beta$-function.
Within the present $\beta$ range we find the points to fall
quite well on straight lines
(see Fig.~\ref{Figscale})\footnote{However, for small $a$, the leading order
correction should be proportional to $g^2\propto 1/\ln a$, instead.
So, it is not surprising that the linear {\em
effective} parametrizations do not extrapolate to the same continuum
limit.}; so we  use a linear interpolation
of our data in the region  $2.4\leq\beta\leq 2.85$,
according to the parametrization
\begin{equation}
\Lambda^{-1}_L(a)=\Lambda^{-1}_L(0)+\eta a \quad.
\end{equation}
{}From the  fitted slope, $\eta$, we obtain the relation
\begin{equation}
a=\Lambda_L^{-1}(a=0)\frac{f(\beta)}{1-\eta f(\beta)}
\end{equation}
and arrive at
\begin{equation}
\tilde{B}^{-1}(\beta)=\frac{\partial\ln
f(\beta)}{\partial\ln\beta}\left(1-\eta f(\beta)\right)^{-1}
=\left(-\frac{\beta}{4Nb_0}+\frac{b_1}{2b_0^2}\right)
\left(1-\eta f(\beta)\right)^{-1}\quad.
\end{equation}

The resulting values for $-\tilde{B}^{-1}$, obtained in the two loop
approximation and by the above fit, are displayed in the fourth and
fifth column of Tab.~\ref{Tabscale}, respectively. Depending on how
many of the points we include  into our fit, we obtain values
$52.2<\eta/\sqrt{\kappa}<59.5$.
This systematic uncertainty is incorporated into the
errors of the $\tilde{B}^{-1}$ values in the
last column. Watch the difference between the
``measured'' $\beta$-function and the corresponding perturbative
expression decrease with the lattice spacing, as it should be!

\subsection{Colour field distributions}
\subsubsection{General features}

\begin{figure}[htb]
\begin{center}
\leavevmode
\epsfxsize=14cm\epsfbox{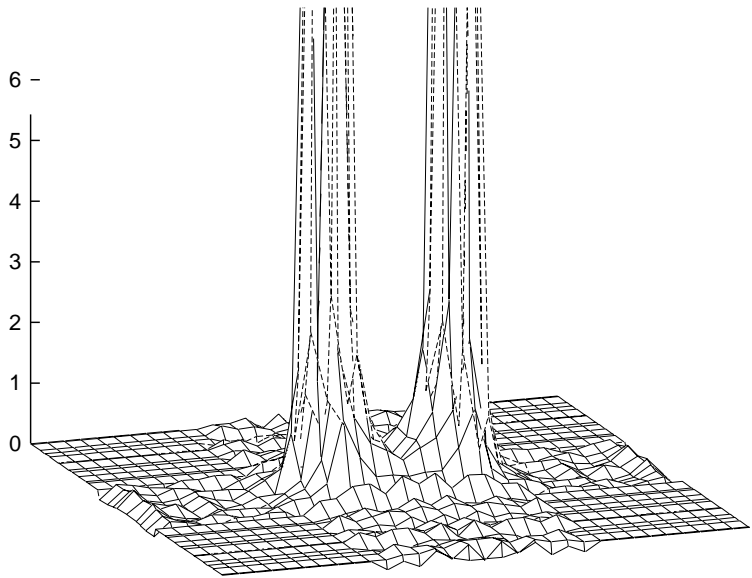}
\end{center}
\vspace{-1.5cm}
\caption
{\em The energy
density distribution at $\beta=2.5$, $R=8$ ($r\approx .7$~fm)
in units of the string tension.}
\label{Fige1}
\end{figure}

\begin{figure}[htb]
\begin{center}
\leavevmode
\epsfxsize=14cm\epsfbox{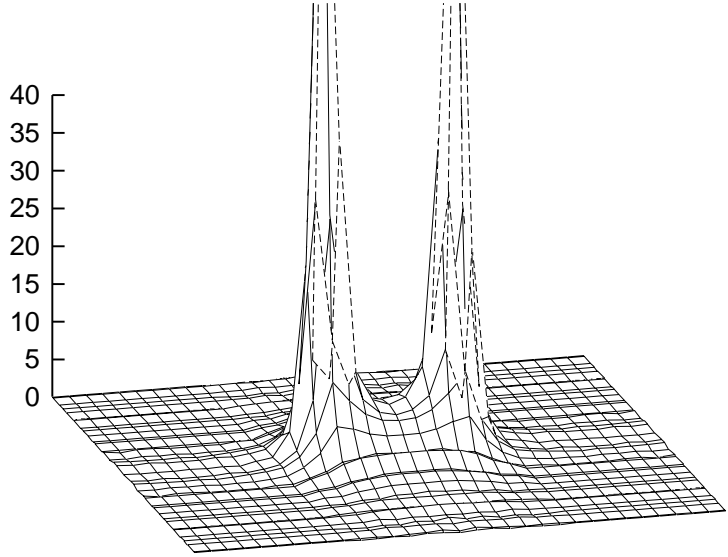}
\end{center}
\vspace{-1.5cm}
\caption
{\em The action density distribution at $\beta=2.5$, $R=8$
($r\approx .7$~fm) in units of
the string tension.}
\label{Figs1}
\end{figure}

\begin{figure}[htb]
\begin{center}
\leavevmode
\epsfxsize=14cm\epsfbox{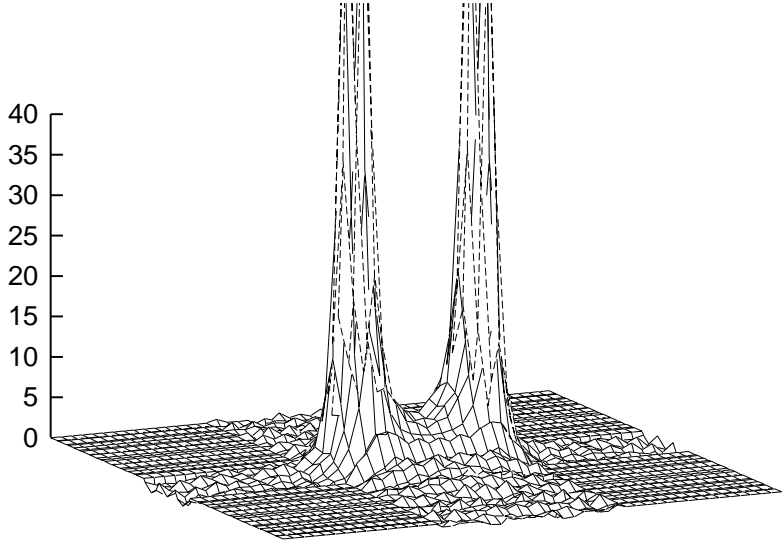}
\end{center}
\vspace{-1.5cm}
\caption
{\em The action density distribution at $\beta=2.635$, $R=12$
($r\approx .7$~fm) in units of
the string tension. Relative
to Fig.~\protect\ref{Figs1}
the vertical axis has been rescaled by the ratio of the
corresponding $\tilde{B}(\beta)$ values.}
\label{Figs4}
\end{figure}

\begin{figure}[htb]
\begin{center}
\leavevmode
\epsfxsize=12cm\epsfbox{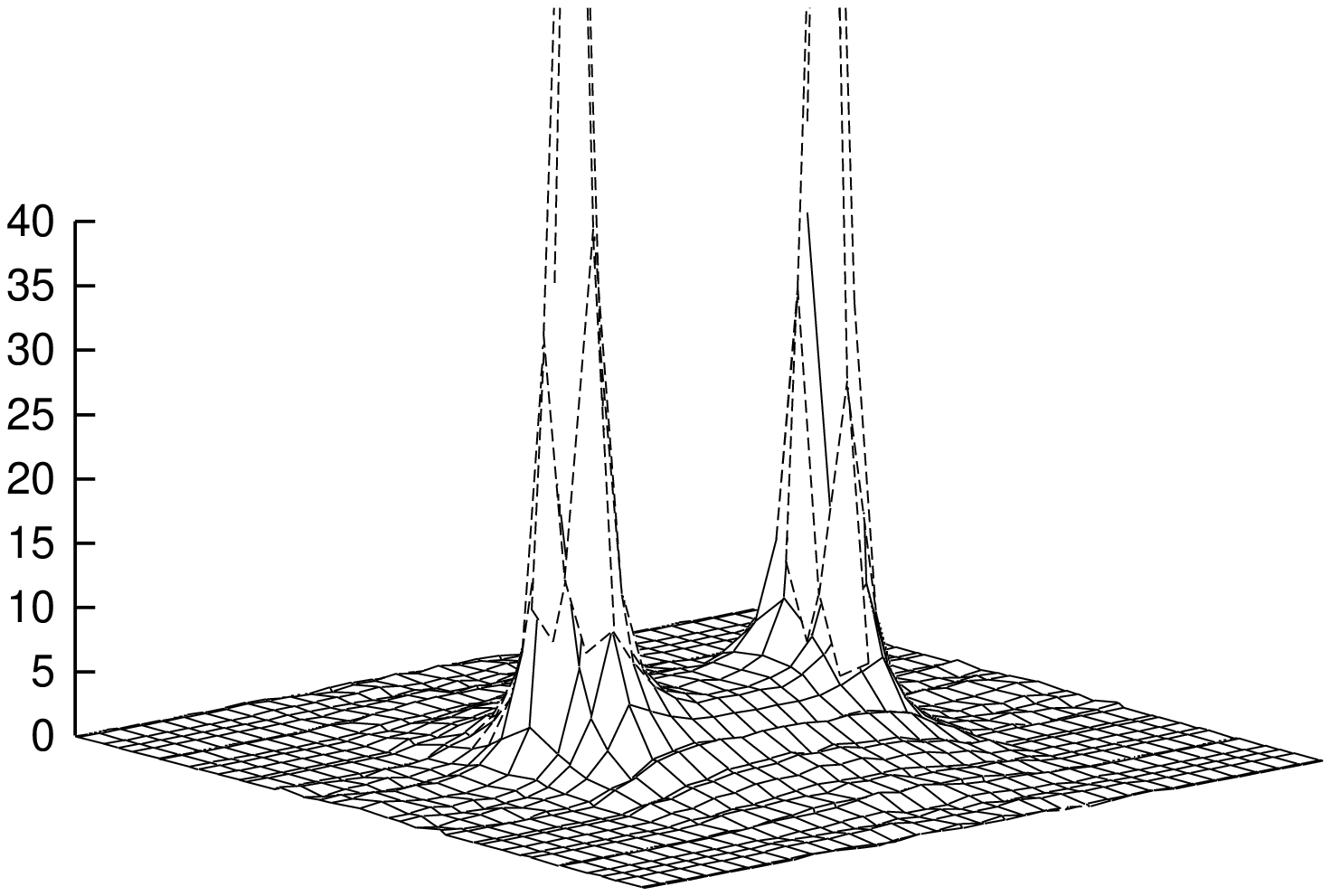}
\end{center}
\vspace{-2.5cm}
\begin{center}
\leavevmode
\epsfxsize=12cm\epsfbox{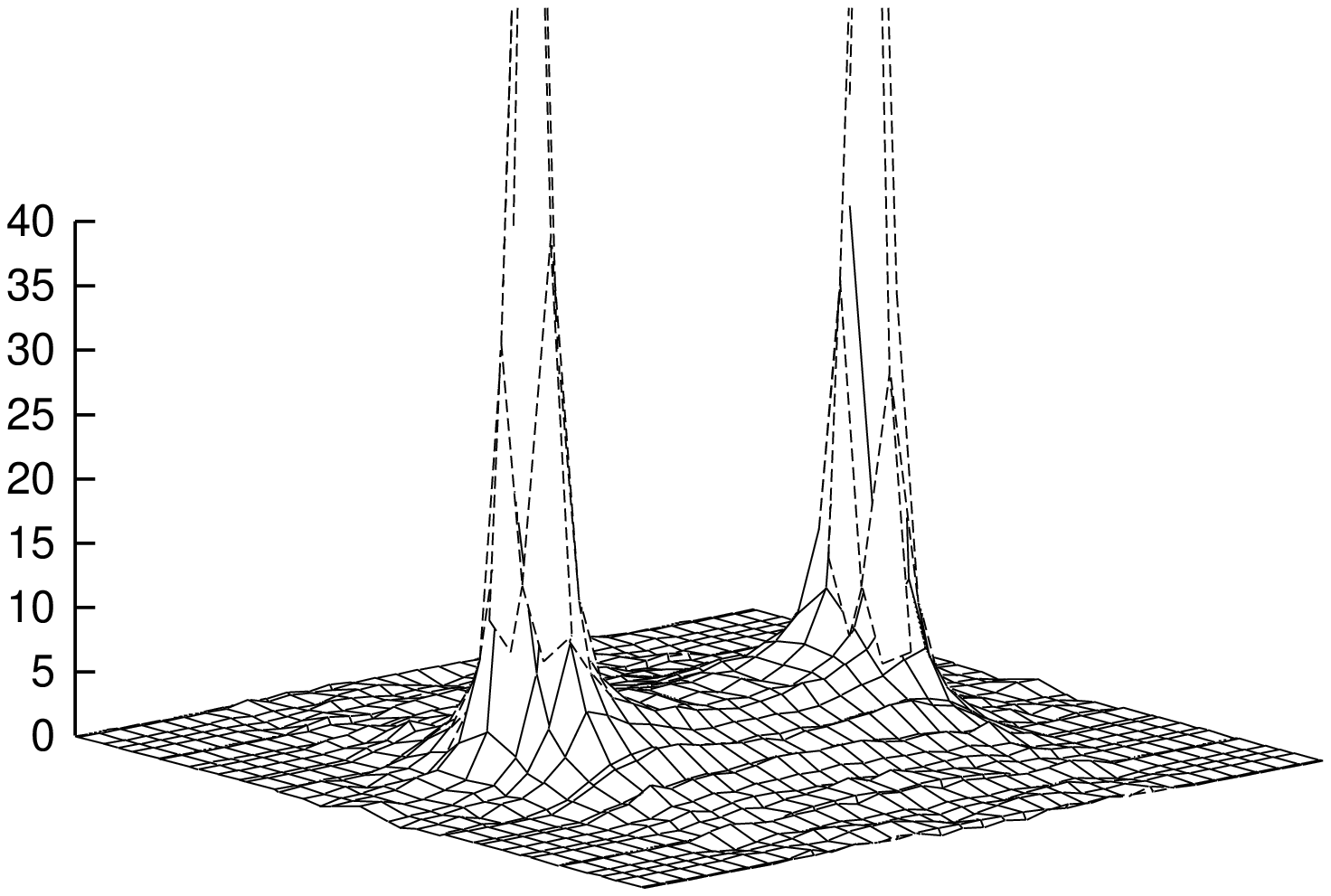}
\end{center}
\vspace{-1.5cm}
\caption
{\em Same as Fig.~\protect\ref{Figs1} for $R=12$ ($r\approx 1$~fm)
and $R=16$ ($r\approx 1.35$~fm).}
\label{Figs2}
\end{figure}

\begin{figure}[htb]
\begin{center}
\leavevmode
\epsfxsize=12cm\epsfbox{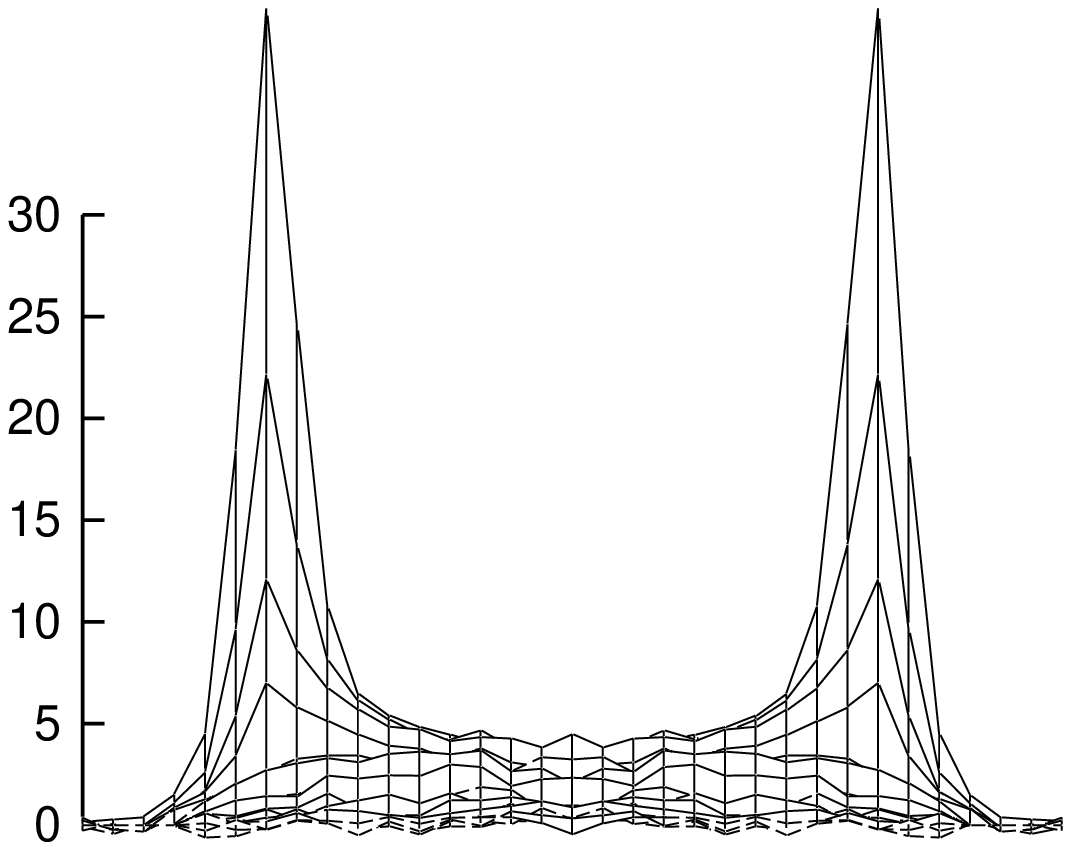}\\
\end{center}
\vspace{-3cm}
\begin{center}
\leavevmode
\epsfxsize=12cm\epsfbox{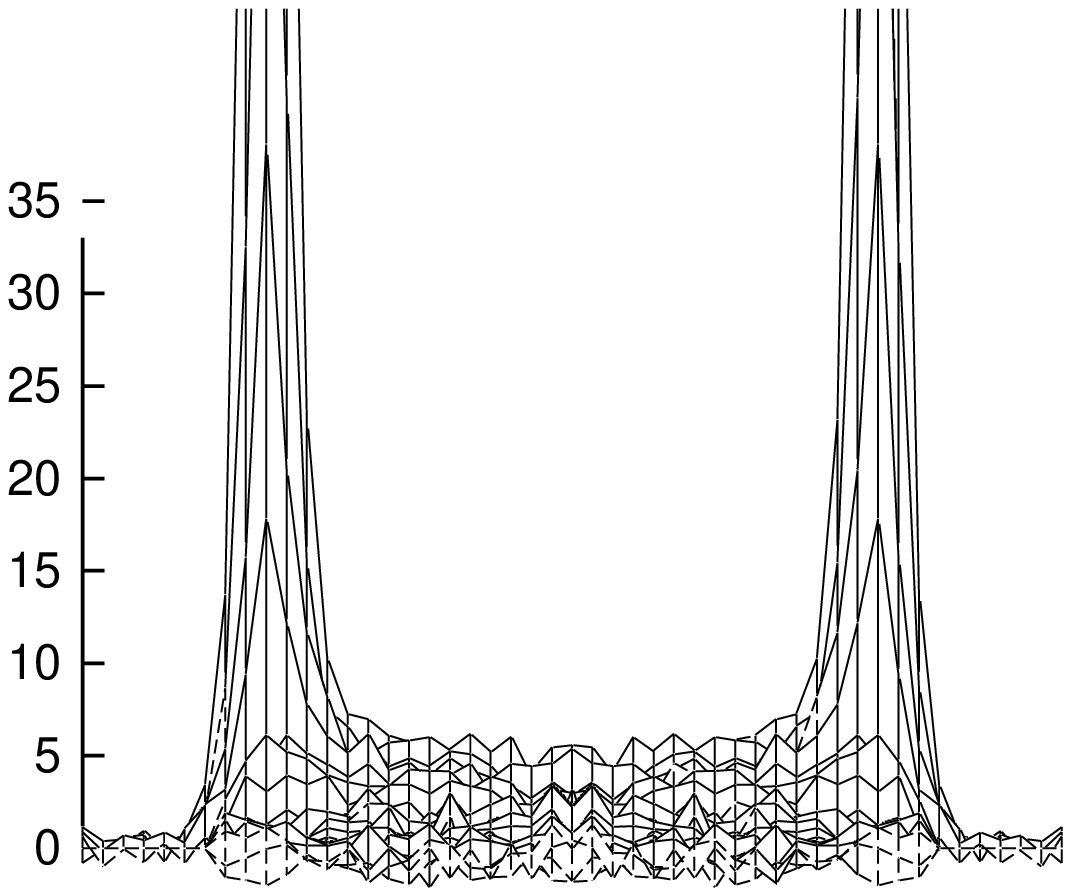}
\end{center}
\vspace{-2cm}
\caption
{\em The action density distributions for quark separations $R=20$ at
$\beta=2.5$ and $R=30$ at $\beta=2.635$, corresponding to
$r\approx 1.7$~fm. The $z$ axis of the first
plot is expanded by the ratio of the two
$\tilde{B}(\beta)$ values
in respect to the second plot to
account for the anomalous dimension.}
\label{Figfa1}
\end{figure}

We are now in the position to present a survey on the flux distributions
and watch the formation of flux tubes with increasing distance 
between the static sources. 

In Figs.~\ref{Fige1} and
\ref{Figs1}, we display the situation at $\beta = 2.5$
and $Ra = 8a \approx .7$~fm, for the energy and action densities,
respectively, in units of the string tension.
Notice, that the mesh is not equidistant in the
perpendicular direction
because the off-axis separations $n_{\perp}\propto (1,1)$ are included.
We confirm the earlier observation~\cite{sommer}
that magnetic and electric
field energies are of similar size (within $20 \%$), i.e.\ definitely
dominated by higher order contributions in $g^2$.
This 
results in a small energy density
in the middle of the flux tube, which in the case of
Fig.~\ref{Fige1} is nevertheless well  above noise.
The vertical axis of Fig.~\ref{Fige1} is expanded by a factor
$-\tilde{B}^{-1}=-\partial\ln a/\partial\ln\beta\approx 7.36$,
relative to Fig.~\ref{Figs1}. This is suggested from the
form of the sum rules (see Eqs.~\ref{actsum3} and \ref{ensum3}).
The figures show that this is indeed  a reasonable choice.
The electric flux tube looks distinctly broader
around the sources: contrary to the values in the physical region, the
(self-interaction) values at the peaks of the action and
energy density distributions
roughly equal each other.
This observation, which is in accord with the
sum rule prediction, causes the above optical impression.

Fig.~\ref{Figs4} illustrates the action density distribution
at equal physical geometry as Fig.~\ref{Figs1}
but with finer lattice spacing ($R=12$ on the
$\beta=2.635$ lattice).
The vertical
scale of Fig.~\ref{Figs4} is contracted
relative to Fig.~\ref{Figs1}
by the ratio of the corresponding two $\tilde{B}$ values. 
We observe  nice scaling of the fields outside the peaks. 
The latter diverge by the expected
additional factor $a_{2.5}^4/a_{2.635}^4\approx 5$.

The elongation of the flux distribution into a tube is traced in
Figs.~\ref{Figs1}, \ref{Figs2}, and \ref{Figfa1}
for the action density. The
physical source separations correspond to
.7, 1, 1.35, and 1.7 fm, respectively.
Our data representation avoids use of a smoothing
procedure, as has been applied in previous work~\cite{wiener,haynew}.
We are in the position to judge the significance of
the actual data from its intrinsic fluctuations. In this
way, we are less bound to be deceived by the beauty of some graphic
interpolating algorithm. Indeed, given the quality
of our data, we can follow --- for the first time in a lattice simulation ---
the flux tube along distances of up to $1.7$ fm or 30 lattice sites at
$\beta=2.635$ (Fig.~\ref{Figfa1})!

\begin{figure}[htb]
\begin{center}
\leavevmode
\epsfxsize=12cm\epsfbox{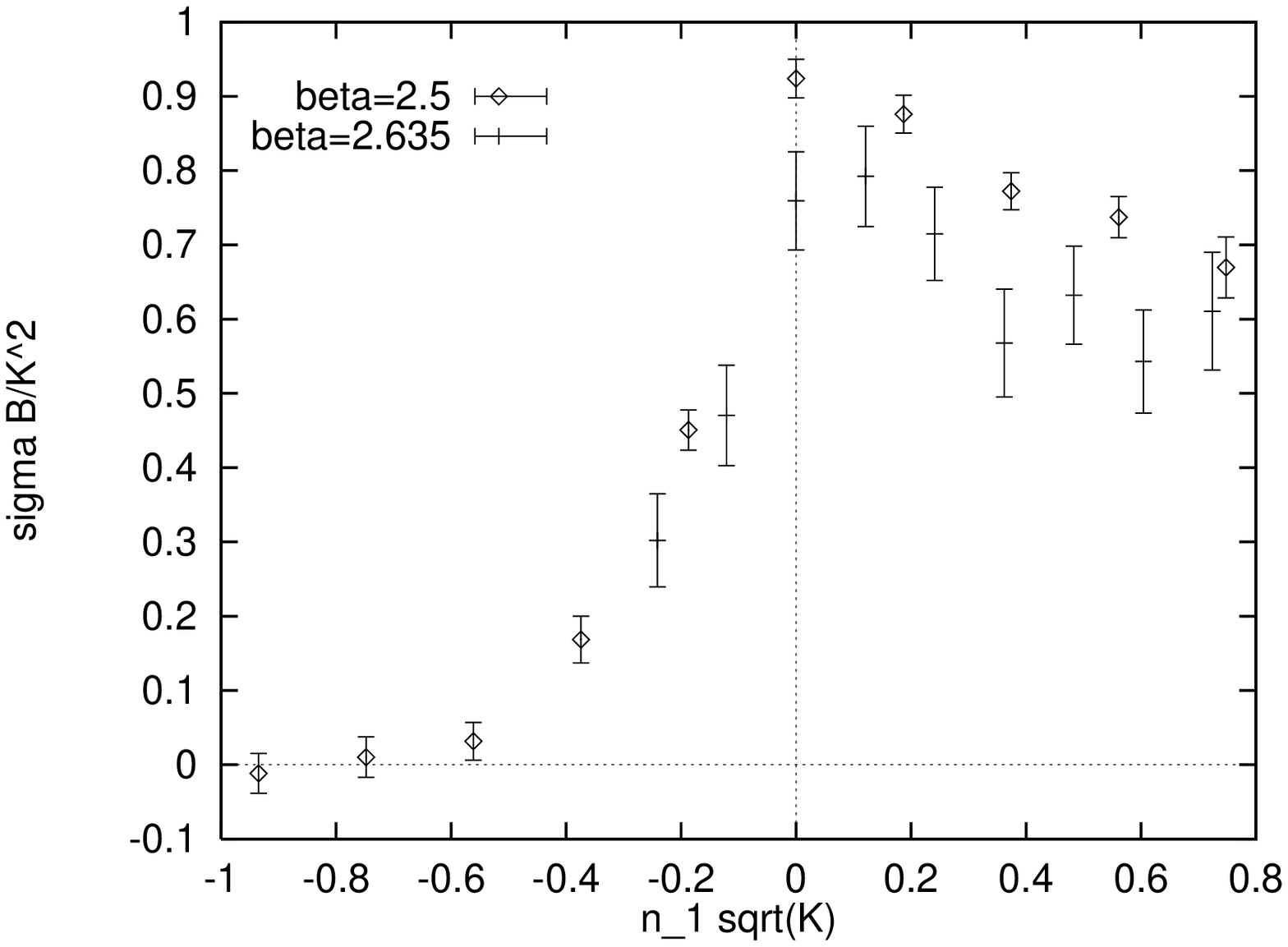}
\end{center}
\caption[ ]
{\em
Comparison of longitudinal action
density profiles at $r=.7$~fm between the $\beta=2.5$
data ($R=8$) and the $\beta=2.635$ data ($R=12$)
at $n_{\perp}a=.17$~fm
in units of the string tension.
The vertical axis has been multiplied by $\tilde{B}(\beta)$.
One source is placed
at position $(n_1,n_{\perp})=0$. The second source is located at
$n_1\sqrt{K}\approx 1.6$ (outside the visible range).}
\label{Figfa2}
\end{figure}

\begin{figure}[htb]
\begin{center}
\leavevmode
\epsfxsize=12cm\epsfbox{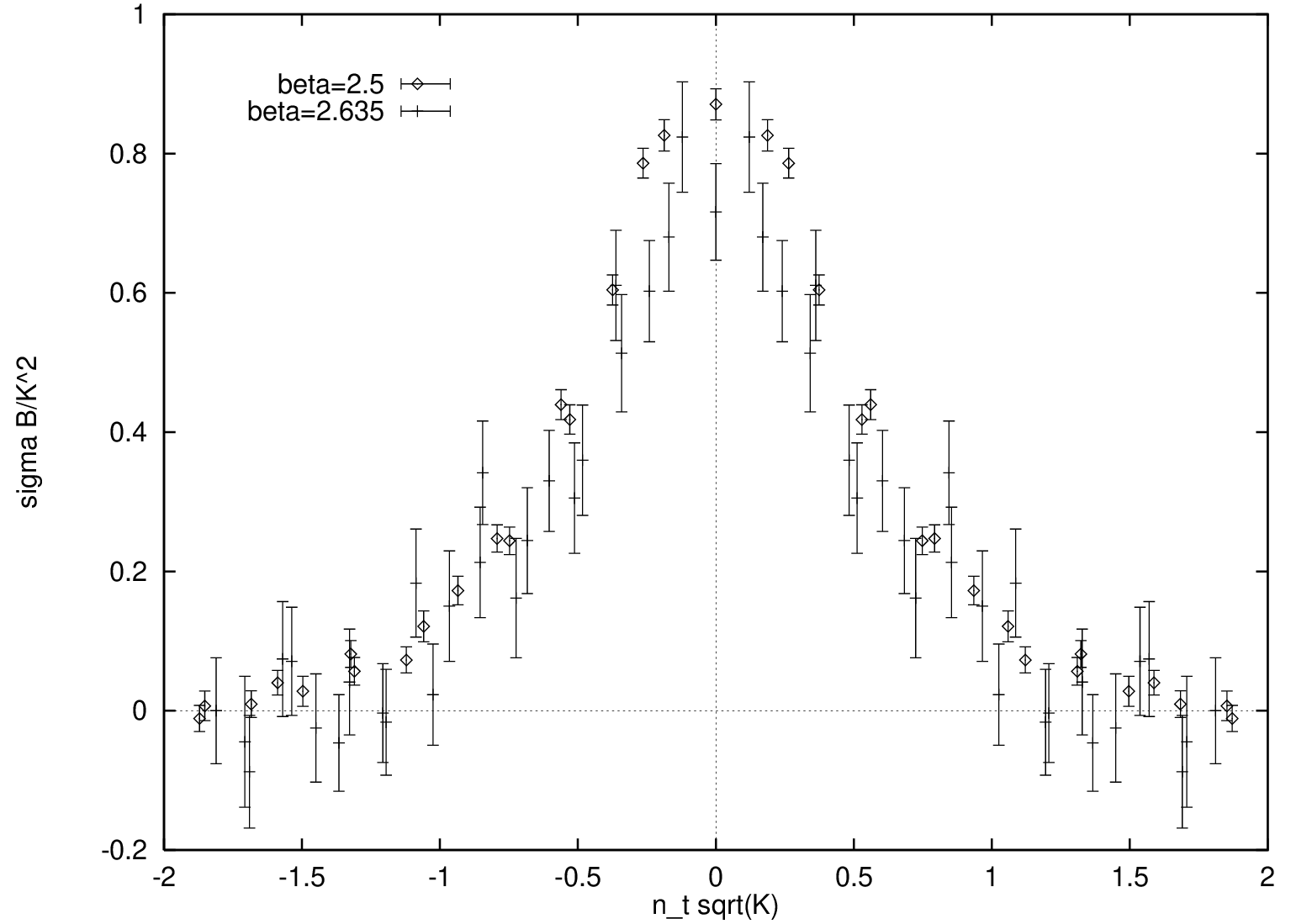}
\end{center}
\caption
{\em
Comparison of transverse action
density profiles at the center plane of two sources, separated by
$r=.7$~fm between the $\beta=2.5$
data ($R=8$) and the $\beta=2.635$ data ($R=12$)
in units of the string tension.
The vertical axis has been multiplied by $\tilde{B}(\beta)$.}
\label{Figfa3}
\end{figure}

\subsubsection{Finite $a$ effects}
In this subsection we will start to discuss the systematic
errors on our field measurements. 
A prominent effect would be expected from the limitation of the
lattice resolution $a$ to which we will
turn first.

The comparison shown in Fig.~\ref{Figfa1} between the action density
distribution at a quark separation of 1.7 fm, obtained
at two different lattice spacings, indicates scaling of the results
outside the self energy region.
The same holds qualitatively true for the situation at a distance of
$.7$~fm as can be seen from Figs.~\ref{Figs1} and Fig.~\ref{Figs4}.
Thus, we are driven to the conclusion that continuum results can be
obtained from quark distances as small as {\em eight}
lattice spacings, at least
at positions, separated by more than two lattice sites from the
sources.

Let us investigate the situation in some more detail. In
Fig.~\ref{Figfa2} we compare longitudinal action flux tube
profiles obtained at
$\beta=2.5$ and $\beta=2.635$ for $r\approx .7$ fm to each other.
One source is placed at the origin of the coordinate system.
The data are appropriately scaled in units of the string
tension and in addition divided by the expected anomalous dimensions
{}from Tab.~\ref{Tabscale}. The situation is displayed for
$x_{\perp}\approx .17$~fm. The latter
distance
corresponds to $n_{\perp}=2$ and $n_{\perp}=3$ for the two $\beta$
values, respectively. Though the data are compatible to each other
within errors, the
values obtained at the finer resolution tend to be systematically
below the corresponding $\beta=2.5$ values.
The same is found in a comparison of transverse profiles
obtained on the two data sets at
the center plane between the sources (Fig.~\ref{Figfa3}).
However, due to the errors on the string tension
values and the $\beta$-function (Notice, that the vertical axis has been
scaled by a factor $K^2$!), a relative overall scale error between the
two data sets of about $8 \%$ is expected which easily explains
systematic deviations from our expectation.

We conclude that at separations $R\geq 8$ continuum action and energy
density distributions can indeed be
observed  on the lattice. This conclusion is further 
supported by the fact that cylindrical
rotational invariance is restored  (within errors)
as can be gathered  from Fig.~\ref{Figfa3} where the values obtained
at plane diagonal sites (multiples of $\sqrt{2}$) neatly interpolate
between the values, measured along a lattice axis.
As we will see in Section~\ref{TS} violations of this
rotational invariance are encountered  for $R\leq 6$,
even at the center plane.
However, for sufficiently small lattice resolution, these violations
can be understood in terms of lattice perturbation theory and
eventually corrected to obtain the corresponding continuum
expressions. This is beyond the scope of the present paper,
where we are mainly interested in non-perturbative large distance
effects.

\subsubsection{Finite size effects}
\label{FSE}
Lattice results for the heavy quark potential and colour flux
distributions are subject to finite size effects (FSE).
The impact of FSE onto (smeared) Wilson loops is twofold. The
ground state potential $V(R)$ itself
might depend on the finite volume, due
to the infra red cut-off. Contrary to the perturbative expectation,
previous lattice studies of the confined phase of $SU(2)$ and $SU(3)$
gauge theories~\cite{michperan,baliref} show that this effect already
becomes negligible for lattice extents as small as $L_S\approx 1$~fm.
In the present simulation we are able to confirm
this observation by comparing the $16^4$ and $32^4$ potential data at
$\beta=2.5$. 

In addition one might worry about the impact of
mirror sources, due
to the toroidal structure of the lattice:
if one places sources at the positions
${\mathbf 0}$ and
${\mathbf R}$, the corresponding state is
virtually indistinguishable from a state created by
so called mirror sources. Thus, in the case of the (self adjoint) fundamental
representation of $SU(2)$, 
one expects --- in addition to a non-vanishing overlap of the
creation operator with
a $Q\bar{Q}$ state with separation
${\mathbf D}({\mathbf 0})={\mathbf R}$ ---
overlaps with states of internal separation ${\mathbf D}({\mathbf
n}) = {\mathbf R}+{\mathbf m}L_S$ with $m_i$ being (not necessarily
positive) integer numbers.
Let us consider for the moment the ``perfectly''
smeared Wilson loop (with no overlap whatsoever
to excited
states). One would thus anticipate
\begin{equation}
\label{hjh}
W({\mathbf R},T)=\sum_{\mathbf m} c_{\mathbf m}e^{-V(D({\mathbf m}))T}\quad.
\end{equation}

In strong coupling these mirror copy effects are exponentially suppressed
as the linear size grows in all directions. For a large planar Wilson
loop the leading order behaviour is:
\begin{equation}
W(R,T)=e^{-KRT}\left(1+e^{-K(L_SL_T-2RT)}+\cdots\right)\quad.
\end{equation}
However, in weak coupling, perturbation theory yields
$W(R,T)=W(L_S-R,L_T-T)$ in the non zero momentum sector. At
least to the lowest order (${\cal O}(g^2/(L_S^3L_T))$) the zero modes
do not obey this behaviour~\cite{Coste}. Their influence might become
even more important to higher orders, especially in the infra red
regime of large $R$.

The na\"{\i}ve geometrical expectation of Eq.~\ref{hjh}
is not borne out by the data, neither for the potential nor for
the action and energy densities.
This can be inferred from selection rules due to symmetries
of the creation operator.
As we shall show in the following,
this indeed happens in case of the Wilson loop, due to a symmetry under
transformations by center group elements of
$SU(2)$ in the fundamental representation: $Z_2=\{-1,1\}$.

Let us introduce a non-trivial
center transformation to all spatial links that  point into
direction $i$ and  cross the hyperplane $n_i=k+1/2$:
\begin{equation}
\tau^i_k:U_i(n)\rightarrow -U_i(n)\quad\mbox{for all $n$ with}\quad
n_i=k\quad.
\end{equation}
Obviously, the action is invariant under this transformation
since each plaquette crossing
the transformation plane contains two such rotated links.

\begin{figure}[p]
\begin{center}
\leavevmode
\epsfxsize=12cm\epsfbox{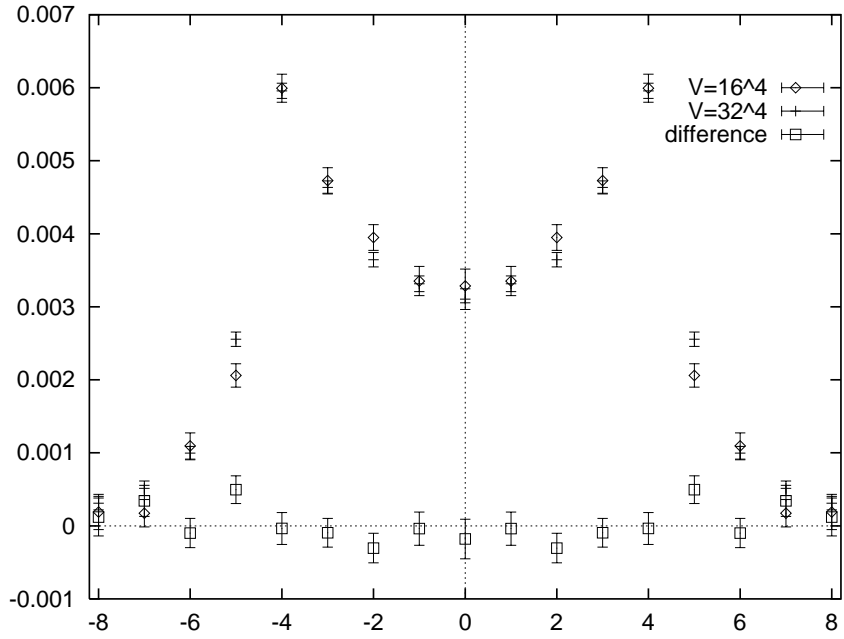}
\end{center}
\caption
{\em Differences between the longitudinal
action density profiles measured on
$16^4$ and $32^4$ lattices at $\beta=2.5$ for $R=8$ and $n_{\perp}=2$.}
\label{Fig9}
\end{figure}

\begin{figure}[p]
\begin{center}
\leavevmode
\epsfxsize=12cm\epsfbox{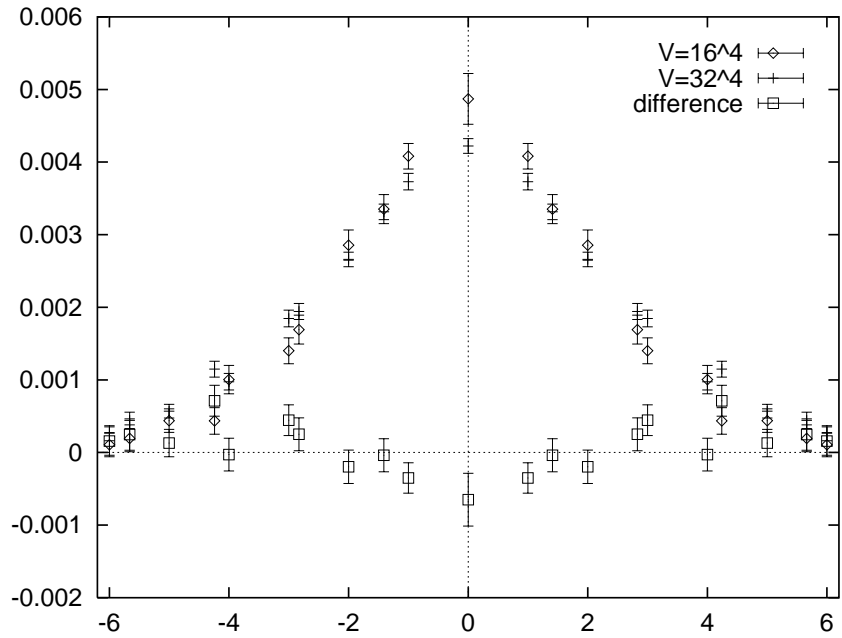}
\end{center}
\caption
{\em Differences between the transverse
action density profiles measured on
$16^4$ and $32^4$ lattices at $\beta=2.5$ for $R=8$ and $n_1=0$.}
\label{Fig9b}
\end{figure}

Our creation operator $\Gamma_{\mathbf R}^{\dagger}=Q({\mathbf
0})U({\mathbf 0}\rightarrow {\mathbf R})Q^{\dagger}({\mathbf R})$
(Eq.~\ref{create})
contains the spatial transporter $U({\mathbf 0}\rightarrow {\mathbf
R})$ that is a combination of various paths connecting the two quarks.
The smearing algorithm (Eq.~\ref{sme})
only permits continuous deformations of the
straight path. Thus, all paths cross the hyper planes
$n_i=0,\ldots,R_i-1$ an odd number of times while all other
planes are crossed an even number of times.
Therefore, $\Gamma^{\dagger}_{\mathbf
R}|0\rangle$ is an eigenstate of $\tau^i_k$:
\begin{equation}
\tau^i_k\Gamma^{\dagger}_{\mathbf R}=\left\{\begin{array}{c}
-\Gamma^{\dagger}_{\mathbf R}\quad,\quad 0\leq k < R_i\\
\Gamma^{\dagger}_{\mathbf R}\quad,\quad\mbox{elsewhere}\end{array}\right.\quad.
\end{equation}
As the eigenvalues remain invariant
under the Hamiltonian evolution they serve as conserved quantum
numbers.  Consequently,
in case of the gauge group $SU(2)$, only coefficients
$c_{\mathbf m}$ with {\it even} $m_i$ are different from {\em zero} in
Eq.~\ref{hjh}.
Therefore, the effective ``periodicity''\footnote{Obviously, the
coefficients $c_{\mathbf m}$ can differ from each other, depending on
the path combination, appearing in the transporter $U({\mathbf
0}\rightarrow {\mathbf R})$, unless $R_i=mL_S$.} is $2L_S$ rather than
$L_S$. For example, the leading order ``pollution'' for the on-axis
separation $R$ carries the decay constant $V(2L_S-R)$. For a linearly
rising long range potential, these large exponents cause a strong
suppression of fake states. This explains why such effects
remain unseen  in the present simulation, even at $R$ as large as 
$3/4L_S$.

The above arguments can  be generalized to the
local field strength measurement
operators $O_R({\mathbf n})=\epsilon_R({\mathbf n}), \sigma_R({\mathbf
n})$. $O_R$ has no overlap to a $Q\bar{Q}$ state, separated by
$L_S-R$.
The only relevant finite size effects stem from the periodicity
\begin{equation}
O_R({\mathbf n})=O_R(L_S-n_1,L_S-n_2,L_S-n_3)\quad.
\end{equation}
For ${\mathbf n}$ taken along the $Q\bar{Q}$ axis, energy and
action densities are strongly suppressed outside the sources:
in case of a dipole field 
(the leading order perturbative expectation, Eq.~\ref{a17})
the action and energy densities fall off
like $(|n_1|-R/2)^{-4}$ for $|n_1|>R/2$.
Thus, FSE into the longitudinal direction are negligible.
The $n_{\perp}$ distributions are more sensitive to FSE as will be
explained in Appendix~\ref{ap4} (see also Fig.~\ref{Fig10}).

A comparison of the potential computed on $16^4$ and $32^4$ lattices
at $\beta=2.5$ shows no statistically
significant bias due to the volume\footnote{The ground
state overlaps tend to be smaller on the larger lattice,
though the same smearing procedure has been applied. We suspect
that the number of
smearing steps  needs to be increased when working on larger
lattices due to a more extended wave function.
However, within our statistical accuracy, this is not
in the colour field distributions.}.
The same holds true for the action and energy density
distributions as a comparison between the two
lattice volumes shows for the largest $Q\bar{Q}$ separation
 realized on the smaller
lattice, $R=8$ (where FSE should be strongest).
As an example, the two data
sets are displayed in Fig.~\ref{Fig9} for the longitudinal slice
$n_{\perp}=2$. Fig.~\ref{Fig9b} shows the corresponding
transverse distributions for $n_1=0$.

Since the $\beta=2.635$ lattice is comparable in physical size to the
$32^4$ lattice at $\beta=2.5$ while the $\beta=2.74$ lattice has about
the same physical extent as the $16^4$ lattice, we conclude that all
our lattices are sufficiently large for the present purpose and that
FSE are below the statistical accuracy of the present investigation.

\subsubsection{T-stability}

\begin{figure}[htb]
\begin{center}
\leavevmode
\epsfxsize=12cm\epsfbox{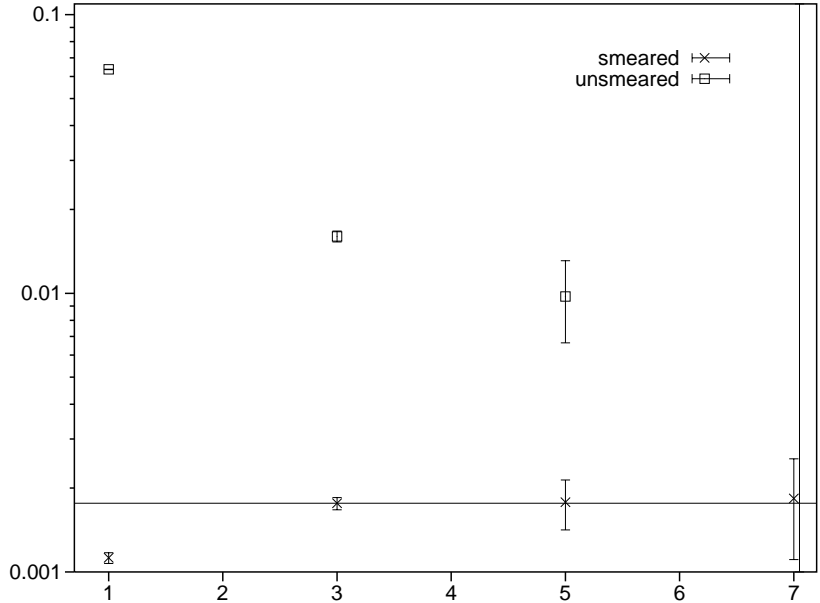}
\end{center}
\caption
{\em Comparison between smeared and unsmeared finite $T$
energy density approximants
in the center ($n_1=n_{\perp}=0$) between two sources,
separated by $R=4$ at $\beta=2.74$. The horizontal axis is
the time $T$.}
\label{Fig6}
\end{figure}

In the two preceding subsections
limitations in the lattice geometry
have been discussed to substantiate the
relevance of our lattice results to continuum
physics. Here, we address the reliability
of our ground state results
in view of the (necessarily)
limited temporal extent of the lattice operators. 
We will explain in 
some detail how the
($T\rightarrow\infty$) results shown above have been
obtained.

For the potential from unsmeared Wilson loops,
one has to take $T\gg R$ in order to obtain asymptotic
results as a consequence of Eq.~\ref{abg}.
In the case of field strength operators
this amounts to $T\gg 2R$ since excitations are damped by a decay
constant $\Delta V(R)/2$ only (instead of $\Delta V(R)$).
For large $R$ values, in which we are interested, it is practically
impossible to obtain signals at sufficiently large $T$.
However,
this situation is considerably improved by the smearing procedure, 
described in Section~\ref{GSE}.
This is evident from Fig.~\ref{Fig6} that presents a comparison between
smeared and unsmeared results for the energy density in the center
between two sources, separated by $R=4$ at $\beta=2.74$.
Notice the logarithmic scale!

To all our (smeared) data we have performed four parameter fits
according to Eq.~\ref{largeT} as well as
two- and three-parameter fits of the
form
\begin{eqnarray}
\langle\Box(S)\rangle_{\cal W}&=&\langle\Box\rangle_{|0,R\rangle-|0\rangle}
\nonumber\\
&+&c_1e^{-\pi
T/R}\cosh(2\pi S/R)\\\nonumber
&+&c_2e^{-2\pi (T-2S)/R}\quad,
\end{eqnarray}
where $\langle\Box\rangle_{|0,R\rangle-|0\rangle}$ and $c_i$ are the
fit parameters. In case of two parameter fits, $c_2$ is constrained
to {\em zero}.
We note that, because of the different temporal positions of magnetic and
electric insertions (i.e.\ different values of $S$ at fixed $T$), the
fits have been performed separately before combining
the expectation values for ${\cal E}$
and ${\cal B}$ to the energy and action densities, $\epsilon$ and
$\sigma$.

\begin{figure}[p]
\begin{center}
\leavevmode
\epsfxsize=12cm\epsfbox{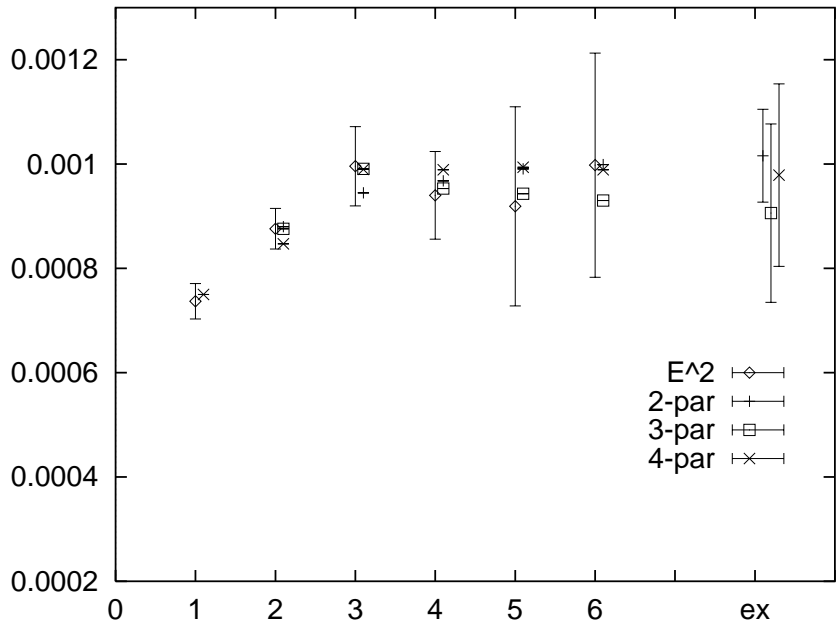}
\end{center}
\caption
{\em Finite $T$ approximants
to the electric plaquette expectation value ${\cal E}_6(0,3)$ in
presence of two quarks (diamonds), separated by $R=6$ at $\beta=2.5$ at the
position $n_1=0$, $n_{\perp}=3$ in lattice units.
Two-, three- and four-parameter 
fits are indicated, together with the corresponding
extrapolated asymptotic values (rightmost points with error bars).}
\label{Fig7}
\end{figure}

\begin{figure}[p]
\begin{center}
\leavevmode
\epsfxsize=12cm\epsfbox{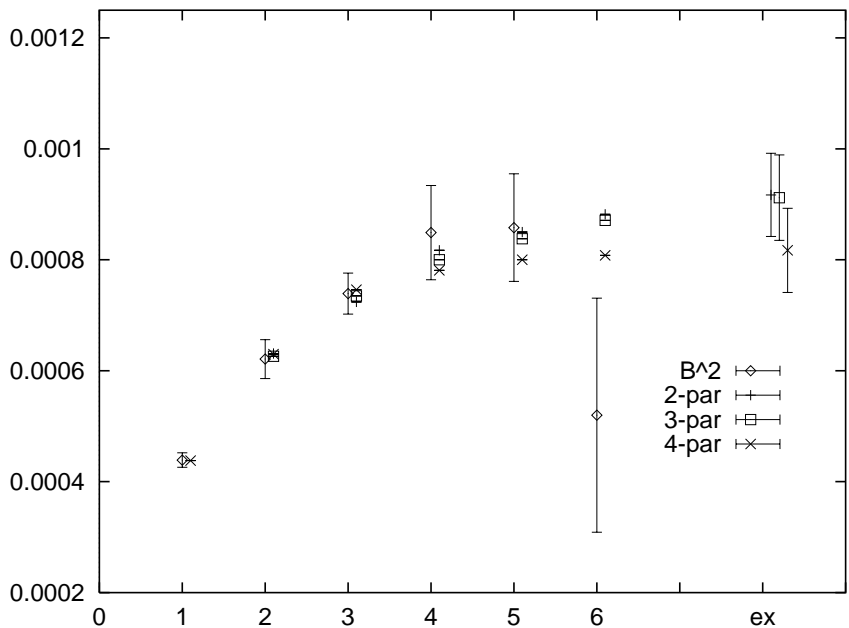}
\end{center}
\caption
{\em  Same as Fig.~\protect\ref{Fig7}, but for the magnetic plaquette
expectation, ${\cal B}_6(0,3)$.}
\label{Fig8}
\end{figure}

In all cases, the agreement with our data was remarkable with
$\chi^2/N_{DF}$ values close to {\em one}. For the two
parameter fits we had to exclude the $T=1$ data point.
The best results have been obtained with the three-parameter fits.
In case of four parameters, $c_3$ was found to agree with {\em zero}
within the (large) statistical uncertainty. Within errors,
the $T\rightarrow\infty$ extrapolated values coincided
with the $T=3$ value for large $R$
and the $T=4$ value for small $R$ in all cases. All our results refer
to the extrapolated values whose errors have been obtained by the
bootstrap method~\cite{bootstrap}.

In Figs.~\ref{Fig7} and \ref{Fig8} we exemplify
the time dependence of the electric and
magnetic energy density estimates, ${\cal E}$ and ${\cal B}$,
at $\beta=2.5$ for a quark separation $R=6$
at the position $n_1=0$, $n_{\perp}=3$.
The corresponding two-, three-, and four-parameter fits are included,
together with the $T\rightarrow\infty$ extrapolated values.
Due to the early ground state dominance, the fits
yield fairly stable results. Notice, that due to the fact
that the distance $S$ of the plaquette insertions from the central time
slice alternates with $T$, the parametrizations are
discontinuous. For this reason,
the fit values are just indicated at integer values of $T$.
In case of integrated quantities, needed for computation of the width
of the flux tube and comparison with the energy and action sum rules, 
the summation was first performed over 
the electric and magnetic energy densities for fixed $T$, separately,
and the $T$-extrapolation was carried out subsequently, before
combining the components to the
energy and action densities.

\section{Physics analysis}
Having presented and substantiated our numerical results
we are now ready to enter the physics analysis.

\subsection{Transverse shape}
\label{TS}

\begin{figure}[p]
\begin{center}
\leavevmode
\epsfxsize=12cm\epsfbox{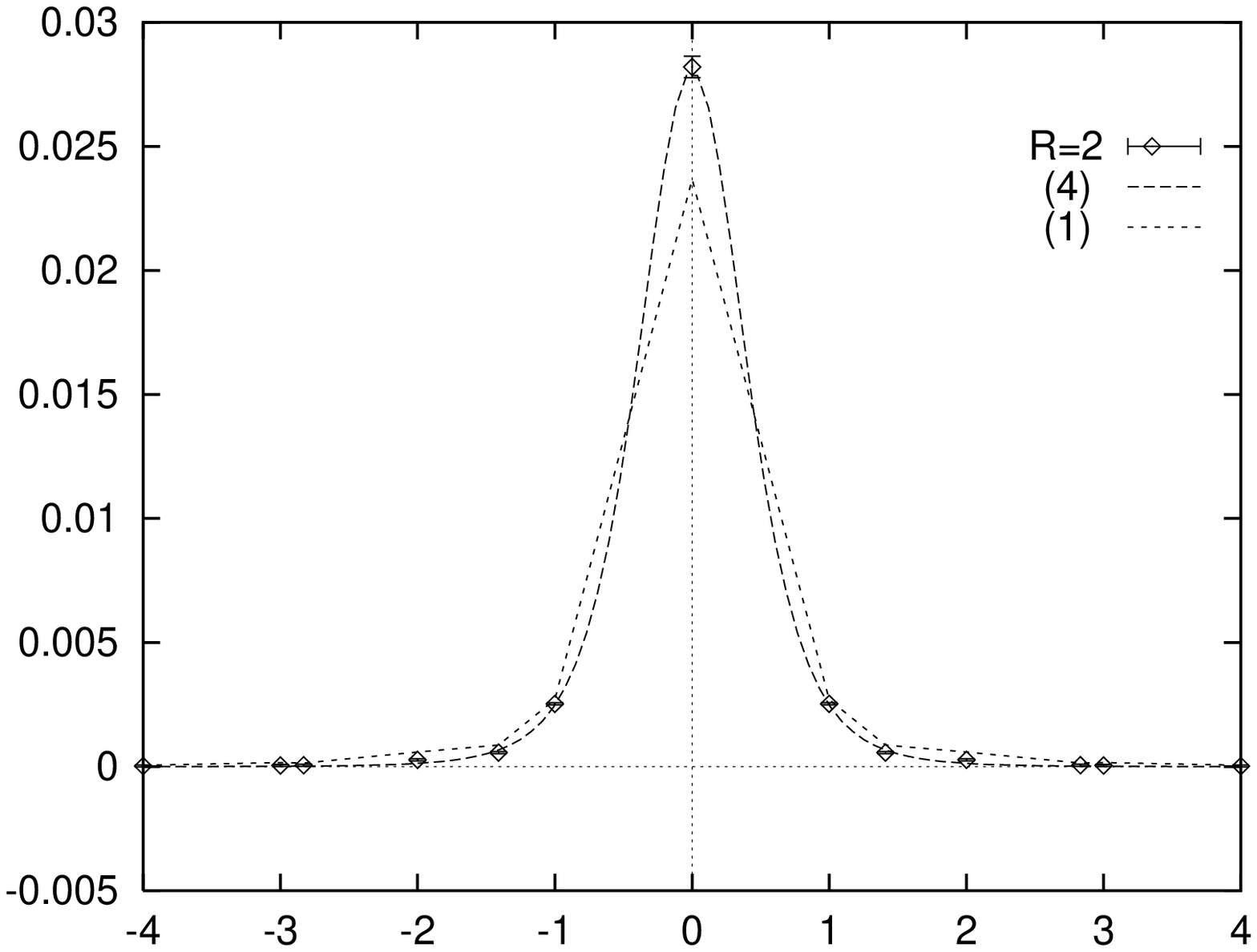}
\end{center}
\caption
{\em The central transverse energy density profile at $R=2$,
$\beta=2.5$ together with fit curves of methods (1) and (4).}
\label{Fige2t}
\end{figure}

\begin{table}[p]
\begin{center}
\begin{tabular}{|c|c|c|c|c|c|c|}\hline
$R$&method&$c_1$&$c_2$&$\delta$&$A$&$\chi^2/N_{DF}$\\\hline
 2 & 1 & ---      &0.043 (1)&1      &0.0223(2)&16.0\\
 2 & 2 &0.0711( 8)& ---     &1      &0.0356(4)&17.4\\
 2 & 3 &0.0296( 8)&0.0252(5)&1      &0.0280(5)&15.6\\
 2 & 4 &0.0570(10)& ---     &0.90(1)&0.0354(8)& 1.6\\ 
 2 & 6 &0.0310( 3)& ---     &0.65(1)&0.0368(5)& 7.4\\\hline
 4 & 1 & ---    &0.082( 2)& 2     &0.0142( 3)&1.2\\
 4 & 2 &0.162(3)& ---     & 2     &0.0205( 4)&5.5\\
 4 & 3 &0.017(3)&0.074( 2)& 2     &0.0150( 6)&1.2\\
 4 & 4 &0.065(7)& ---     &1.55(5)&0.0136(17)&0.9\\
 4 & 5 &0.001(2)&0.077( 8)&1.09(1)&0.0134(14)&1.3\\
 4 & 6 &0.023(2)& ---     &1.01(3)&0.011 ( 1)&1.6\\
 4 & 7 &0.011(4)&0.040(10)&1.03(6)&0.012 ( 3)&0.8\\\hline
 6 & 1 & ---     &0.110( 6)& 3      &0.0086( 4)&1.9\\
 6 & 2 &0.166( 9)& ---     & 3      &0.0095( 5)&1.5\\
 6 & 6 &0.090(30)& ---     &2.2 ( 2)&0.0090(30)&1.6\\\hline
 8 & 1 & ---     &0.210(20)& 4      &0.0082( 8)&0.8\\
 8 & 2 &0.260(30)& ---     & 4      &0.0087(10)&0.8\\\hline
\end{tabular}\end{center}
\caption{\em Results of fits to the central transverse profile of the energy
density distribution at $\beta=2.5$. $c_1$, $c_2$, and $\delta$ are
fit parameters. $A$ is the integrated area.}
\label{et1}
\end{table}

\begin{figure}[p]
\begin{center}
\leavevmode
\epsfxsize=12cm\epsfbox{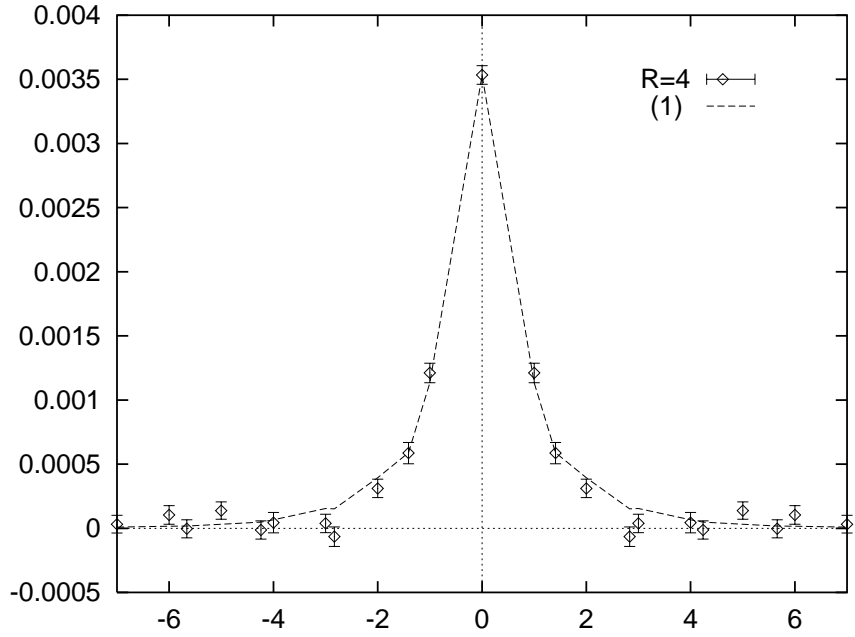}
\end{center}
\caption
{\em The central transverse energy density profile at $R=4$,
$\beta=2.5$ together with a fit to the lattice expression, $f_l$ 
(method (1)).}
\label{Fige4t}
\end{figure}

\begin{table}[p]
\begin{center}
\begin{tabular}{|c|c|c|c|c|c|c|}\hline
$R$&method&$c_1$&$c_2$&$\delta$&$A$&$\chi^2/N_{DF}$\\\hline
  2 & 1 & ---      &0.0271( 5)& 1      &0.0143( 3)&3.7\\
  2 & 2 &0.0446( 8)& ---      & 1      &0.0223( 4)&1.3\\
  2 & 3 &0.0383( 8)&0.0039( 5)& 1      &0.0211( 5)&1.3\\
  2 & 6 &0.0230( 9)& ---      & 0.83(3)&0.0160(10)&1.6\\\hline
  4 & 1 & ---     &0.052( 1)& 2      &0.0093(2)&1.0\\
  4 & 2 &0.102( 2)& ---     & 2      &0.0128(3)&3.3\\
  4 & 3 &0.019( 2)&0.043( 1)& 2      &0.0101(4)&0.9\\
  4 & 4 &0.045( 5)& ---     & 1.59(5)&0.009 (1)&0.5\\
  4 & 5 &0.040(10)&0.010(10)& 1.59(6)&0.010 (3)&0.5\\
  4 & 6 &0.016( 2)& ---     & 1.03(3)&0.008 (1)&0.9\\
  4 & 7 &0.009( 3)&0.025( 8)& 1.06(5)&0.007 (2)&0.5\\\hline
  6 & 1 & ---    &0.063(4)& 3     &0.0054(4)&0.5\\
  6 & 2 &0.096(6)& ---    & 3     &0.0054(4)&0.4\\
  6 & 3 &0.090(6)&0.004(4)& 3     &0.0053(4)&0.4\\
  6 & 5 &0.000(1)&0.068(4)& 0.3(8)&0.0058(4)&0.5\\\hline
  8 & 1 & ---     &0.13(1)& 4     &0.0063( 5)&1.1\\
  8 & 2 &0.170(20)& ---   & 4     &0.0055( 6)&1.3\\
  8 & 5 &0.000( 1)&0.13(2)& 1  (3)&0.0063(10)&1.2\\
  8 & 7 &0.000( 1)&0.13(2)& 0.9(9)&0.0063(10)&1.2\\\hline
\end{tabular}\end{center}
\caption{\em
Same as Tab.~\protect\ref{et1} for the energy density profile
at $\beta=2.635$.}
\label{et2}
\end{table}

\begin{table}[p]
\begin{center}
\begin{tabular}{|c|c|c|c|c|c|c|}\hline
$R$&method&$c_1$&$c_2$&$\delta$&$A$&$\chi^2/N_{DF}$\\\hline
  2 & 1 & ---     &0.1395(6)& 1       &0.0734(3)&437.7\\
  2 & 2 &0.234( 1)& ---     & 1       &0.1171(5)&443.6\\
  2 & 3 &0.105( 1)&0.0774(6)& 1       &0.093 (1)&455.0\\
  2 & 4 &0.468( 6)& ---     & 1.381(8)&0.123 (2)& 41.6\\
  2 & 5 &0.560(10)&0.052 (2)& 1.77 (2)&0.114(3) &  3.9\\
  2 & 6 &0.217( 3)& ---     & 1.034(7)&0.101(2) &156.0\\
  2 & 7 &0.247( 6)&0.064 (1)& 1.35 (1)&0.101(2) & 12.1\\\hline
  4 & 1 & ---    &0.304(2)& 2      &0.0527(4)&174.4\\
  4 & 2 &0.62(1) & ---    & 2      &0.0783(5)& 65.5\\
  4 & 4 &1.81(7) & ---    & 2.69(3)&0.128 (6)& 11.8\\
  4 & 6 &0.64(2) & ---    & 1.75(2)&0.103 (5)& 32.7\\
  4 & 7 &1.15(7) &0.131(5)& 2.37(5)&0.125 (9)&  5.4\\\hline
  6 & 1 & ---    &1.11(2)& 3      & 0.087(2)&19.0\\
  6 & 2 &1.67(3) & ---   & 3      & 0.096(2)&14.7\\
  6 & 3 &1.65(3) &0.01(2)& 3      & 0.096(2)&15.7\\
  6 & 4 &5.7 (5) & ---   & 4.2 (1)& 0.17 (2)& 1.9\\
  6 & 6 &2.2 (2) & ---   & 2.82(6)& 0.13 (1)& 3.7\\\hline
  8 & 1 & ---    &2.40(3)&  4      &0.093( 1)&12.7\\
  8 & 2 & 3.03(4)& ---   &  4      &0.101( 1)&20.0\\
  8 & 4 &10.1 (7)& ---   &  5.6 (1)&0.184(14)& 1.2\\
  8 & 6 & 3.7 (2)& ---   &  3.72(7)&0.13 ( 1)& 2.9\\\hline
 10 & 4 &14  (1)&---&6.3(2)&0.21(2)&1.5\\
 10 & 6 & 5.1(4)&---&4.2(1)&0.15(1)&2.5\\\hline
 12 & 4 &18  (1)&---&  6.8(4) &0.23(3)&1.0\\
 12 & 6 & 6.5(7)&---&  4.5(1) &0.16(2)&1.2\\\hline
 14 & 4 &27(1)&---&7.8(2)&0.27(2)&1.4\\
 14 & 6 &10(1)&---&5.2(2)&0.18(3)&1.0\\\hline
 16 & 4 &19(1)&---&7.4(2)&0.21(2)&1.0\\
 16 & 6 & 7(1)&---&4.9(3)&0.15(3)&0.8\\\hline
 18 & 4 &38(2)&---&9.4(2)&0.26(2)&1.1\\
 18 & 6 &13(7)&---&6.1(3)&0.17(1)&1.0\\\hline
 20 & 4 &16(1)&---&7.2(2)& 0.19(2)&0.6\\
 20 & 6 & 6(1)&---&4.7(3)& 0.13(2)&0.6\\\hline
 22 & 6 &12(1)&---&5.8(2)& 0.18(1)&1.4\\
 24 & 6 &13(2)&---&6.1(3)& 0.18(2)&0.7\\\hline
\end{tabular}\end{center}
\caption{\em Same as Tab.~\protect\ref{et1} for the action density
profile at $\beta=2.5$.}
\label{et3}
\end{table}

\begin{table}[htb]
\begin{center}
\small
\begin{tabular}{|c|c|c|c|c|c|c|}\hline
$R$&method&$c_1$&$c_2$&$\delta$&$A$&$\chi^2/N_{DF}$\\\hline
  2 & 1 & ---    &0.1027(5)& 1      &0.0544(3)& 7.3\\
  2 & 2 &0.166(1)& ---     & 1      &0.0830(4)&20.1\\
  2 & 3 &0.050(1)&0.0721(5)& 1      &0.0632(5)& 4.7\\
  2 & 4 &0.261(1)& ---     & 1.38(1)&0.0682(5)& 5.4\\
  2 & 5 &0.15 (1)&0.047 (5)& 1.42(4)&0.062 (4)& 0.5\\
  2 & 6 &0.133(3)& ---     & 1.09(1)&0.055 (2)&15.7\\
  2 & 7 &0.058(5)&0.062 (3)& 1.13(2)&0.054 (3)& 0.5\\\hline
  4 & 1 & ---    &0.111(1)& 2      &0.0199(2)&55.6\\
  4 & 2 &0.226(2)& ---    & 2      &0.0285(3)&10.8\\
  4 & 4 &0.35 (1)& ---    & 2.27(2)&0.034 (1)& 4.3\\
  4 & 6 &0.118(4)& ---    & 1.46(1)&0.028(1) &12.9\\
  4 & 7 &0.17 (1)&0.053(3)& 1.90(5)&0.033(2) & 3.5\\\hline
  6 & 1 & ---    &0.264(5)& 3      &0.0227(4)&6.0\\
  6 & 2 &0.394(7)& ---    & 3      &0.0226(4)&3.5\\
  6 & 4 &0.86 (9)& ---    & 3.7 (1)&0.032 (4)&1.2\\
  6 & 6 &0.31 (3)& ---    & 2.46(7)&0.026 (3)&2.2\\\hline
  8 & 1 & ---   &0.57(2)&  4     &0.0275(9)&2.5\\
  8 & 2 &0.73(2)& ---   &  4     &0.0243(6)&4.0\\
  8 & 4 &2.9 (5)& ---   &  5.9(3)&0.044 (9)&1.1\\
  8 & 6 &1.0 (2)& ---   &  3.9(2)&0.035 (6)&1.5\\\hline
 10 & 6 &1.2(2)&---&4.3(2)&0.032(7)&0.9\\
 12 & 6 &2.9(7)&---&5.7(4)&0.04 (1)&1.0\\
 14 & 6 &1.8(5)&---&5.1(4)&0.03 (1)&0.7\\
 16 & 6 &2.7(9)&---&5.8(6)&0.04 (2)&0.9\\
 18 & 6 &3  (1)&---&6.0(6)&0.04 (2)&0.7\\
 20 & 6 &4  (1)&---&6.5(6)&0.05 (2)&0.6\\
 22 & 6 &3  (1)&---&6.0(6)&0.04 (2)&0.9\\
 24 & 6 &2  (1)&---&5.3(6)&0.03 (2)&0.6\\\hline
\end{tabular}\end{center}
\caption{\em Same as Tab.~\protect\ref{et1} for the action density
profile at $\beta=2.635$.}
\label{et4}
\end{table}

\begin{figure}[htb]
\begin{center}
\leavevmode
\epsfxsize=12cm\epsfbox{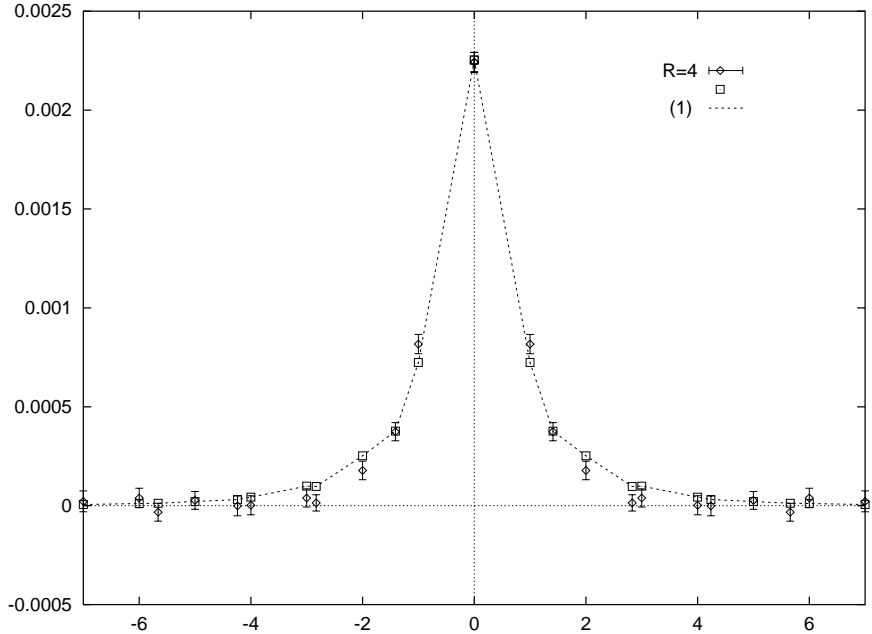}
\end{center}
\caption
{\em Same as Fig.~\protect\ref{Fige4t} at $\beta=2.635$}
\label{Fige42t}
\end{figure}

\begin{figure}[htb]
\begin{center}
\leavevmode
\epsfxsize=12cm\epsfbox{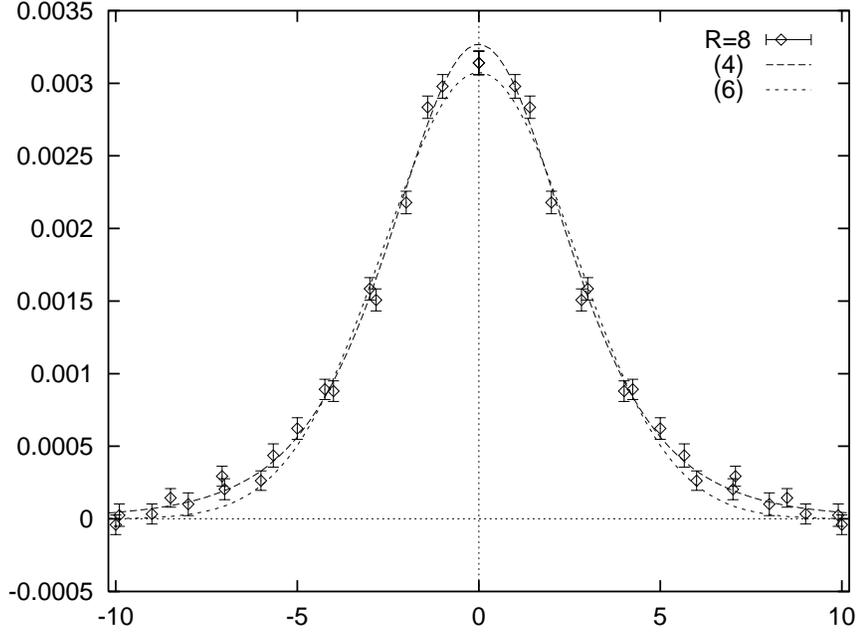}
\end{center}
\caption
{\em The central transverse action density profile at $R=8$,
$\beta=2.5$ together with fit curves of methods (4) and (6)
(dipole and Gauss).}
\label{Figa8t}
\end{figure}

\begin{figure}[htb]
\begin{center}
\leavevmode
\epsfxsize=12cm\epsfbox{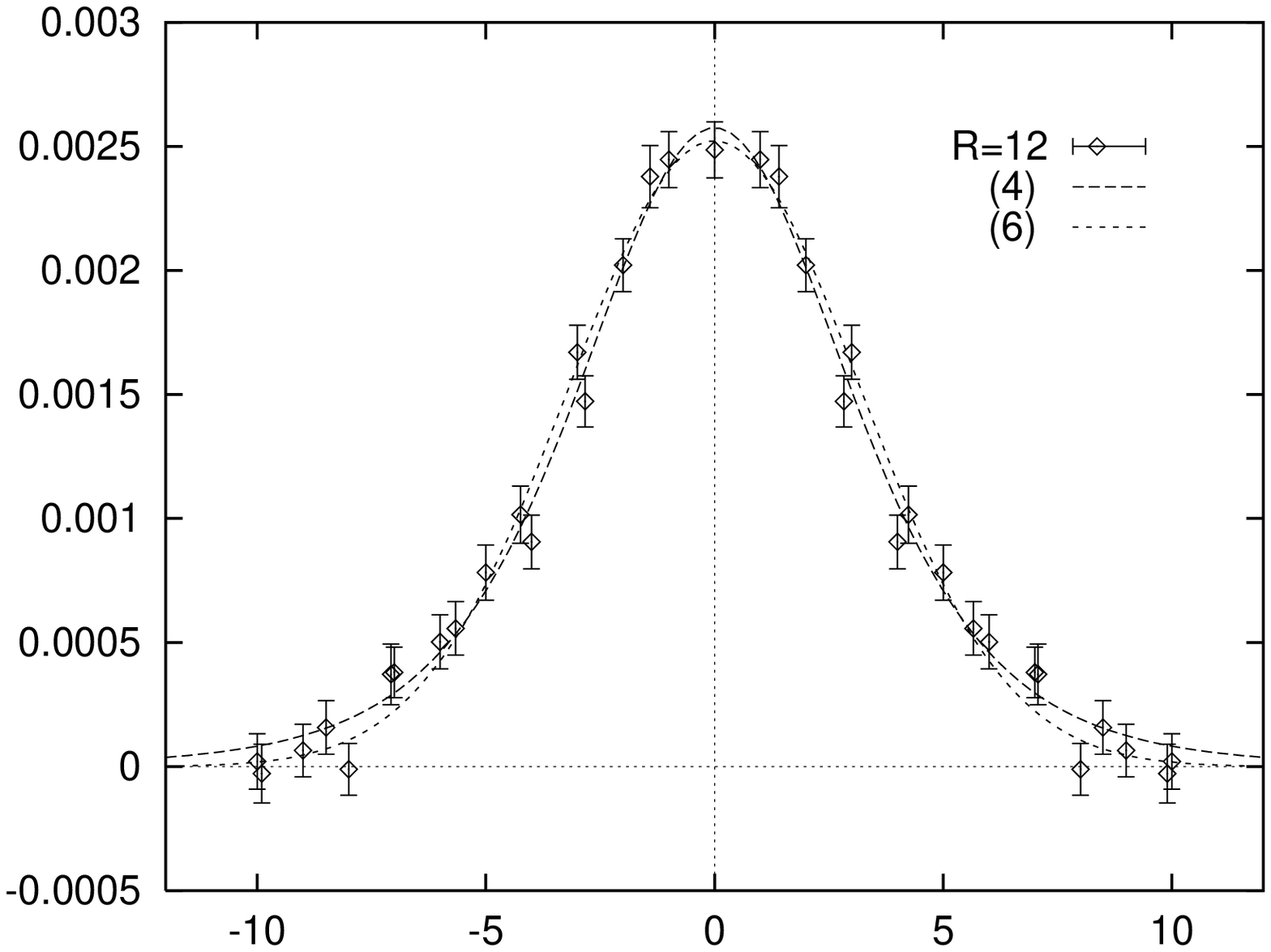}
\end{center}
\caption
{\em Same as Fig.~\protect\ref{Figa8t} for $R=12$.}
\label{Figa12t}
\end{figure}

We will focus on the transverse profile of field distributions
in the center plane of the flux tube.
For small separation of the sources, $r$, perturbation theory is
likely to apply, and one might thus
expect the
(energy and action) distributions to follow the shape
of the dipole field (see Eq.~\ref{a17})
\begin{equation}
\label{dipole}
f_d(n_{\perp})=\frac{1}{(4\pi)}
\frac{4\delta^2}{\left(\delta^2+n_{\perp}^2\right)^3}\quad,
\end{equation}
where the width of the flux tube would increase linearly
with $R$: $\delta=R/2$. 
In lowest order 
this would be multiplied by $C_F\alpha$ with $\alpha=g^2/(4\pi)$.

For small $R$, the continuum  form of Eq.~\ref{dipole}
needs be replaced
by a lattice sum, $f_l(n_{\perp})$,
that can be computed from Eqs.~\ref{a8} and \ref{a9}.
Remember, that this is only the leading order perturbative expectation.
The data reveals that
restoration of rotational invariance takes place
at unexpectedly small separations, especially in the action density.
To account for these higher order effects, that cancel
lattice artefacts in a subtle way, we will also
allow for a mixture of both, lattice and continuum
expressions.

As the source separation becomes large, compared to the
transverse size of the object, the string picture comes
into play and we might expect (at least for small $n_{\perp}$) the flux
distributions to be proportional to
\begin{equation}
f_g(n_{\perp})=\frac{1}{2\pi\delta^4}
\exp\left(-\frac{n_{\perp}^2}{\delta^2}\right)\quad.
\end{equation}
The normalization has been chosen such that
\begin{equation}
\sum_{n_{\perp}} f_l(n_{\perp})
\approx\int\!d^2n_{\perp} f_c(n_{\perp})
=\int\!d^2n_{\perp} f_g(n_{\perp})
=\frac{1}{2\delta^2}
\end{equation}
to allow for a direct comparison of the fitted coefficients.
The question arises how the lattice data might connect between
the two regimes.
We will attempt to model the transition region
by fits to the dipole parametrization,
$f_c(n_{\perp})$, with  $\delta$ treated  as free 
parameter. This is motivated by the idea that due to
antiscreening of the colour sources, their effective charge increases
when viewed at increasing  distance from the $Q\bar{Q}$
axis, $n_{\perp}$, which is tantamount to a rescaling of $R$.

In this heuristic spirit, a wide variety of  
one-, two- and three-parameter fits
(with free parameters
$c_1,c_2$, and $\delta$)
have been performed on the data,
which are listed here in shorthand notation:
\begin{enumerate}
\item $c_2f_l(n_{\perp})$,
\item $c_1f_d(n_{\perp})$ with $\delta=R/2$,
\item $c_1f_d(n_{\perp})+c_2f_l(n_{\perp})$ with $\delta=R/2$,
\item $c_1f_d(n_{\perp};\delta)$,
\item $c_1f_d(n_{\perp};\delta)+c_2f_l(n_{\perp})$,
\item $c_1f_g(n_{\perp};\delta)$, and
\item $c_1f_g(n_{\perp};\delta)+c_2f_l(n_{\perp})$.
\end{enumerate}
The stable fit results
are collected in Tabs.~\ref{et1}--\ref{et4} which 
also contain the integrated area,
\begin{equation}
A=\frac{c_1}{2\delta^2}\quad\left(+\frac{2c_2}{R^2}\right)\quad.
\end{equation}
The formula is only exact in the infinite volume limit.
The second term has been corrected by numerical computations of the
corresponding lattice sums. In case of the Gaussian and dipole
parametrizations, additional fits, according to the
finite volume expressions, derived in Appendix~\ref{ap4}
(Eq.~\ref{fform}),
have been performed. Subsequently, the
results have been corrected for the finite volume in the way,
described in the appendix.
For the Gaussian profile the finite size
corrections on the integrated area are negligible (below $.2\%$).
However, in case of a dipole distribution,
though $\delta$ is little affected by FSE (up to $4\%$), the
impact on the area is substantial (up to $25\%$ at large $R$ !).

In case of the dipole fits to the energy density, the combination
$c_1+c_2=C_F\alpha_c(R)$ amounts to a kind of
effective coupling on the scale $R$.
Notice, that the odd numbered ans\"atze incorporate the lattice
expression, $f_l$, while the forms (2),
(4), and (6) only involve continuum formulae. Ans\"atze (1) and (2)
require one parameter only, while (3), (4), (6) are based on two
and (5), (7) on three parameters.

{\bf Energy profile}
We start with the  discussion of the energy density data.
We will concentrate the analysis mainly on the
preformation of flux tubes, along the guidelines
of perturbative prejudice. For our
energy density resolution is not yet high enough
to map out the proper string region, $r>.75$ fm.

At $\beta=2.5$ and $R=2$ the 
data is very precise and excludes the
one-parameter fits (1) and (2) 
as well as the two parameter fit with constrained width (3).
The first acceptable results are reached
with ansatz (4), yielding
a width\footnote{The deviation from the expected value, $\delta=1$,
can be attributed to the fact that for small $R$ the lattice dipole
tends to be more narrow than its continuum counter part
(Fig.~\ref{Fig11}).}
$\delta\approx .9$
The situation is visualized in
Fig.~\ref{Fige2t} where we compare ans\"atze (1) and (4) against the
data\footnote{Excluding
the point $n_{\perp}=0$, we also find acceptable fits with
methods (2) and (3). The same is the case at $\beta=2.635$
where, due to the link integration procedure,
no data point is available at this position.
It is interesting to see from the large coefficient $c_1$, that
the data prefers the continuum dipole
over the lattice dipole.}.

This situation changes at $R=4$ where ansatz (1) leads to good
results at both $\beta$ values (while (2) fails), as can be seen from 
Figs.~\ref{Fige4t} and \ref{Fige42t}. Ans\"atze (3) and (5) yield
results  of equal quality with
$c_1\ll c_2$ which fits very nicely into the 
lattice perturbative picture.
The data can also be
parametrized by a continuum dipole with width $\delta\approx 1.55$
(ansatz (4)). However, the result of ansatz (5) ($c_1\ll c_2$)
shows that the data
prefers the (one-parameter) lattice expression to the (two-parameter)
continuum expression. 
At $R=6$ statistical errors 
allow for all parametrizations (apart from occasional
numerical instabilities).

We conclude that qualitatively the $R=2$ data is described by leading
order lattice perturbation theory while the $R=4$ and $R=6$ data can
be quantitatively understood along this line.

It is gratifying to see that
the effective coupling parameter $\alpha_c(R)$
increases
with $R$, as is expected from asymptotic freedom.
For $\beta=2.5$ we obtain values
ranging from $.06<\alpha_c^2(2a)<.075$ (under exclusion of $n_{\perp}=0$) over
$.105<\alpha_c^2(4a)<.125$ up to $.145<\alpha_c^2(6a)<.225$
while at $\beta=2.635$ we find the ranges $.055<\alpha_c^2(2a)<.060$,
$.065<\alpha_c^2(4a)<.085$ and $.085<\alpha_c^2(6a)<.13$, respectively.
We note, that $4a_{2.5}\approx 6a_{2.635}$. Thus, these numbers give a
consistent picture and should be
put in perspective to
the bare couplings $\alpha=2N/(4\pi\beta)\approx .127$
and $\alpha\approx .121$ for
$\beta=2.5$ and $\beta=2.635$, respectively.

{\bf Action profile}
In case of the action density,
a pure lattice Coulomb ansatz is expected to
fail since the action density is largely due to higher order
effects. Nonetheless, it would be interesting to see whether an
admixture of this term within the parametrization
remains necessary to account
for lattice artefacts.

It turns out that this heuristic approach looks little promising
as all fits to the $R=2$ and $R=4$ action data yield values $\chi^2\gg
N_{DF}$\footnote{The $\beta=2.635$, $R=2$ data is exceptional since
in this case we have omitted the $n_{\perp}=0$ point from our fits.}.
Among the fits, the three-parameter forms (5) and (7)
come closest to being successful.
The fits are not good enough, however,  to decide whether this gives
genuine evidence for 
perturbative lattice effects or trivially reflects 
the higher flexibility of a three-parameter ansatz.

{}From $R=6$ up to $R=10$ ($R=8$ at $\beta=2.635$) the dipole fits with
unconstrained width (4) appear to be the best parametrization of the data. 
Beyond these $R$-values, we observe the data being equally
well described by ansatz (4) and (6) at $\beta=2.5$, while
at $\beta=2.635$ from
$R=10$ onwards the Gaussian parametrization turns out to be more
robust than the dipole ansatz against statistical fluctuations.
{}From $R=12$ onwards the other fitting methods also started
yielding $\chi^2\approx N_{DF}$ values. Since these fits are
unphysical in this region we have not included them into the table.

The quality of the dipole (4) and  Gauss (6) fits
is exhibited for source separations $R=8$ and $R=12$
at $\beta=2.5$ in Figs.~\ref{Figa8t} and \ref{Figa12t},
respectively.

In conclusion, leading order lattice perturbation theory is found to
describe the energy density data well at small $R$. The fitted
amplitude is in accord with asymptotic freedom.
The lattice dipole term helps finding
a parametrization for action density data at small $R$, though it
is not a dominant term. Continuum dipole fits to the action yield acceptable
results from distances of about 0.5~fm onwards. Up to
1~fm this continuum parametrization has a width larger than $R/2$.
This effect is at variance with  the antiscreening picture of
colour sources and might well 
be a lattice artefact since a lattice dipole
is broader than a continuum dipole in this $R$ region (see Fig.~\ref{Fig10}).
For larger $R$, the combination
$2\delta/R$ decreases to values substantially below $1$. 
{}From a separation of $1$~fm onwards the Gaussian parametrization yields
an equally good (and occasionally superior) description of the data.

\subsection{String formation}
\label{DFT}

\begin{figure}[p]
\begin{center}
\leavevmode
\epsfxsize=15cm\epsfbox{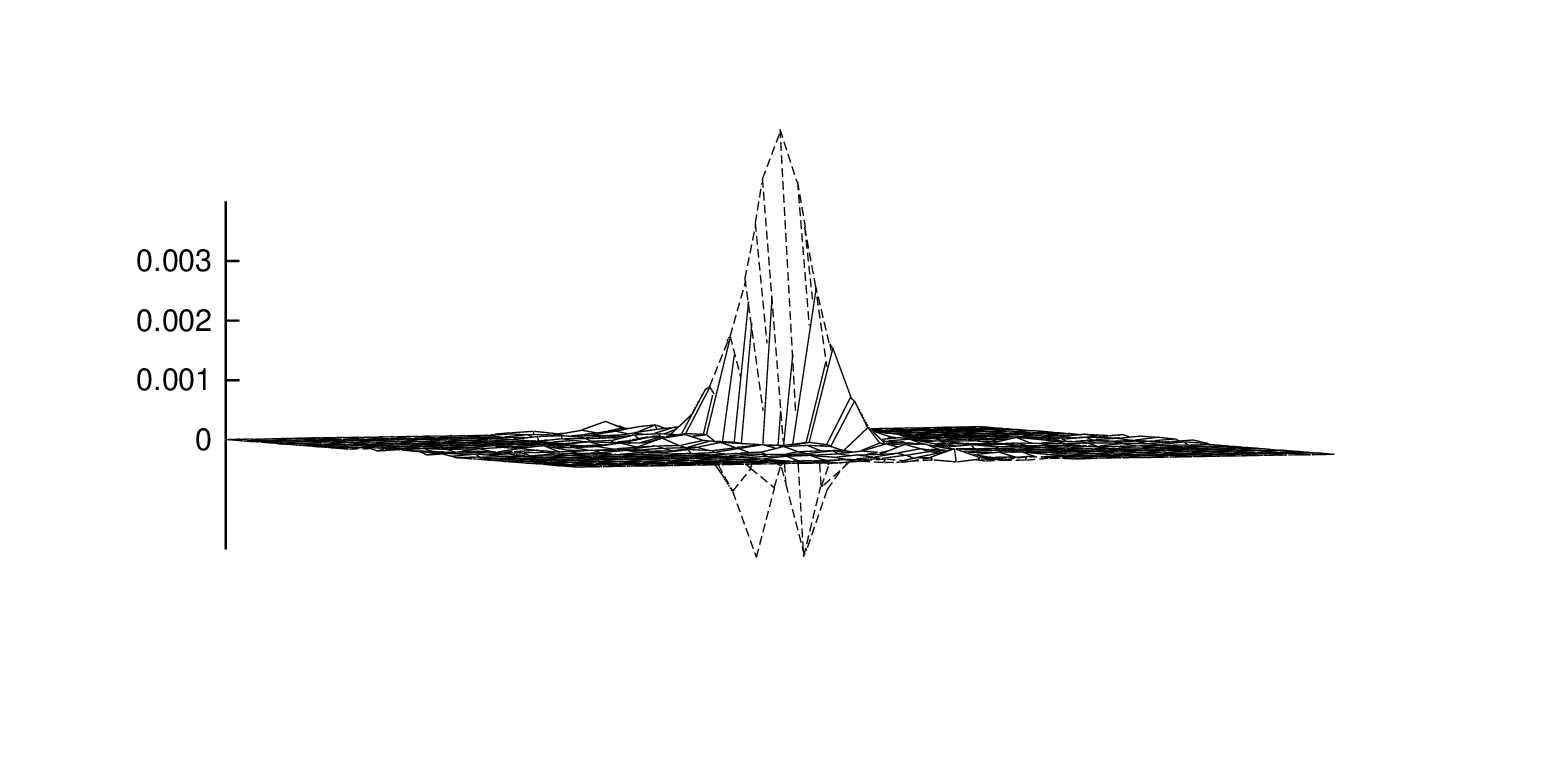}
\end{center}
\vspace{-3cm}
\begin{center}
\leavevmode
\epsfxsize=15cm\epsfbox{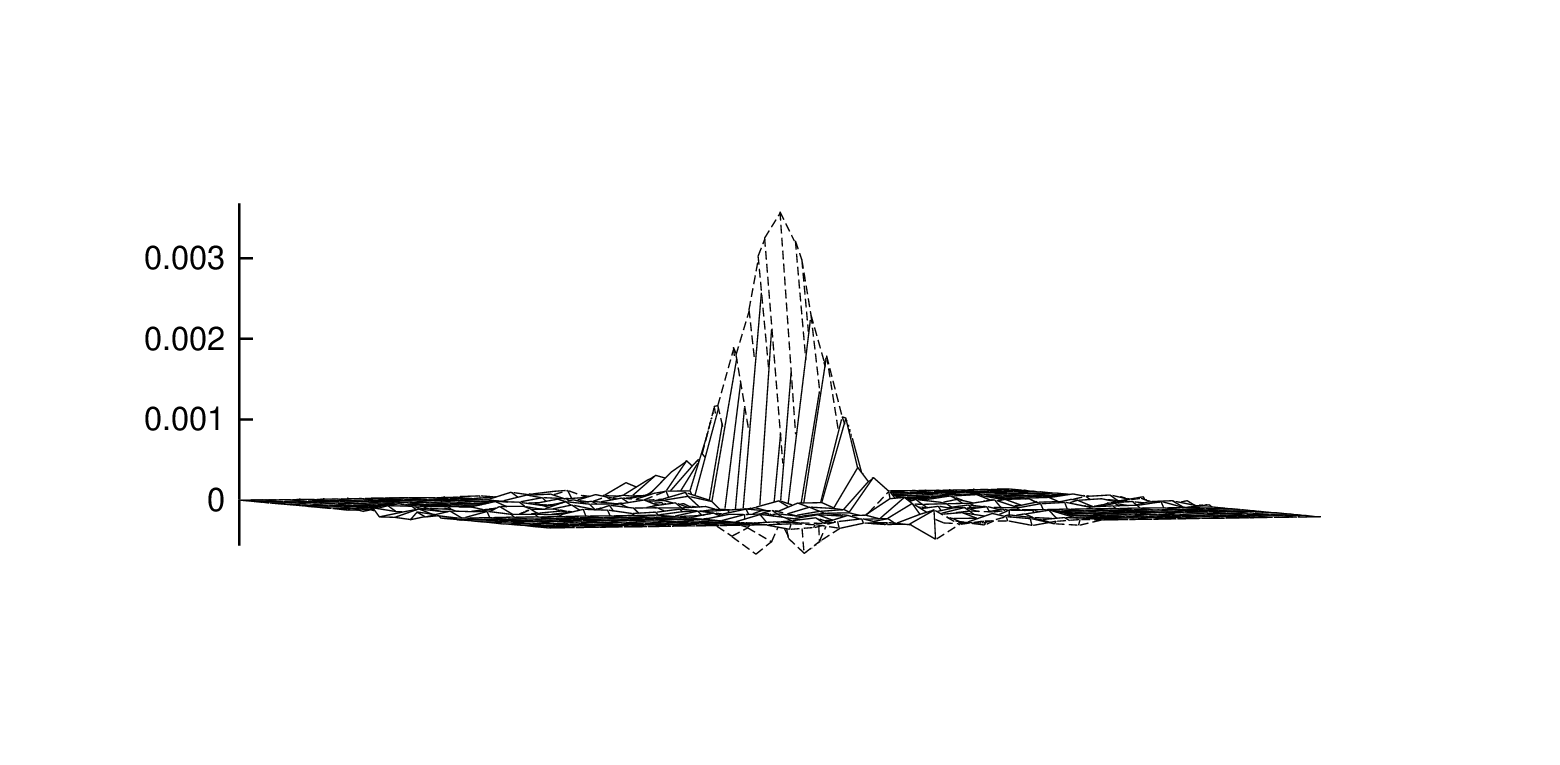}
\end{center}
\vspace{-3cm}
\begin{center}
\leavevmode
\epsfxsize=15cm\epsfbox{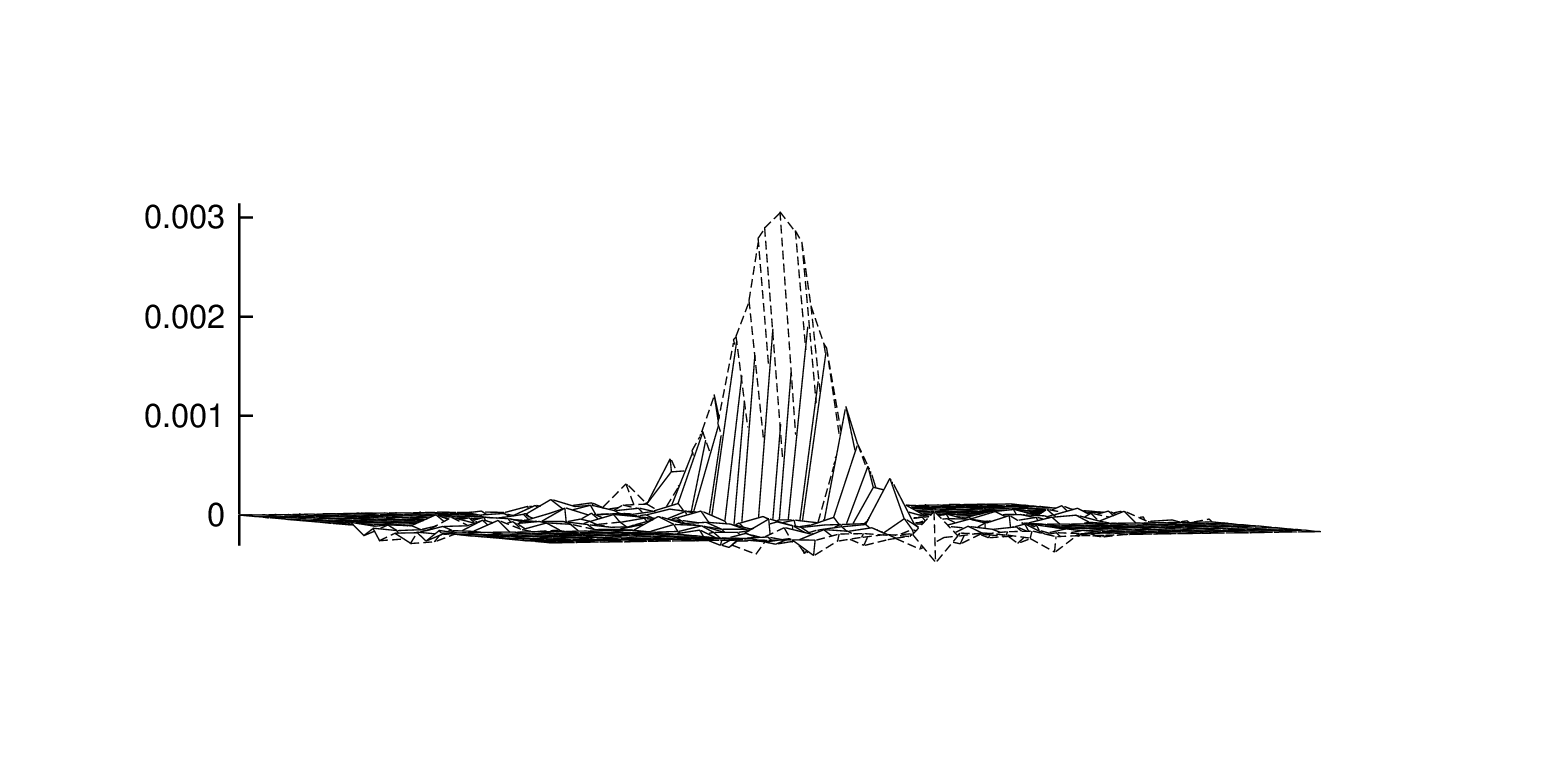}
\end{center}
\vspace{-1.5cm}
\caption
{\em Differences, $\Delta\sigma_R$,
between the action density distributions at $R$ and
$R-2$ for $R=6, 8,$ and 10 ($\beta=2.5$).
The sources have been aligned.
The labelling corresponds to lattice units.}
\label{Figp1}
\end{figure}

When the sources 
are adiabatically pulled apart, 
the accretion of action density, $\Delta\sigma$,  should
--- according to the string picture --- 
be strictly localized
in the center plane between the sources.
This holds for $R$ large enough
compared to the other inherent length scales in the problem, i.e.
the transverse width of the tube
and the size of the Coulomb dominated region. It is not
{\it a priori} obvious when this $R$-asymptotia sets in. 
The accretion phenomenon will be exploited  to determine
this transition point to genuine string formation.

For this purpose, we differentiate the action density distributions
with respect to an increase in the source separation.
This is done by computing the change,
$\Delta\sigma_R=\sigma_R-\sigma_{R-2}$, 
under stretches $(R-2) \rightarrow R$.
 
In Fig.~\ref{Figp1} we display 
the results for $\beta = 2.5$ and $R=6$, 8, and 10, respectively.
At $r=10a\approx .85$~fm, we find impressive
evidence that $\Delta\sigma$ is in fact zero outside the center plane!
This does not hold at smaller separations, where 
$\Delta\sigma$ exhibits a net flow of
action into the center plane from the next neighbour planes.
This latter feature is in accord with the dipole picture described
in Section~\ref{PS}.
This action flow is a substantial effect at $R=6$ and
decreases  to the $5\%$ level at $R=8$. Within our resolution
we thus conclude, that the
transition point to string formation is located at\footnote{Strictly
speaking, there is of course no
transition point into the asymptotic regime, since the
transition is smooth!} $R\approx 9$.

We might view the above analysis as a
differential diagnosis
of flux tube formation, which provides much more sensitivity than
the more conventional global tool offered by the potential. In fact, the
$R = 9$ transition point into the string regime
appears to be rather deep in the asymptotia
of the linear part of the confining potential: it corresponds to the point
$R\sqrt{K} = 1.7$ in Fig.~\ref{FIGP}, which in physical units is
$.75$~fm!

\subsection{Sum rules}

\begin{figure}[htb]
\begin{center}
\leavevmode
\epsfxsize=12cm\epsfbox{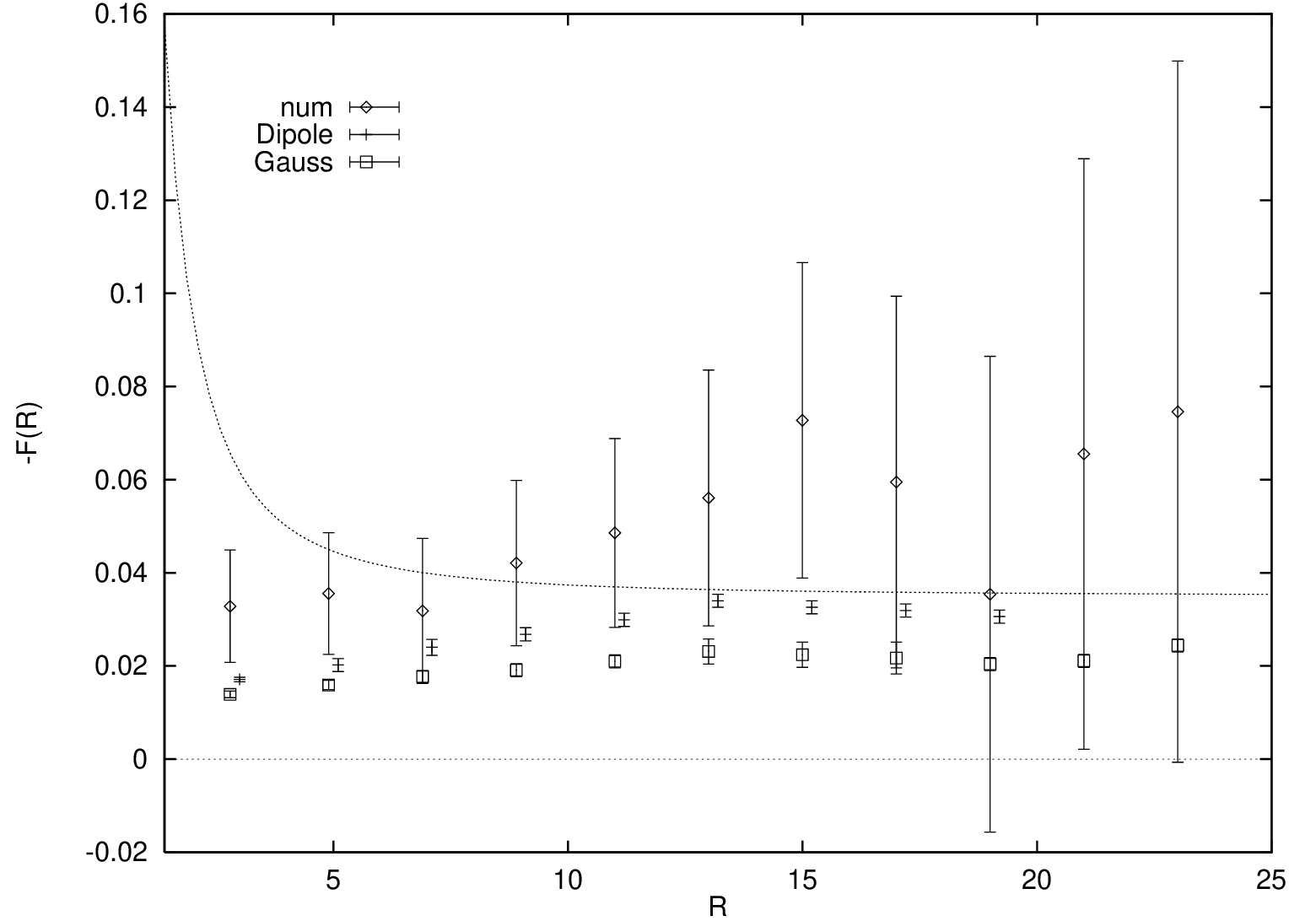}
\end{center}
\caption
{\em The $Q\bar{Q}$ force, obtained from the potential at $\beta=2.5$,
compared to the integrated center plane action density, scaled by the
anomalous dimension $\tilde{B}$ in lattice units.
In addition to the numerically integrated data (num),
the dipole and Gauss fit results from Section~\protect\ref{TS} are
displayed.}
\label{Figsumrule}
\end{figure}

Within the string regime, $Ra > .75$~fm,
one can write a differential
version of the sum rules:
\begin{eqnarray}
F(\sqrt{R^2-1})&=&\frac{1}{2}\left(V(R-1)-V(R+1)\right)\\
&=&-\frac{a^3}{2}\sum_{n_2,n_3}\left(\epsilon_{R-1}(0,n_2,n_3)
+\epsilon_{R+1}(0,n_2,n_3)\right)+{\cal O}(\beta^{-1})\\\label{ediff}
&\approx&
-a\pi\int_0^{x_{cut}}\!\!\!dx\,x\left(\epsilon_{R-1}(0,x)
+\epsilon_{R+1}(0,x)\right)\quad.
\end{eqnarray}
For the second equality we have assumed rotational invariance and
large $x_{cut}$.
Likewise we can write a  differential
action sum rule\footnote{The anomalous dimension of the action
density outside the sources and adjacent sites equals
$\tilde{B}(\beta)$, independent of the position, ${\mathbf n}$.}:
\begin{equation}
\label{adiff}
F(\sqrt{R^2-1})\approx
a\pi\tilde{B}(\beta)\int_0^{x_{cut}}\!dx\,x\left(\sigma_{R-1}(0,x)
+\sigma_{R+1}(0,x)\right)\quad.
\end{equation}

Notice, that the fundamental assumption made for the differential
version of the sum rules is only justified for $R+1\geq 9$ at
$\beta=2.5$. Also, the data has to exhibit approximate
rotational invariance and $x_{cut}$ has to be chosen sufficiently large.
We start from numerically integrating the data. In varying
$x_{cut}$ we try to find a plateau. For $R+1\leq 10$ a
clear plateau is established while for the
$R+1>10$ data, $x_{cut}=10a$, the maximal distance
for which we have performed measurements, had to be chosen.
Thus, these values are only lower limits on the
r.h.s.\ of Eqs.~\ref{ediff} and \ref{adiff}.

As can be seen from Fig.~\ref{Figsumrule} the action sum rule is consistent
with our data for $R+1\geq 6$. This lends further support
to  the asymptotic character of our data (in $T$).
Violations of rotational invariance appear to be small
beyond the two directions on which we have
performed our measurements. Consistency 
of the energy density data with the sum rule is found,
albeit within reduced statistical accuracy.

The fitted integrated action density, $A$ (rescaled by the factor
$\tilde{B}$), obtained in 
Section~\ref{TS} (Tab.~\ref{et3}), is also
shown in Fig.~\ref{Figsumrule} for a Gaussian and a dipole transverse
shape of the flux tube. The Gauss values
are substantially smaller than suggested by the force. This
discrepancy can only be due to a slower $n_{\perp}\rightarrow\infty$
fall-off of the data
than assumed by the Gauss ansatz.
Notice, that the string picture is only
applicable for small transverse fluctuations while a large portion of
the integral stems from the area of large $n_{\perp}$.
The dipole values are consistent with the force for large separations
$R$. It remains to be clarified
whether this is just a lucky coincidence. At least for
$.75$~fm$ <r<1.5$~fm the large $n_{\perp}$
data also seems to be underestimated by
a dipole shape.
The functional form of the profile can be studied in more
detail by varying the transverse spatial volume and exploiting the
observed FSE (Appendix~\ref{ap4}).

\subsection{Width of the flux tube}

\begin{figure}[htb]
\begin{center}
\leavevmode
\epsfxsize=12cm\epsfbox{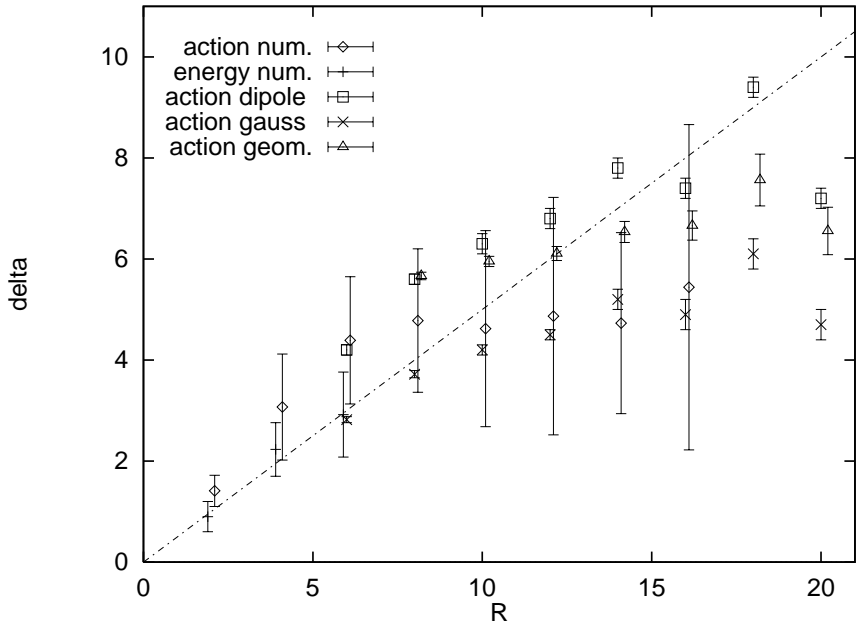}
\end{center}
\caption
{\em The width of the flux tube, $\delta$, against $R$ at $\beta=2.5$
(in lattice units). The dashed line corresponds to $\delta=R/2$.
In addition to the numerically integrated results from the energy and
action density distributions, fit results to the action
density from Tab.~\protect\ref{et3}, and
results from the geometric method (with $\gamma$=1) are displayed.}
\label{Figw1}
\end{figure}

\begin{figure}[htb]
\begin{center}
\leavevmode
\epsfxsize=12cm\epsfbox{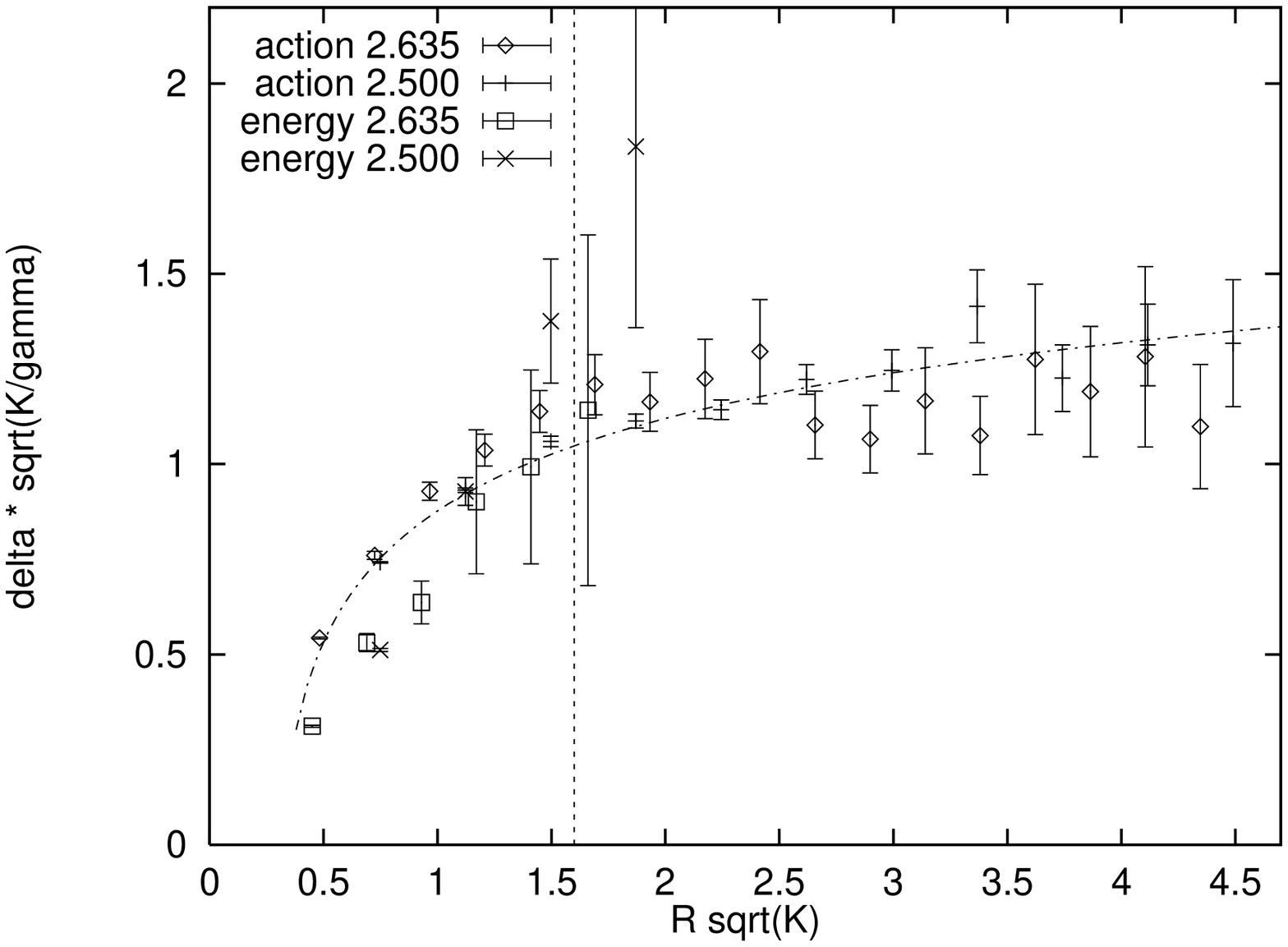}
\end{center}
\caption[ ]
{\em The width of the flux tube, $\delta$, against $R$ in units of the
string tension, obtained by use of the geometric method ($\gamma=1$)
for the energy and action density distributions at
$\beta=2.5$ and $\beta=2.635$. The vertical line indicates the
lower limit of applicability of the geometric method.
The dashed-dotted curve is the string picture expectation
(Eq.~\protect\ref{a18}) for
$R_0\sqrt{K}=1/3$, $\delta_0\sqrt{K/\gamma}=0.5$.}
\label{Figw2}
\end{figure}

In addition to the (parametrization dependent)
results on the width of the flux tube of Section~\ref{TS}
(Tabs.~\ref{et1}--\ref{et4}), we attempt to compute
this important parameter by direct
numerical integration:
\begin{equation}
\label{b12}
\delta_{\sigma}^2
=\frac{\int_0^{n_{cut}}\!dn_{\perp}\,n_{\perp}^3\sigma(0,n_{\perp})}
      {\int_0^{n_{cut}}\!dn_{\perp}\,n_{\perp}\sigma(0,n_{\perp})}\quad.
\end{equation}
The results, including their systematic errors from varying $n_{cut}$, are
displayed in Fig.~\ref{Figw1}, together 
with the expectations from the above dipole and Gauss
fits. We realize that
this method is not a viable way to determine
the $R$ dependence of $\delta$: the relative error,
$\Delta\delta$,
of the numerical integration is intolerably
large and the two fit
results also differ by a factor of about $1.5$.
This, of course, is related to the large weight with which 
large $n_{\perp}$ points contribute
to Eq.~\ref{b12}. For $R\geq 10$ the data
is well described, both by a dipole and by a
Gaussian parametrization
for our $n_{\perp}$ window within statistical
errors, and yet the two
parametrizations differ substantially at large $n_{\perp}$.

The data on the numerically integrated
widths for physical distances below $.5$~fm (the largest separation at
which numerical integration of the energy density data could be
performed) exhibits that the energy density values fall onto the line
$\delta=R/2$ while the action density values are significantly larger.
This tendency has also been observed in Ref.~\cite{sommer} and is
consistent with our fit results (Tabs.~\ref{et1} and \ref{et3}).

An alternative approach
to study the functional dependence,
$\delta(R)$, is to constrain the center plane analysis
to the results of the differential action sum rule.
In addition, we apply a geometric method that will
correlate results from different $R$-values to the extent that we
end up with reduced relative errors and all uncertainty cast
into a large overall scale error.
To quote the assumptions:
\begin{enumerate}
\item
Accretion of additional energy and action when pulling the
sources apart is localized in
the center plane.
\item
At sufficiently large $R$, the change of the transverse shape
under variations of $R$ can be absorbed into two (independent) scale
transformations.
\end{enumerate}
Assumption (1) has been verified from our data
in Section~\ref{DFT} for distances
$r>.75$~fm and $\beta\geq 2.5$. Within our statistical errors,
assumption (2) is also fulfilled in this region, according to
the fit results in Tabs.~\ref{et3} and \ref{et4}.

In this case, we can define:
\begin{equation}
\label{aaa1}
\delta^2=\gamma\frac{A}{\pi h}\quad.
\end{equation}
$A$ is the area below the curve. It can be fixed by the sum rules
with high accuracy. For the action density we obtain
$A$ from
\begin{equation}
A=\tilde{B}^{-1}(\beta)(V(R)-V(R-1))\quad.
\end{equation}
Note, that here we have taken an asymmetric derivative of the
potential.
In case of the energy density, $A$ directly equals the
force, up to a renormalization constant.
$h$ denotes the action/energy density
in the middle of the tube (${\mathbf n}={\mathbf 0}$).
$\gamma$ is a geometry factor. Depending on
the parametrization it can take the following values\footnote{Note,
that these values only apply to the infinite volume case. At large
$\delta/L_S$, they tend to be smaller.}:
\begin{itemize}
\item $\gamma=\frac{1}{2}$ for a distribution, constant for
$n_{\perp}<n_{\max}$ and {\em zero} outside of this circle,
\item $\gamma=1$ for a Gaussian shape,
\item $\gamma=2$ for a dipole shape and
\item $\gamma=3$ for an $\exp(-c|n_{\perp}|)$ shape.
\end{itemize}
By employing the definition Eq.~\ref{aaa1},
a large portion of the error on the width is cast into the
(overall) uncertainty of the geometry factor $\gamma$.

In addition to data, obtained by use of the other methods, we have
included the data from this geometric method into Fig.~\ref{Figw1}
for the case $\gamma=1$ (triangles). The differences between these
points and the Gauss fits (crosses) reflect the fact that the large
$n_{\perp}$ data is not well approximated by the Gaussian form.
Remember, that this very effect has also led to
an underestimation of the force in Fig.~\ref{Figsumrule}. 

In Fig.~\ref{Figw2}, we display our geometric results
($\gamma=1$) for the action and energy densities at $\beta=2.5$ and
$\beta=2.635$, scaled in units of the string tension. The dashed
vertical line denotes the distance $.75$~fm above which the
geometrical approximation is justified. As can be seen, the data
exhibits scaling even below this limit. The width of the energy
density starts out to be smaller than the width of the action density,
as has been observed in Ref.~\cite{sommer},
but increases faster. It then reaches
the same magnitude as the action density
width, before it disappears under the noise level
at about $.8$~fm.

Above $r=1$~fm the action density data is in agreement with a constant
value
\begin{equation}
\delta_{\sigma}a\approx 1.17\sqrt{\gamma/\kappa}\approx
0.52\:\mbox{fm}\sqrt{\gamma}\quad,
\end{equation}
where we expect $\gamma$ to take values
between {\em one} and {\em two}.
Logarithmic fits to the $r>1$~fm data according to
the string picture expectation,
Eq.~\ref{a18},  yield values
\begin{equation}
R_0\sqrt{K}<1/3\quad.
\end{equation}
No lower limit is imposed on the cut-off since the data is also in
agreement with a constant. A curve with parametrization
$R_0\sqrt{K}=1/3$ and $\delta_0\sqrt{K/\gamma}=0.5$ is indicated in
Fig.~\ref{Figw2}.

The strong parametrization dependence of the
r.m.s.\ width, $\delta$, is reflected in the difference by a factor of
about 1.5 between the Gaussian and the dipole parametrizations (and by
the different geometry factors, $\gamma$). 
The half width,
$\rho$, is much less sensitive to the parametrization:
both forms are
valid interpolations of the data in the small $n_{\perp}$ region.
For the two
parametrizations, $\rho$ can be connected to $\delta$ by the
following relations:
\begin{eqnarray}
\rho&=&2\delta\sqrt{\ln 2}\approx
1.67\,\delta\quad\qquad\mbox{(Gauss)}\quad,\\\label{dcon}
\rho&=&2\delta\sqrt{2^{1/3}-1}\approx
1.02\,\delta\quad\mbox{(dipole)}\quad.
\end{eqnarray}

\begin{figure}[htb]
\begin{center}
\leavevmode
\epsfxsize=12cm\epsfbox{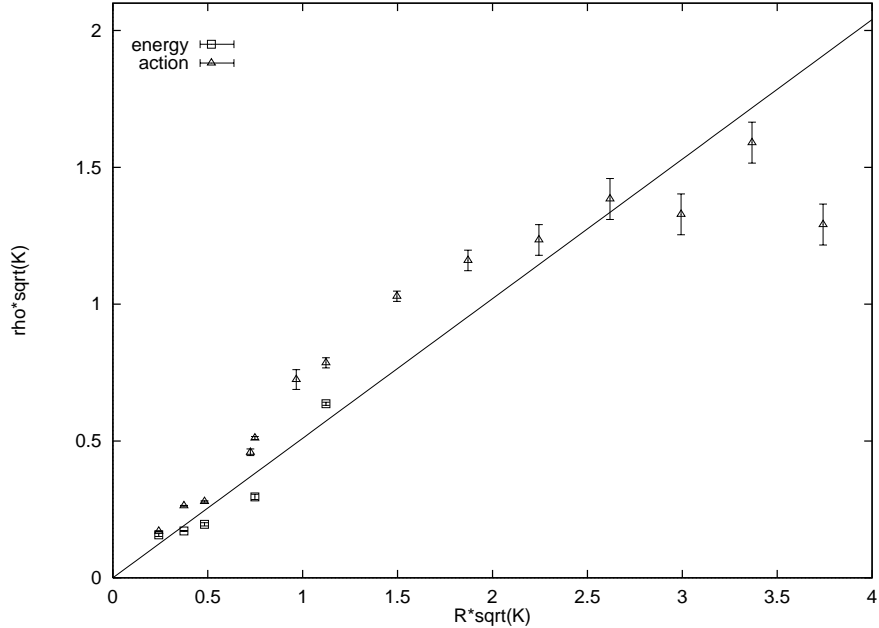}
\end{center}
\caption[ ]
{\em The half widths of the 
energy and action flux tubes
at $\beta=2.5$ and $\beta=2.635$, $\rho$, against $R$ in units of the
string tension. The line corresponds to the small $R$ dipole expectation.}
\label{Figlast}
\end{figure}

The resulting half widths for energy and action
densities in units of the string tension
are displayed in Fig.~\ref{Figlast}.
We have attempted a finite volume
correction to the dipole results by fitting the data to the functional
form, described in Appendix~\ref{ap4} (Eq.~\ref{fform}),
and subsequently converting the resulting
$\delta_{\infty}$ values into $\rho$ via Eq.~\ref{dcon}.
This amounts to a reduction of $\delta$ by less than $10\%$.
Differences between the uncorrected dipole data and the
Gauss results (up to $6\%$) reflect the systematic uncertainty
due to the form of
the interpolating curve.
We observe nice scaling between both $\beta$
values. We also confirm the width of the energy flux tube to be smaller
than the width of the action flux tube for distances below $.5$~fm.
Both densities increase till $r\approx 1.1$~fm.
The action density saturates at the level $\rho\approx .7$~fm.

We conclude that the data beyond $1$~fm is in agreement with a
constant but does not contradict the expected string picture
behaviour either. However, the ultra violet cut-off, $r_0^{-1}$, of
the effective string
theory is found to be larger than $3\sqrt\sigma$ or $1.3$~GeV. This
has to be compared with a lattice resolution of $2.35$ or $3.64$ GeV
at $\beta=2.5$ and $\beta=2.635$, respectively. Thus, we are not too
far away from the limit of applicability of the string theory.

\section{Summary and conclusions}

We have demonstrated that Wilson loop plaquette correlations offer a
viable access to a lattice study of the flux tube problem on the required
length scale of one to two fm.  

Prior to a workable application of
this tool one must ascertain an essential improvement of the lattice
observation technique:
the crucial ingredient of our method is the smearing of the parallel
transporter within the bilocal $Q\bar{Q}$ creation operator. This
secures a controlled ground state preparation of long flux tubes
within few lattice time slices.
Smearing is combined with integration on the
time-like links of the Wilson loop to cut noise.

As a result, we can observe flux formation in the action density over
lengths well beyond 1.5~fm with spatial resolution .05~fm. We find
that --- due to a center group symmetry of the Wilson loop ---
finite size effects remain well below the level of accuracy reached
in the present simulation, at least as long as $L_S$ is kept larger than 
1.3 fm and $R\leq \frac{3}{4}L_S$. In particular, there are
manifestly no 
effects of $L_S$-periodic distortions of the field distribution
or potential due to mirror
sources\footnote{It goes without saying
that an approach based on the measurement of Polyakov lines would neither 
be amenable to such an improvement programme towards ``early T
asymptotics'' nor would it be safe from $L_S$-periodic effects
{}from mirror sources.}.
This implies that we can safely
accommodate a flux tube of length $1.9$~fm on our largest lattice of
volume $(2.7$~fm$)^4$!  The energy and action densities exhibit the expected
scaling behaviour, and are consistent with the potential measurements
through Michael's sum rules.

At small distances the flux tube is corrupted by lattice artefacts,
which can be understood in terms of lattice perturbation theory. This
holds in particular for the self energy peak around the sources, whose
non-scaling behaviour is well in accord with perturbative expectations.

The transverse r.m.s.\ width of the action flux distribution
in the midplane between the sources rises with
source separation, $r$, until it reaches a rather constant level 
for separations between $1$ and $2$~fm.
The physical value for this constant remains model dependent
and ranges between $.5$ and $.75$~fm, as
we estimated from a set of transverse profiles,
supplementing our measurements with various plausible assumptions on
the large
$n_{\perp}$ behaviour.
For the half width we find a plateau value of $\rho\approx .7$~fm.
In the preasymptotic domain, the action width is observed to
rise by a factor
six, distinctly majoring the width of the energy distribution, before
both reach their (common ?) plateau values. 
A logarithmic increase
as suggested by string pictures 
for the ``asymptotic'' $R$ region is consistent with
our data, suggesting a rather large ultra violet cut-off on
inverse wavelengths in effective string models, $r_0^{-1}>1.3$~GeV.

In the range $r\geq .75$~fm,
we observe a
remarkable stability of long flux tubes in the sense that field
accretion exclusively occurs in the center plane of the tube,
as the sources are
further pulled apart. This is another important quantitative support
for the flux tube picture. The issue of establishing a definite
tube profile (like Gaussian transverse shape) will remain a rather
elusive subject for any numerical approach like ours. 

The present research can readily be generalized to the situation of
more than two static sources, like three quark sources in
$SU(3)$~\cite{sommer_wosiek} or the case of ``nuclear chemistry'', with two
quark-antiquark pairs in $SU(2)$. The latter has been studied recently by
Green and coworkers~\cite{green_michael} in the context of hadronic
potentials, while the former will help to answer interesting questions
related to the three-body character of colour forces in the proton.
Work along this line is in progress.

The methods described here should also be useful in the quantitative
studies of the confinement mechanism in the maximal Abelian
gauge~\cite{suzukirev,suzuki2,stack1}.

During completion of this work, we have received a preprint of Haymaker
et al.~\cite{haynew}. They work at $\beta$-values up to 2.5 and
refrain 
{}from applying ground state enhancement techniques. Instead, they
attempt $T$-extrapolations on derived flux tube properties. This enforces
a smaller $R$ range and implies less control on systematic effects.

\vskip .8 cm {\bf Acknowledgements.} 
G.B.\ thanks R.~Sommer and B.~Bunk for
inspiring and helpful discussions. M.~L\"uscher is acknowledged  for
suggesting the center group idea. We are grateful to the Deutsche
Forschungsgemeinschaft for providing the Datavault to
our 256 MByte CM-2 parallel computer, on which the $16^4$ and $32^4$
lattices were computed. Moreover, we thank Prof.  Rollnik and Dr.
V\"olpel for their support to simulate the $48^3\times 64$ lattice on
the 2 GByte CM-5 at the Gesellschaft f\"ur Mathematik und
Datenverarbeitung within the HLRZ setup.

\newpage
\begin{appendix}
\noindent {\LARGE\bf Appendix}
\section{Weak coupling expansion of field distributions}
\label{ap1}
In this appendix, we recall the one gluon exchange approximation to
the lattice potential and compute the leading order perturbative
contribution to the electric energy distribution.

With the lattice gluon propagator in Feynman-t'Hooft gauge,
\begin{equation}
\label{a6}
a,\mu\stackrel{k}\longrightarrow b,\nu
\quad
:\quad\frac{\delta_{ab}\delta_{\mu\nu}}{\sum_{\mu}\bar{k}_{\mu}^2}\quad,\quad
\bar{k}_{\mu}=2\sin k_{\mu}/2\quad,
\end{equation}
a weak coupling expansion of the Wilson loop yields
\begin{equation}
W({\mathbf R},T)=1+C_Fg^2\sum_{\tau,\tau'=0}^{T}
\left(G({\mathbf R},\tau'-\tau)-G({\mathbf 0},\tau'-\tau)\right)\quad,
\end{equation}
where just terms, extensive in $T$, have been kept and the leading
term {\em one} is the expectation value of the loop with the interaction
switched off. Note, that we have neglected the zero momentum
contribution in the calculation that is suppressed by 
a factor $1/(L_S^3L_T)$.
The colour factor $C_F$ can be calculated by contracting the colour
indices of the $SU(N)$ generators $T_a$ ($a=1,\ldots,N^2-1$):
\begin{equation}
C_F=\frac{1}{N}\tr\{T_aT_b\}\delta_{ab}=\frac{N^2-1}{2N}\quad.
\end{equation}

Fourier transforming Eq.~\ref{a6} yields
\begin{eqnarray}
\label{a11}
G(n)&=&\sum_{k\ne 0}\frac{e^{ikn}}{\bar{k}_{\mu}^2}\quad,\\\nonumber
k_i &=& \frac{2\pi}{L_S} m_i\quad, \qquad
m_i=-\frac{L_S}{2}+1,\ldots,\frac{L_S}{2}\quad,\\\nonumber
k_4 &=& \frac{2\pi}{L_T} m_4\quad, \qquad
m_4=-\frac{L_T}{2}+1,\ldots,\frac{L_T}{2}
\end{eqnarray}
for the real space gluon propagator on a finite lattice.
With
\begin{equation}
\label{a10}
G_L({\mathbf R})=\sum_{\tau}G({\mathbf R},\tau)=
\sum_{{\mathbf k}\ne {\mathbf 0}}\frac{e^{i{\mathbf k}{\mathbf R}}}
{\sum_i\bar{k}_i^2}
\end{equation}
and $V({\mathbf R})=-\lim_{T\rightarrow\infty}\ln\left(W({\mathbf
R},T)\right)/T$ one
obtains
\begin{equation}
V({\mathbf R})=-C_Fg^2\left(G_L({\mathbf R})-G_L({\mathbf 0})\right)\quad.
\end{equation}

By construction, the weak coupling expansion of
the quantity (Eq.~\ref{wdef}),
\begin{equation}
\label{a5}
\langle\Box\rangle_{\cal W}=\langle\Box\rangle\left(
\frac{\langle\Box{\cal W}\rangle}{\langle\Box\rangle\langle{\cal W}\rangle}
-1\right)\quad,
\end{equation}
involves only interactions between the plaquette and the Wilson
loop. All self-interactions of the plaquette
and Wilson loop are cancelled
by the denominator.

An expansion of the plaquette yields
\begin{equation}
\langle\Box\rangle=1-c_1g^2+\cdots
\end{equation}
with $c_1=2C_fg^2(G(0)-G(1))=C_Fg^2/4$ on symmetric lattices.
If we are interested in the leading order behaviour only, the
plaquette can be approximated by {\em one}. However, at realistic values of
the coupling higher order corrections are large. At $\beta=2.5$
we find for example $\langle\Box\rangle=.65198$.
This observation inspired Parisi to formulate a programme of mean field
improved lattice operators~\cite{parisi}. The idea is to split every
lattice operator into a part that corresponds to (discretization
dependent) fluctuations on the
ultra violet lattice scale 
and a {\em physical} infra red part. 

More recently, the deviations of
lattice results from perturbative expansions in terms of the bare
lattice coupling parameter have been explained as being due
to large contributions
{}from tadpole diagrams~\cite{mackenzie}. This circumstance has revived
the interest in mean field or tadpole improved lattice
perturbation theory and operators. The hope is to
suppress ultra violet contaminations
by dividing every
link in a given lattice operator by its Monte Carlo mean field value $u_0$.
This is supposed to procure early asymptotic
scaling and reliable perturbation theory predictions.

A popular choice of $u_0$ is the fourth root of the
average plaquette. Following this procedure, we should divide the
expression on the r.h.s.\ of Eq.~\ref{a5} by the average plaquette.
However, in the end, we are interested in the combination
$\beta\langle\Box\rangle_{\cal W}$ only. Since the plaquette in the action
$S_W-\beta U$ has also to
be divided by its mean field value, $\beta$ is replaced by a mean
field coupling $\beta_{MF}=\beta\langle\Box\rangle$. Performing both
replacements, the $\langle\Box\rangle$ contributions cancel. In this
spirit, the definition of action and energy densities in
Section~\ref{CF} represents already in itself
a tadpole improved definition.
Keeping in mind that in the last step our
operator will be multiplied by $\beta$, it is justified
to neglect the multiplicative $\langle\Box\rangle$ factor in
Eq.~\ref{a5} even in a region where $g^2$ depending deviations from
{\em one} are not small.

The two loops being disconnected in
colour space, only singlets can be exchanged. Thus, 
we expect an exchange of two gluons as the leading order contribution.
Technically, this can be seen as follows:
for computation of the product of two (real) traces, both possible
relative orientations of the Wilson loop and the plaquette have to be
averaged over. Thus, exchanges of single (bare or dressed) gluons
cancel. The same holds for a
triple gluon vertex that can only be attached with two legs to one loop and
with one leg to the other. 
Due to the Lorentz structure of the
propagator Eq.~\ref{a6}, magnetic plaquettes cannot interact
by a direct
exchange of gluons with the timelike links of the Wilson loop.

The colour factor of the two gluon exchange between the
disconnected loops turns out to be:
\begin{equation}
\frac{1}{N}\tr \{T_aT_b\}\frac{1}{N}\tr\{T_cT_d\} \delta_{ac}\delta_{bd}
 =\frac{1}{2N}\frac{1}{N}\tr\{T_aT_b\}\delta_{ab}=\frac{C_F}{2N}
\end{equation}
By squaring the one-gluon exchange contribution, dividing
the expression by a factor
{\em two} to avoid an overcounting of gluon exchanges, and
performing the $T$ integration, we
obtain (Again, only terms extensive in $T$ have been kept.):
\begin{eqnarray}
\langle U({\mathbf n})_{i4}\rangle_{\cal W}&=&y_i({\mathbf n})\nonumber\\
&=&\frac{C_Fg^4}{4N}\left(G_L({\mathbf n}-{\mathbf r}_1+{\mathbf e}_i)
-G_L({\mathbf n}-{\mathbf r}_1)\right.\label{a8}\\\nonumber
&&\quad\quad -\left.G_L({\mathbf n}-{\mathbf r}_2+{\mathbf e}_i)
+G_L({\mathbf n}-{\mathbf r}_2)\right)^2\quad,
\end{eqnarray}
where the sources are placed at the positions
${\mathbf r}_1=\frac{R}{2}{\mathbf e}_1$.
and ${\mathbf r}_2=-\frac{R}{2}{\mathbf e}_1$.

After averaging over the two plaquettes used for construction of the
electric field operator and multiplying by $2\beta/a^4$, we end up
with
\begin{equation}
\label{a9}
a^4\langle E_i^2({\mathbf n})\rangle_{|0,R\rangle-|0\rangle}
=g^2C_F\frac{y_i({\mathbf n})
+y_i({\mathbf n}-{\mathbf e}_i)}{2}
\end{equation}
while
\begin{equation}
a^4\langle B_i^2({\mathbf n})\rangle_{|0,R\rangle-|0\rangle}
={\cal O}(g^4)\quad.
\end{equation}
Many of the higher order diagrams contribute to ${\cal B}$ as well as to
${\cal E}$. Thus, we would expect a partial
cancellation of higher order effects
in the energy density
\begin{equation}
\epsilon_R({\mathbf n})=
\frac{{\cal E}_R({\mathbf n})-{\cal B}_R({\mathbf n})}{2}\quad.
\end{equation}

{}From
\begin{equation}
\label{a7}
\sum_{{\mathbf n},i}y_i({\mathbf n})=2\left(G_L(0)-G_L(R)\right)\quad,
\end{equation}
we obtain
\begin{equation}
a^4\sum_{\mathbf n}\epsilon_R({\mathbf
n})=g^2C_F(G_L(0)-G_L(R))+{\cal O}(g^4)\approx V(R)\quad.
\end{equation}
Note, that to order $g^2$ the action density equals the energy
density. However, the action density is expected to deviate
much more from the leading order perturbative expectation since 
higher order electric and magnetic
contributions are added and no cancellations of diagrams
occur.

Perturbation theory yields (up to a divergent self-energy
part)
\begin{equation}
v(r)=-C_Fg^2\frac{1}{4\pi r}
\end{equation}
for the continuum potential.
The associated electric field is given (up to a colour factor) by
$g^{-1}\nabla\left(v({\mathbf x}+r/2{\mathbf e}_1)-v({\mathbf x}
-r/2{\mathbf e}_1)\right)$. In the continuum limit the differences in
Eq.~\ref{a8} will be replaced by derivatives, yielding exactly this
expression. After squaring and expressing the result in lattice units,
one obtains:
\begin{equation}
\label{a13}
\epsilon_R^{(c)}({\mathbf n})=
g^2C_F\frac{1}{(4\pi)^2}
\sum_i\left(\frac{n_i-\delta_{i1}R/2}{|{\mathbf n}-{\mathbf
e}_1R/2|^3}-\frac{n_i+\delta_{i1}R/2}{|{\mathbf n}+{\mathbf
e}_1R/2|^3}\right)^2+\cdots
\end{equation}
which is just the continuum limit of Eq.~\ref{a9}.

\section{Action sum rule}
\label{ap2}
In order to derive the action sum rule we start from the definition
(${\cal W}=W(R,T)$, Dirac indices and spatial position are suppressed):
\begin{equation}
\label{defi}
\langle \Box\rangle_{{\cal
W},2}=\frac{1}{T}\sum_{\tau=0}^{L_T-1}
\left(\frac{\langle \Box(\tau){\cal W}\rangle}{\langle{\cal W}\rangle}
-\langle\Box\rangle\right)=
\frac{2}{T}\sum_{S=0}^{L_T/2-1}\langle\Box(S)\rangle_{\cal W}\quad.
\end{equation}
In Eq.~\ref{decom}, the spectral decomposition of the argument of the
sum for $0\leq S< T/2$ has been carried out. For the plaquette position
outside the loop, i.e. $S\geq T/2$, we obtain:
\begin{equation}
\label{decom2}
\langle\Box(S)\rangle_{\cal W}=\mbox{const.}\times
e^{-E_1(L_T-T)/2}\cosh\left(E_1(L_T/2-S)\right)+\cdots\quad.
\end{equation}
The signal is suppressed with the temporal distance of the
plaquette insertions from the Wilson loop, $S-T/2$, times the mass gap,
$E_1=m_{A_1^+}a\approx 3\sqrt{K}$,~\cite{michaelglue} in the exponent. 

After summing over all $S$, we obtain:
\begin{eqnarray}
\label{expa2}
\langle\Box\rangle_{{\cal
W},2}&=&\langle\Box\rangle_{|0,R\rangle-|0\rangle}\nonumber\\
&+&\frac{b}{T}+\frac{|d_1|^2}{|d_0^2|}e^{-\Delta VT}+{\cal O}
\left(\frac{e^{-\Delta VT}}{T}\right)\quad.
\end{eqnarray}
We have made use of the fact that the off-diagonal (i.e.\ $S$ dependent)
pollutions, inside and outside the loop are incomplete geometric
series. The summation gives, apart from the constant parts,
only multiplicative $1-e^{-n\Delta V(T+1)}$ or $1-e^{-nE_1(L_T-T+1)}$
factors. The constant $b_{\mu\nu}({\mathbf n})$
is the sum of all off-diagonal coefficients
{}from the expansions Eq.~\ref{decom} and Eq.~\ref{decom2} of
$\langle\Box\rangle_{\cal W}$, weighted by corresponding
coefficients $(1-e^{-\Delta V_i})^{-1}$ or $(1-e^{-E_i})^{-1}$.

Together with 
\begin{equation}
\langle{\cal W}\rangle=\int\!{\cal D}U\,{\cal W}e^{\beta
U}\quad,\quad U=\sum_{n,\mu>\nu}U_{\mu\nu}(n)\quad,
\end{equation}
and the decomposition of the Wilson loop (Eq.~\ref{expa1}) we
obtain from Eq.~\ref{expa2}:
\begin{eqnarray}
\sum_{{\mathbf n},\mu>\nu}
\langle U_{\mu\nu}({\mathbf n})\rangle_{{\cal W},2}
&=&
\frac{1}{T}\left(\frac{\langle U{\cal
W}\rangle}{\langle{\cal W}\rangle}-\langle U\rangle\right)\nonumber\\
&=&
\frac{1}{T}\frac{\partial}{\partial\beta}\ln\langle{\cal W}\rangle\\\nonumber
&=&\frac{1}{T}\frac{\partial\ln|d_0|^2}{\partial\beta}
-\frac{\partial V}{\partial\beta}
-\frac{\partial\Delta
V}{\partial\beta}\frac{|d_1|^2}{|d_0|^2}e^{-\Delta VT}
-\cdots
\end{eqnarray}
Note, that the equalities are exact!
Thus, in the expansion of the
derivative of the Wilson loop the same terms appear as in
Eq.~\ref{expa2}.

A comparison of the $1/T$ coefficients between the above equation
and $b$ of Eq.~\ref{expa2} yields
\begin{equation}
B=\sum_{{\mathbf n},\mu>\nu}b_{\mu\nu}({\mathbf n})
=\frac{\partial\ln|d_0|^2}{\partial\beta}\quad.
\end{equation}
{}From the estimate for ground state overlaps of unsmeared operators
Eq.~\ref{abg} we obtain
\begin{equation}
B=-R\frac{\partial V_0}{\partial\beta}\approx\beta^{-1}V_0R
\end{equation}
for large $R$ and weak coupling. The monotonous increase
of the ground state overlap 
at fixed $R$ with $\beta$ is confirmed
in the present simulation. Therefore, $B$ is positive.
For smeared operators, $V_0$ is
replaced by some constant $f$ that is small compared to all $R^{-1}$,
such that the exponential can be expanded and the ground state
overlaps
$|d_0(R)|^2$ exhibit the linear behaviour of Fig.~\ref{Fig1}.
$f$ is expected to be
proportional to $g^2$ to the lowest order such that $B\approx\beta^{-1}fR$.
Under the
assumption that $d_1$ dominates other excited state overlaps,
we obtain
$|d_1(R)|^2\approx fR$.
Eq.~\ref{defi} can also serve as a definition for colour field
measurement operators. However, we have preferred to use
$\langle\Box\rangle_{\cal W}$ instead, due to the better asymptotic
behaviour: excited states are suppressed by factors proportional to
$\sqrt{R}e^{-\Delta VT/2}\approx\sqrt{R}e^{-\frac{\pi}{R}T}$
instead of $R/T$ only.

{}From
Eqs.~\ref{act1}, \ref{act2}, \ref{wdef}, \ref{act3} and \ref{act7},
one obtains:
\begin{equation}
\sum_{{\mathbf n},\mu>\nu}
\langle U_{\mu\nu}({\mathbf n})\rangle_{{\cal W},2}\\
\stackrel{T\rightarrow\infty}{\longrightarrow}
\frac{a^4}{\beta}\sum_{\mathbf n}\frac{1}{2}\left({\cal
E}({\mathbf n}) - {\cal B}({\mathbf n})\right)
=\frac{a^4}{\beta}\sum_{\mathbf
n} \sigma_R({\mathbf n})\quad.
\end{equation}
By carefully comparing the coefficients of the expansions one finds
many (exact) ``sum rules''. In the following we list three such
examples.
\begin{eqnarray}
\label{actsum}
\sum_{\mathbf n}a^3\sigma_R({\mathbf n})&=&
-\frac{\beta}{a}\frac{\partial V}{\partial\beta}\quad,\\
\sum_{\mathbf n}a^3\left(\sigma'_R({\mathbf n})
-\sigma_R({\mathbf n})\right)&=&
-\frac{\beta}{a}\frac{\partial\Delta V}{\partial\beta}\quad,\\
\sum_{\mathbf n}a^3\sigma_{\!A_1^+}({\mathbf n})&=&
-\frac{\beta}{a}\frac{\partial E_1}{\partial\beta}\quad.
\end{eqnarray}
$\sigma'_R$ denotes the
action density distribution of the first excited $Q\bar{Q}$
state without angular momentum, $\sigma_{\!A_1^+}$ is the action density
distribution in presence of the lightest glueball state.

The ground state potential consists of a constant physical part
$v(R)$ and a self energy contribution $V_0$ which diverges in the
continuum limit:
\begin{equation}
V(R)=av(R)+V_0
\end{equation}
By using this decomposition, we obtain from Eq.~\ref{actsum} the
action sum rule
\begin{equation}
\label{a3}
\sum_{\mathbf n}a^3\sigma_R({\mathbf n})=
-\frac{\partial\ln a}{\partial\ln\beta}v(R)-\frac{1}{a}\frac{\partial
V_0}{\partial\ln\beta}\quad.
\end{equation}

\section{Energy sum rule}
\label{ap3}
The derivation of the energy sum rule, though the more intuitive one,
turns out to be more complicated. We start
{}from the decomposition of the Wilson
action into a spatial and a temporal part
\begin{equation}
S_W=-\beta_tU_t-\beta_sU_s\quad.
\end{equation}
In the following, the spatial and temporal lattice spacings will be
called $a_s$ and $a_t$, respectively. The asymmetry is defined by
$\xi=a_s/a_t$.
Following the steps of the previous section, one finds:
\begin{eqnarray}
\frac{a_s^4}{2\beta}\sum_{\mathbf n}{\cal E}_R({\mathbf n})
&\stackrel{T\rightarrow\infty}{\longleftarrow}&
\xi\sum_{{\mathbf n},i}\langle
U_{i4}({\mathbf n})\rangle_{{\cal
W},2}
=\frac{\xi}{T}
\frac{\partial}{\partial\beta_t}\ln\langle{\cal W}\rangle\quad,\\ 
\frac{a_s^4}{2\beta}\sum_{\mathbf n}
{\cal B}_R({\mathbf n})
&\stackrel{T\rightarrow\infty}{\longleftarrow}&
-\sum_{{\mathbf n},i>j}\langle
U_{ij}({\mathbf n})\rangle_{{\cal
W},2}=-\frac{1}{T}\frac{\partial}{\partial\beta_s}\ln\langle{\cal W}\rangle
\quad.
\end{eqnarray}

A weak coupling expansion~\cite{karsch}
relates the anisotropic lattice couplings to
the isotropic coupling $\beta(a_s)$:
\begin{eqnarray}
\label{a53}
\beta_s\xi&=&\beta+2Nc_s(\xi)+{\cal O}(\beta^{-1})\quad,\\
\label{a54}
\beta_t\xi&=&\beta+2Nc_t(\xi)+{\cal O}(\beta^{-1})\quad.
\end{eqnarray}
The coefficients fulfil the relations:
\begin{equation}
c_s(1)=c_t(1)=0\quad,\quad c_s'(1)+c_t'(1)=b_0=\frac{11N}{48\pi^2}
\quad.
\end{equation}
The derivatives of the coefficients
have been calculated by Karsch in Ref.~\cite{karsch}. For $SU(2)$
the result is
\begin{equation}
c_s'=c_s'(1)=.11403\ldots\quad,\quad c_t'=c_t'(1)=-.06759\ldots\quad.
\end{equation}

After expressing the derivatives in respect to the asymmetric
couplings by derivatives in respect to $\beta$ and $\xi$
and taking $\xi=1$, we end up
with:
\begin{eqnarray}
\label{a1}
a^4\sum_{\mathbf n}{\cal E}_R({\mathbf n})
&=&
-\frac{\partial
V}{\partial\ln\beta}\left(1-\frac{N}{\beta}b_0\right)+V
\left(1-\frac{N}{\beta}(c_t'-c_s')\right)\\\label{a2}
a^4\sum_{\mathbf n}{\cal B}_R({\mathbf n})
&=&\frac{\partial
V}{\partial\ln\beta}\left(1+\frac{N}{\beta}b_0\right)+V
\left(1-\frac{N}{\beta}(c_t'-c_s')\right)
\end{eqnarray}
Subtracting both expression yields the (exact) action sum rule
Eq.~\ref{actsum}.

Adding Eqs.~\ref{a1} and \ref{a2} and dividing by a factor {\em two}
yields the energy sum rule
\begin{eqnarray}
\sum_{\mathbf n}a^3\epsilon_R({\mathbf n})&=&\frac{1}{a}\left(
V\left(1-\frac{N}{\beta}(c_t'-c_s')\right)+\frac{\partial
V}{\partial\ln\beta}\frac{N}{\beta}b_0+\cdots\right)\\
&=&\left(v+\frac{V_0}{a}\right)\left(1-\frac{N}{\beta}
(c_t'-c_s')\right)\nonumber\\\label{a55}
&+&\left(v\frac{\partial\ln a}{\partial\ln\beta}
+\frac{1}{a}\frac{\partial
V_0}{\partial\ln\beta}\right)
\frac{N}{\beta}b_0
+\cdots
\end{eqnarray}
The energy sum rule is not exact due to the perturbative origin of the
relation between the symmetric and asymmetric couplings
(Eqs.~\ref{a53}, \ref{a54}). Of course, it would be preferable to
measure the corresponding derivatives of $\ln a$
directly on the lattice instead.

Note that the coefficient of the last term in Eq.~\ref{a55}
is identical to
the action sum Eq.~\ref{a3}. The factor
$\partial\ln a/\partial\ln\beta$
appearing in front of $v$ within this term carries an anomalous
dimension (as the action does), cancelling a $\beta^{-1}$ factor.
Thus, an additional factor $-v/4$ seems to
survive the continuum limit $a\rightarrow 0$.
Its origin is an incomplete resummation of the series: the
non-perturbatively determined coefficients contribute to
all orders in $\beta^{-1}$.
The order $\beta^{-1}$ term yielding the above $-v/4$ contribution
has to be cancelled by other anomalous terms appearing in
higher orders of the expansion.
However, if consistently cutting the expansion at order $\beta^{-2}$ by
expanding the potential perturbatively, the coefficient {\em one}
is reproduced in accord to the expected continuum limit.

\section{Finite Size Corrections}
\label{ap4}
In this appendix, we elaborate details on the computation
of finite size corrections on the action/energy density
distributions within the center plane. These FSE are mainly
due to the periodicity of the measurement operator,
\begin{equation}
O_R(0,n_2,n_3)=O_R(0,L_S-n_2,L_S-n_3)\quad,
\end{equation}
at a given $Q\bar{Q}$ separation, ${\mathbf R}=R{\mathbf e}_1$. This
mirror source effect should not be confused with contributions from the
replacement $R\rightarrow R\pm nL_S$ which are negligible (as shown in
Section~\ref{FSE}). We also neglect effects from
mirror sources along the $Q\bar{Q}$ axis
which are almost completely screened from the center plane
since the colour field
densities fall off at least as fast as $(|n_1|-R/2)^{-4}$ into the
longitudinal direction.
The effect from mirror copies placed along the transverse directions can be
substantial (depending on the ratio $\delta/L_S$).

We perform
our calculations for two models, namely a dipole transverse shape,
\begin{equation}
f_d(x_{\perp};\infty)=
\frac{c}{\pi}\frac{\delta^2}{\left(\delta^2+x_{\perp}^2\right)^3}\quad,
\end{equation}
and a Gaussian shape
\begin{equation}
f_g(x_{\perp};\infty)=\frac{c}{2\pi\delta^4}
\exp\left(-\frac{x_{\perp}^2}{\delta^2}\right)\quad.
\end{equation}
$f_d(x_{\perp};\infty)$ and $f_g(x_{\perp};\infty)$
are the corresponding (infinite volume) center plane
energy/action density distributions.

As argued above it is justified to neglect
interactions between different pairs of mirror sources.
We also assume that the chromo electric and magnetic
fields on the finite volume can be obtained by superimposing the
(infinite volume) fields of all (pairs of) mirror sources.
Note, that we have to add the fields rather than the action and energy
densities themselves.
{}From the geometry it is clear that the
perpendicular electric and longitudinal magnetic field components
vanish in the center plane. Under the assumption that the
(perpendicular) magnetic
field component is proportional to the
(longitudinal) electric component, we find:
\begin{equation}
\label{fform}
f(x_2,x_3;L_S)=\left(\sum_{n_2,n_3}
g\left(x_2+n_2L_S,x_3+n_3L_S\right)\right)^{\!2}
\end{equation}
with $g(x_2,x_3)=g(x_{\perp})=\sqrt{f(x_{\perp};\infty)}$.
The integrated area can be computed in the following way:
\begin{eqnarray}
A(L_S)&=&\int_{0}^{L_S}\!dx_2dx_3\,f(x_2,x_3;L_S)\\
&=&
\sum_{n_2n_3}\sum_{m_2m_3}
\int_{m_2L_S}^{(m_2+1)L_S}\!\!\!\!\!\!\!\!\!dx_2
\int_{m_3L_S}^{(m_3+1)L_S}\!\!\!\!\!\!\!\!\!dx_3
\,g\left(x_{\perp}\right)
g\left(x_2+n_2L_S,x_3+n_3L_S\right)\nonumber\\
&=&
\sum_{n_2n_3}\int\!d^2\!x_{\perp}\,
g\left(x_2-n_2L_S/2,x_3-n_3L_S/2\right)\nonumber\\
&\quad&\qquad\quad\quad\times
g\left(x_2+n_2L_S/2,x_3+n_3L_S/2\right)\quad.
\end{eqnarray}

In case of a dipole field with (infinite volume) r.m.s.\ width, $\delta$,
we have
\begin{equation}
g_d(x_{\perp})\propto\left(\delta^2+x_{\perp}^2\right)^{-3/2}
\end{equation}
and obtain for the area,
\begin{equation}
A(L_S)\propto\sum_{n_2n_3}\int_0^{2\pi}\!\!\!d\phi\int_0^{\infty}\!\!\!dr\,r
\left(d^2(n_2,n_3)
+r^2\left(1-L_S^2(n_2\cos\phi+n_3\sin\phi)^2\right)\right)^{-3/2}
\end{equation}
with
\begin{equation}
d^2(n_2,n_3)=\delta^2+\frac{L_S^2}{4}(n_2^2+n_3^2)\quad.
\end{equation}
After performing the radial integration, we arrive at:
\begin{eqnarray}
A(L_S)
&\propto&
\sum_{n_2n_3}\frac{1}{d^2}\int_0^{2\pi}\!d\phi\,
\frac{1}{4d^2-L_S^2(n_2\cos\phi+n_3\sin\phi)^2}\\
&=&\frac{\pi}{2}\sum_{n_2n_3}\frac{1}{d^3\sqrt{d^2-L_S^2(n_2^2+n_3^2)/4}}\\
&=&A(\infty)\sum_{n_2n_3}
\left(1+\frac{L_S^2}{4\delta^2}(n_2^2+n_3^2)\right)^{-3/2}
\end{eqnarray}
with $A(\infty)=c/(2\delta^2)$.

For the Gauss fits we have
\begin{equation}
g_g(x_{\perp})\propto\exp\left(-\frac{x_{\perp}^2}{2\delta^2}\right)\quad.
\end{equation}
Thus, we end up with
\begin{eqnarray}
A(L_S)&=&\sum_{n_2n_3}\int\!d^2\!x_{\perp}\,
\exp\left(-\frac{x_{\perp}^2}{\delta^2}\right)
\exp\left(-\frac{L_S^2}{4\delta^2}(n_1^2+n_2^2)\right)\\
&=&A(\infty)\sum_{n_2n_3}
\exp\left(-\frac{L_S^2}{4\delta^2}(n_1^2+n_2^2)\right)\quad.
\end{eqnarray}

In conclusion, the results for both transverse profiles read
\begin{eqnarray}
A(L_S)&=&A(\infty)\sum_{n_2n_3}
\frac{g_d\left(L_S^2(n_1^2+n_2^2)/4\right)}{g_d(0)}
\quad\mbox{and}\\
A(L_S)&=&A(\infty)\sum_{n_2n_3}
\frac{g_g\left(L_S^2(n_1^2+n_2^2)/2\right)}{g_g(0)}
\quad,
\end{eqnarray}
respectively, with $g(x_{\perp})=\sqrt{f(x_{\perp};\infty)}$
and $A(\infty)=c/(2\delta^2)$.
For the typical dipole r.m.s.\ width $\delta/L_S=6/32$ we find an
increase of the area by $43\%$ due to the finite volume
while the corresponding Gaussian
result ($\delta/L_S\approx 4/32$) is only affected by
$5\times 10^{-7}$. Notice, that the infinite volume $\delta$ can be
obtained by fits of the form Eq.~\ref{fform} from finite volume data.
\end{appendix}

\end{document}